\documentclass[reprint,amsfonts, amssymb, amsmath, showkeys,aps,pra, superscriptaddress, twocolumn,longbibliography,nofootinbib]{revtex4-2}

\newcommand{\pdftitle}{Moments of Quantum Channel Ensembles}

\providecommand{\pdftitle}{}
\providecommand{\figurespath}{./figures/}
\providecommand{\tikzpath}{./tikz/}
\providecommand{\texpath}{./tex/}

\usepackage[left=16mm,right=16mm,top=16mm,bottom=16mm,columnsep=15pt]{geometry}
\usepackage{adjustbox}
\usepackage{placeins}
\usepackage[toc,page,titletoc]{appendix}

\usepackage{xparse}

\usepackage{float}
\makeatletter
\let\newfloat\newfloat@ltx
\makeatother
\usepackage{graphics}
\usepackage{graphicx}
 \usepackage{grffile}
\usepackage{minibox}
\usepackage{slashed}
\usepackage{tabularray}

\usepackage[convert-footnotes=true]{notes2bib}
\bibnotesetup{note-name=,use-sort-key=false}

\usepackage[english]{babel}
\usepackage[utf8]{inputenc}
\usepackage[T1]{fontenc}
\usepackage{selinput}
\usepackage[normalem]{ulem}
\usepackage[shortlabels]{enumitem}

\usepackage{lipsum}
\usepackage{csquotes}

\usepackage{xcolor}
\definecolor{highlight}{HTML}{07223E}
\definecolor{highlightone}{HTML}{2a788e}
\definecolor{highlighttwo}{RGB}{235,8,138}
\definecolor{highlightthree}{HTML}{7ad151}
\definecolor{highlightfour}{RGB}{68,1,84}
\definecolor{highlightfive}{HTML}{481567}
\definecolor{highlight}{HTML}{07223E}
\colorlet{outline}{black}

\usepackage[colorlinks=true,citecolor=blue,linkcolor=magenta]{hyperref}
\usepackage{url}

\graphicspath{{\figurespath}{\tikzpath}{\texpath}}
\makeatletter
\def\input@path{{\figurespath}{\tikzpath}{\texpath}}
\makeatother

\usepackage{tikz}

\usepackage{bbm}
\usepackage{amsmath}
\usepackage{amssymb}
\usepackage{accents}
\usepackage{mathtools}
\usepackage{amsthm}
\usepackage{bbm}
\usepackage{scalerel}
\usepackage{relsize}
\usepackage{bm}
\usepackage{empheq}
\usepackage{stackengine}
\usepackage{thmtools}
\allowdisplaybreaks

\newtheorem{theorem}{Theorem}
\newtheorem{lemma}{Lemma}
\newtheorem{corollary}{Corollary}

\newtheorem{conjecture}{Conjecture}
\newtheorem{proposition}{Proposition}

\newtheorem{definition}{Definition}

\newtheorem{claim}{Claim}

\usepackage{physics}

\newcommand{\rrangle}{\rangle\!\rangle}
\newcommand{\llangle}{\langle\!\langle}

\makeatletter

\DeclarePairedDelimiter\@ceil{\lceil}{\rceil}
\DeclarePairedDelimiter\@floor{\lfloor}{\rfloor}

\DeclareRobustCommand\ceil[1]{\@ceil*{#1}}
\DeclareRobustCommand\floor[1]{\@floor*{#1}}

\DeclareRobustCommand\bra[1]{%
	\@ifnextchar\ket{\br@k@t{#1}}{\br@{#1}}%
}
\newcommand\br@[1]{{\langle{#1}\lvert}}

\DeclareRobustCommand\ket[1]{%
	\@ifnextchar\bra{\k@t{#1}\!}{\k@t{#1}}%
}
\newcommand\k@t[1]{{\lvert{#1}\rangle}}

\newcommand\br@k@t[1]{{\langle{#1}}}

\DeclareRobustCommand\bbra[1]{%
	\@ifnextchar\kket{\bbr@kk@t{#1}}{\bbr@{#1}}%
}
\newcommand\bbr@[1]{{\llangle{#1}\lvert}}

\DeclareRobustCommand\kket[1]{%
	\@ifnextchar\bbra{\kk@t{#1}\!}{\kk@t{#1}}%
}
\newcommand\kk@t[1]{{\lvert{#1}\rrangle}}

\newcommand\bbr@kk@t[1]{{\llangle{#1}}}

\makeatother

\DeclarePairedDelimiter{\tnorm}{\lVert}{\rVert}
\DeclarePairedDelimiter{\texpval}{\langle}{\rangle}

\newcommand{\spacesys}{\mathcal{H}}
\newcommand{\spaceenv}{\mathcal{E}}
\newcommand{\spaceobs}{\mathcal{W}}
\newcommand{\spaceobsbar}{\overline{\spaceobs}}

\newcommand{\dspace}{d}
\newcommand{\dsys}{\dspace_{\vphantom{\spaceenv}{}}}
\newcommand{\denv}{\dspace_{\spaceenv}}
\newcommand{\dobs}{\dspace_{\spaceobs}}
\newcommand{\dobsbar}{\dspace_{\spaceobsbar}}

\newcommand{\channelunitary}[1][]{\mathcal{U}}
\newcommand{\channelnoise}[1][]{#1{\mathcal{N}}}
\newcommand{\channeldepolarize}[2][]{#1{\mathcal{D}}_{#2}}
\newcommand{\unitary}{U}
\newcommand{\channel}{\Lambda}
\newcommand{\channelother}{\Upsilon}

\newcommand{\ensembleunitary}{\mathcal{U}}
\newcommand{\ensembleunital}{\mathcal{C}_{\mathcal{U}}}
\newcommand{\ensemblenoise}{\mathcal{N}}
\newcommand{\ensemblechannel}{\mathcal{C}}
\newcommand{\ensembledepolarize}{\mathcal{D}}

\newcommand{\ensemblehaar}[1][\dsys]{\ensembleunitary({#1})}
\newcommand{\ensemblechaar}[1][\dsys,\denv]{\ensemblechannel(#1)}
\newcommand{\ensembledep}[1][\dsys]{\ensembledepolarize(#1)}
\newcommand{\ensemblevariable}{\mathcal{G}_{\Theta}}

\newcommand{\constant}{C}
\newcommand{\weingarten}[2][]{\chi_{#2}^{#1}}
\newcommand{\mobius}[2][t]{c(#2)}
\newcommand{\catalan}[2][t]{c_{#2}}
\newcommand{\mobiuscatalan}[2][t]{c_{#2}}

\newcommand{\idsys}{I_{\dsys}}
\newcommand{\idenv}{I_{\denv}}

\newcommand{\idobs}{I_{\dobs}}
\newcommand{\idobsbar}{I_{\dobsbar}}

\newcommand{\idpermutations}{I_{t!}}

\newcommand{\permutations}[1][t]{\mathcal{S}_{#1}}
\newcommand{\representations}[1][t]{\mathcal{S}_{\dsys}^{(#1)}}
\newcommand{\normalizations}[1][t]{\hat{\mathcal{S}}_{#1}}
\newcommand{\localizations}[1][t]{[\normalizations[#1]]}

\newcommand{\represent}[1][\dsys]{V_{#1}}

\newcommand{\representation}[2][\dsys]{\represent[#1](#2)}
\newcommand{\normalization}[2][]{\hat{#2}#1}
\newcommand{\localization}[2][]{[\normalization[#1]{#2}]}

\newcommand{\Tau}{\mathcal{T}}
\newcommand{\Epsilon}{\mathlarger{\mathlarger{\epsilon}}}

\renewcommand{\Tr}{{\rm Tr}}

\begin{document}

\title{\pdftitle}

\author{Matthew Duschenes}
\email{Contact author: mduschen@uwaterloo.ca}
\affiliation{Theoretical Division, Los Alamos National Laboratory, 87545 NM, USA}
\affiliation{Institute for Quantum Computing, University of Waterloo, N2L 3G1 ON, Canada}
\affiliation{Perimeter Institute for Theoretical Physics, Waterloo, N2L 2Y5 ON, Canada}

\author{Diego Garc\'ia-Mart\'in}
\affiliation{Information Sciences, Los Alamos National Laboratory, 87545 NM, USA}

\author{Zo\"e Holmes}
\affiliation{Institute of Physics, \'Ecole Polytechnique F\'{e}d\'{e}rale de Lausanne (EPFL), Lausanne, Switzerland}
\affiliation{Centre for Quantum Science and Engineering, \'Ecole Polytechnique F\'{e}d\'{e}rale de Lausanne (EPFL), Lausanne, Switzerland}

\author{M. Cerezo}
\email{Contact author: cerezo@lanl.gov}
\affiliation{Information Sciences, Los Alamos National Laboratory, 87545 NM, USA}

\begin{abstract}
Moments of ensembles of unitaries play a central role in quantum information theory as they capture the statistical properties of dynamics of systems with some form of randomness. Indeed, concepts such as approximate $t$-designs arise when comparing how close an associated moment operator of a given unitary ensemble is to that of another, reference ensemble. Despite the importance of moment operators, their properties have not been as explored for quantum channels. In this work we develop a theoretical framework to compute moment operators for ensembles of quantum channels, for all moment orders $t$, with a special focus on determining ensembles that can be used as points of reference. By deriving hierarchies between ensembles, via inequalities of their moment operator norms, we give them operational meaning, and define useful concepts such as that of channel $t$-designs. Finally, we perform theoretical and numerical studies which show that different types of noise can decrease the norm of the moment operators (e.g., depolarizing noise), as well as increase it (e.g., amplitude damping noise), and generalize noise-induced concentration phenomena to channel-design-induced phenomena. Along the way, we find a block-orthogonal basis for permutations, which greatly simplifies our analyses, and may be of independent interest in moment calculations.
\end{abstract}

\maketitle

\section{Introduction}\label{sec:introduction}

Quantum experiments, whether intrinsically or by design, can involve randomness. Thus, each repetition samples a distinct evolution from an ensemble and applies it to the system. The statistical properties of such experiments are captured by the moment operator, i.e., the superoperator given by the ensemble-averaged adjoint action (the twirl). Focusing on the moment operator allows one to study ensembles independently of initial states and final measurements. Moment operators arise often in quantum information science, including in quantum-supremacy experiments \cite{boixo2018characterizing,arute2019quantum,wu2021strong,dalzell2022randomquantum,oszmaniec2022fermion,huang2021provably,knill2008randomized}, quantum chaos \cite{nahum2017quantum,von2018operator,nahum2018operator,ho2022exact}, transitions of quantum correlations \cite{li2018quantum,skinner2019measurement,jian2020measurement}, variational quantum algorithms \cite{mcclean2018barren,cerezo2020cost,larocca2024review,holmes2021connecting,wang2020noise,franca2020limitations,napp2022quantifying,ragone2023unified,fontana2023theadjoint,diaz2023showcasing}, concentration phenomena \cite{mcclean2018barren,cerezo2020cost,larocca2024review,holmes2021connecting,wang2020noise,franca2020limitations}, randomized benchmarking \cite{elben2022randomized,huang2020predicting,zhao2021fermionic,wan2017quantum,sauvage2024classical,west2024real}, error mitigation \cite{hu2024demonstration,cai2022quantum}, and as Markov process matrices in analyses of fixed points of ensembles \cite{hayden2007black,sekino2008fast,brown2012scrambling,lashkari2013towards,hosur2016chaos,nahum2018operator,von2018operator,hunter2019unitary,barak2020spoofing,napp2022quantifying,harrow2018approximate,letcher2023tight,hayden2016holographic,belkin2023approximate,mittal2023local,schuster2024random,braccia2024computing,deneris2024exact,garcia2024architectures,west2024random,schatzki2024random,garcia2023deep}.

Within this context, given an ensemble, it is often useful to quantify its proximity to a reference ensemble. Mathematically, this reduces to comparing their moment operators under a chosen distance measure. This problem has been extensively studied for ensembles of unitaries, where we say that an ensemble forms a $t$-design with respect to a reference if their moments match up to the $t$-th order \cite{dankert2009exact,harrow2009random,brandao2016local,hunter2019unitary,haferkamp2022random,haferkamp2021improved,brown2010random,nakata2017efficient,harrow2018approximate,chen2024efficient,chen2024incompressibility,belkin2023approximate,mittal2023local,schuster2024random,yada2025nonhaar,deneris2024exact}. A priori, computing and comparing moment operators are very demanding--often intractable--tasks. The situation simplifies when the reference ensemble is a group as the associated moment operator is the projector onto the group’s commutant, and its properties admit analytic characterization via Weingarten calculus \cite{collins2006integration,collins2022weingarten} (see \cite{mele2023introduction} for a quantum information-oriented overview). Of particular relevance is the ensemble drawn from the Haar measure on the unitary group. Not only is this a natural comparison point for other ensembles, it also enjoys desirable extremal properties: there is a hierarchy between unitary ensembles arising from inequalities between moment-operator Hilbert–Schmidt norms, with the Haar ensemble at the base \cite{gross2007evenly}. Equivalently, the Haar ensemble minimizes frame potentials \cite{sim2019expressibility}, which quantify how evenly spread a set of unitaries is.

The previous studies, however, have been almost entirely restricted to ensembles of strictly unitaries. In contrast, ensembles of completely-positive trace-preserving maps, or quantum channels, have been largely unexplored \cite{bengtsson2006An,collins2009random,collins2016random,bruzda2009random,kukulski2021generating,bai2024primitivity}. For instance, it is still unknown whether hierarchies and desirable properties for reference ensembles of more general operators can be found. Ultimately, exact forms of moment operators are necessary to determine the most appropriate reference ensemble. Indeed, there are many important open questions concerning ensembles that must be addressed, such as: \emph{What is the most natural generalization of the uniform (Haar) distribution of unitaries to channels?}, or, \emph{Is there a reference ensemble of channels that plays a privileged role over other ensembles?}

\begin{figure}[t]
		\centering
		\includegraphics[width=.8\linewidth]{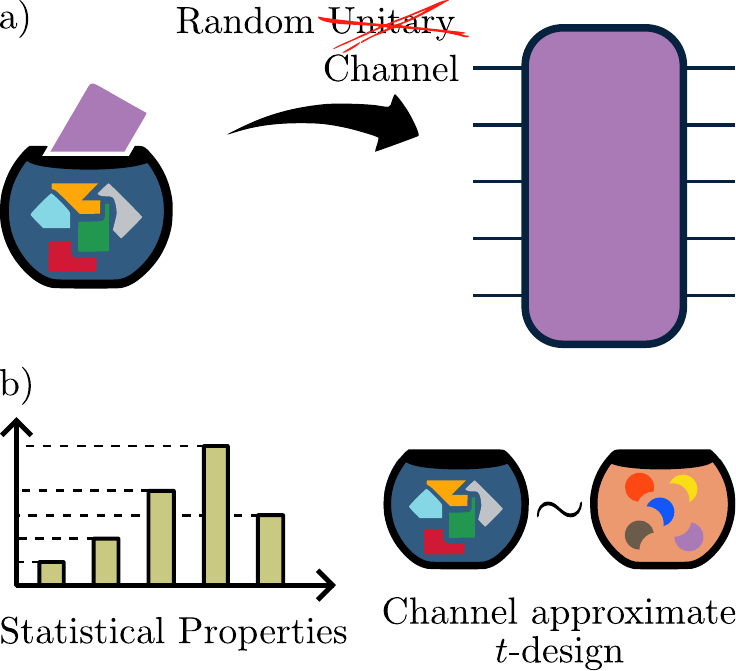}
		\caption{\textbf{Schematic representation of our results.} (a) We present a framework to study the statistical properties of an experimental setup where, at every run, a channel is sampled from some set according to a given probability distribution. (b) Our formalism centers around the moment operator, as it encodes the statistical properties arising from evolving states according to the ensemble, and performing a measurement at its output; and is also the central object used to define $t$-designs.}
		\label{fig:schematic}
\end{figure}

In this work we take steps towards answering the previous questions by presenting a framework to study moments of ensembles of quantum channels (see Fig. \ref{fig:schematic}). In Sec. \ref{sec:preliminaries}, we introduce our formalism and notation, which captures as a special case ensembles of strictly unitaries, thus allowing us to borrow inspiration from the standard literature. In Sec. \ref{sec:results}, we start by generalizing the archetypal reference ensemble of the Haar measure over the unitary group, i.e., sampling uniformly at random from the set of all unitaries, and we consider the ensemble of the Lebesgue measure \cite{bruzda2009random} over quantum channels, i.e., sampling uniformly at random from the set of all quantum channels. This study prompted us to consider a parameterized family of reference channel ensembles \cite{kukulski2021generating} which we dub the "channel Haar" (cHaar) ensemble. The cHaar ensemble is shown to interpolate between two reference ensembles of interest, the Haar reference ensemble of strictly unitaries, and what we dub as the Depolarize reference ensemble, where the maximally depolarizing channel is applied with unit probability. By deriving the exact forms and spectral properties for the associated moment operators, we compute deviations of the cHaar ensemble from the Haar and Depolarize ensembles. Such calculations indicate that random quantum channels are on average, inherently depolarizing, with perturbative non-unital behavior, and interpolate with environment dimension between unitary and depolarizing behavior. Finally, given these insights, we deduce a hierarchy between these ensembles, with the Depolarize ensemble found to be at the base of the hierarchy. Thus, we argue that the Depolarize ensemble is the most appropriate reference ensemble for quantum channels.

Expanding on these studies, we explicitly calculate the extent to which quantum circuits consisting of random unitaries, and fixed unital or non-unital noise, form approximate Depolarize $t$-designs. Here, we observe that certain types of noise can decrease the norm of the moment operator (e.g., unital Pauli noise), or can increase it (e.g., non-unital amplitude damping). Finally, we study concentration of expectation values. We find that what has previously been attributed to be noise-induced concentration phenomena  \cite{wang2020noise}, can be more generally thought of as channel-design-induced concentration phenomena. To complement our theoretical studies, in Sec. \ref{sec:numerics} we conduct numerical studies of several noisy quantum circuits of interest, consisting of layers of parameterized unitary and fixed noise components. In this setting, we demonstrate the convergence of a circuit's statistical properties as a function of circuit depth. Finally, in Sec. \ref{sec:conclusions}, we interpret the derived average behaviors of random quantum channels, and discuss several future applications of our formalisms.

\section{Preliminaries}\label{sec:preliminaries}
In this section we present the notation used throughout this work, with additional details in Appendix \ref{app:spaces_operators_and_permutations}.

\subsection{Moments of Ensembles of Channels}

Here, we denote $\dsys$-dimensional quantum Hilbert spaces as $\spacesys=\mathbb{C}^{\dsys}$, the space of linear operators acting on $\spacesys$ as $\mathcal{B}[\spacesys]$, and the space of linear superoperators acting on $\mathcal{B}[\spacesys]$ as $\mathcal{B}[\mathcal{B}[\spacesys]]$. We also make use of the linear vectorization map which transforms operators to vectors, ${\rm vec}:\mathcal{B}[\spacesys]\rightarrow \spacesys \otimes \spacesys$, and whose action on $X \in \mathcal{B}[\spacesys]$ is
\begin{align}\label{eq:vec-def}
	\hspace{-0.4cm}
	X = \sum_{\alpha,\beta \in [\dsys]} X_{\alpha\beta} \ket{\alpha}\bra{\beta} ~\xrightarrow[{\rm vec}]{}~ \kket{X} = \sum_{\alpha,\beta \in [\dsys]} X_{\alpha\beta} \ket{\alpha\beta}~,
\end{align}
where $[\dsys]\equiv\{0,1,\ldots,\dsys-1\}$. The (Hilbert-Schmidt) inner product between two vectorized operators $X,Y\in\mathcal{B}[\spacesys]$ is $\bbra{X}\kket{Y}=\Tr\left[X^{\dagger} Y\right]$, and given the identity operator $\idsys\in\mathcal{B}[\spacesys]$ on $\spacesys$, we normalize it as
\begin{equation}
	\chi_{\dsys}(X,Y) := \frac{\bbra{X}\kket{Y}}{\dsys}
	\quad,\quad
	\chi_{\dsys}(X) := \chi_{\dsys}(\idsys,X)~.
\end{equation}
The vectorization map induces an action over linear maps, ${\rm vec}:\mathcal{B}[\mathcal{B}[\spacesys]]\rightarrow\mathcal{B}[\spacesys \otimes \spacesys$], which allows us to represent superoperators as matrices. Given a quantum channel $\channel\in\mathcal{B}[\mathcal{B}[\spacesys]]$ (a completely positive, trace-preserving linear map), its action in terms of Kraus operators $\{K_{\channel}\}$ is
\begin{equation}\label{eq:Krauss}
		\channel(\cdot)=\sum_{K_{\channel}} K_{\channel} (\cdot) K_{\channel}^{\dagger} ~\xrightarrow[{\rm vec}]{}~ \widehat{\channel}=\sum_{K_{\channel}} K_{\channel} \otimes K_{\channel}^{*} ~,
\end{equation}
and superoperators satisfy $\kket{\channel(X)}=\widehat{\channel}\kket{X}$. Finally, $t$-copies of channels have corresponding superoperators,
\begin{equation}
		\widehat{\channel^{\otimes t}} = \!\!\!\!\!\!\!\!\sum_{K_{\channel_{0}},\dots,K_{\channel_{t-1}}} \!\!\!\!\!\!\!\!K_{\channel_{0}}\otimes \cdots \otimes K_{\channel_{t-1}} \otimes K_{\channel_{0}}^{*}\otimes\cdots \otimes K_{\channel_{t-1}}^{*}~. \label{eq:kraus_superoperators}
\end{equation}

In this work, we will analyze ensembles of quantum channels $\ensemblechannel \subseteq \mathcal{B}[\mathcal{B}[\spacesys]]$, namely sets of channels $\{\channel \in \mathcal{B}[\mathcal{B}[\spacesys]]\}$ accompanied with a probability distribution $d\channel$, with ensemble averages of channel-dependent quantities $\mathcal{F}(\channel)$,
\begin{align}
	\expval{\mathcal{F}(\channel)}_{\ensemblechannel} =&~ \int_{\ensemblechannel} d\channel ~\mathcal{F}(\channel) ~.
\end{align}
Note that here we have assumed that the ensemble is continuous, leading to an integral over $\ensemblechannel$. The discrete case follows by simply replacing the integral by a summation.

In particular, we will consider scenarios where an initial state $\rho$ is evolved under the action of a channel $\channel$, sampled from $\ensemblechannel$, at whose output we measure an observable $O$. The statistical properties of this setup are captured by $t$-th order moments of this expectation value, of the form
\begin{align}\label{eq:momentsdef}
\expval{\Tr\left[\channel(\rho) O \right]^t}_{\ensemblechannel} ~ .
\end{align}
For example, expectation value means and variances are $\mu_{\ensemblechannel} \!=\! \expval{\Tr\left[\channel(\rho) O \right]}_{\ensemblechannel}$ and $\sigma^{2}_{\ensemblechannel} \!=\! \expval{\Tr\left[\channel(\rho) O \right]^{2}}_{\ensemblechannel}\!\!-\expval{\Tr\left[\channel(\rho) O \right]^{\vphantom{2}{}}}_{\ensemblechannel}^{2}$.\!\!\!

General $t$-th order moments of expectation values can be more conveniently rewritten as
\begin{align}
\expval{\Tr\left[\channel(\rho) O \right]^t}_{\ensemblechannel}&=\expval{\Tr\left[\channel^{\otimes t}(\rho^{\otimes t}) O^{\otimes t} \right]}_{\ensemblechannel}\nonumber\\
&=\Tr\left[\expval{\channel^{\otimes t}(\rho^{\otimes t}) O^{\otimes t}}_{\ensemblechannel}\right]\nonumber\\
		&=\Tr\left[\Tau_{\ensemblechannel}^{(t)}(\rho^{\otimes t})O^{\otimes t} \right]~,\label{eq:moment}
\end{align}
in terms of the $t$-th order twirl operator $\Tau_{\ensemblechannel}^{(t)}:\mathcal{B}[\spacesys^{\otimes t}]\rightarrow \mathcal{B}[\spacesys^{\otimes t}]$, whose action on operators $X \in \mathcal{B}[\spacesys^{\otimes t}]$ is
\begin{align}
		\hspace{-0.3cm}
		\Tau_{\ensemblechannel}^{(t)}(X) &=\expval{\channel^{\otimes t}(X)}_{\ensemblechannel}= \int_{\ensemblechannel} d\channel ~ \channel^{\otimes t}(X)~. \!\!\!\!\label{eq:twirl-t-th}
\end{align}
We can then define the $t$-th order moment operator as the superoperator of the $t$-th order twirl,
\begin{align}
		\widehat{\Tau}_{\ensemblechannel}^{(t)} =\int_{\ensemblechannel} d\channel~ \widehat{\channel^{\otimes t}} ~,\label{eq:twirl-t-th-vect-channel}
\end{align}
such that
\begin{equation}\label{eq:expect-vectorized}
\expval{\left[\Tr\left[\channel(\rho) O \right]^t\right]}_{\ensemblechannel}=\bbra{O^{\otimes t}}\widehat{\Tau}_{\ensemblechannel}^{(t)}\kket{\rho^{\otimes t}}~.
\end{equation}

The moment operators of Eq.~\eqref{eq:twirl-t-th-vect-channel} are central in our theory, as they allow us to study the statistical properties of ensembles independently from the initial state and measurement operator. Moreover, we can use them to study experiments involving $k$ concatenations of several random channels $\channel$, each sampled identically and independently from $\ensemblechannel$ according to $d\channel$. In this case, we obtain expectation values $\bbra{O^{\otimes t}}\widehat{\Tau}_{\ensemblechannel}^{(t)k}\kket{\rho^{\otimes t}}$, with the moment operator replaced by its $k$-th power,
\begin{align}
	\widehat{\Tau}_{\ensemblechannel}^{(t)} \to \widehat{\Tau}_{\ensemblechannel}^{(t)k}~,
\end{align}
and consider ensemble statistics with respect to $d,t,k$.

At this point, we find it important to remark that we can always express the moment operator as
\begin{align}\label{eq:channel-explicit}
	\widehat{\Tau}_{\ensemblechannel}^{(t)} =&~
		\channeldepolarize[\widehat]{\dsys}^{\otimes t} ~+~ \widehat{\Delta}_{\ensemblechannel}^{(t)}~,
\end{align}
where we have defined an explicit trace-preserving term, and extraneous terms mapping to non-identity operators,
\begin{align}
	\channeldepolarize[\widehat]{\dsys}^{\otimes t} =&~ \frac{1}{\dsys^{t}}\kket{\idsys^{\otimes t}}\bbra{\idsys^{\otimes t}} ~,
	\\
	\widehat{\Delta}_{\ensemblechannel}^{(t)} =&~ \frac{1}{\dsys^{t}} \hspace{-0.2cm} \sum_{\substack{P \in \mathcal{S}_{\ensemblechannel}^{(t)} \backslash \{\idsys^{\otimes t}\} \\S \in \mathcal{S}_{\ensemblechannel}^{(t)}}} \hspace{-0.2cm} \tau_{\ensemblechannel}^{(t)}(P,S)~\kket{P}\bbra{S}~,
\end{align}
in terms of an appropriate ensemble-dependent basis of operators $P,S \in \mathcal{S}_{\ensemblechannel}^{(t)}$. The ensemble-dependent coefficients $\tau_{\ensemblechannel}^{(t)}(P,S)$ correspond to the entries of a transfer matrix that represent how an element of the basis is mapped to different elements of the basis by the moment operator.

To finish, we note that analytically evaluating the moment operator can lead to intractable calculations. In these cases, one can instead opt to study how $\varepsilon$-close in norm the moment operator of $\ensemblechannel$ is to that of a reference ensemble $\ensemblechannel^{\prime}$. This allows us to introduce the notion of an (additive) $\varepsilon$-approximate $t$-channel design, with respect to $\ensemblechannel^{\prime}$, as the case when the difference of moment operators satisfies
\begin{align}\label{eq:channeldesign}
	\norm{\widehat{\Tau}_{\ensemblechannel}^{(t)} - \widehat{\Tau}_{\ensemblechannel^{\prime}}^{(t)}} \leq \varepsilon~.
\end{align}
The choice of norm $\norm{\cdot}$ is task-dependent, and unless noted, we use the Hilbert-Schmidt norm as it is overall the simplest to compute. Moreover, we consider $\varepsilon$ to be the additive error, but our definition can readily be adapted to the relative error. Importantly, we will also see below that in many cases of interest, the evaluation of Eq.~\eqref{eq:channeldesign} leads to analysis of the moment operator norm,
\begin{equation}\label{eq:norm-moment}
		\norm{\widehat{\Tau}_{\ensemblechannel}^{(t)}}~.
\end{equation}
We note, however, that our definition can readily be adapted and translated to the diamond norm via the inequalities (please refer to Appendix \ref{app:properties_of_moment_operators} for a proof),
\begin{align}
	\!\!\!\!\!
	\norm{\Tau_{\ensemblechannel}^{(t)} - \Tau_{\ensemblechannel^{\prime}}^{(t)}}_{\diamond}
	\!\leq \dsys^{t} \norm{\widehat{\Tau}_{\ensemblechannel}^{(t)} - \widehat{\Tau}_{\ensemblechannel^{\prime}}^{(t)}}_{2}
	\leq \dsys^{t} \norm{\widehat{\Tau}_{\ensemblechannel}^{(t)} - \widehat{\Tau}_{\ensemblechannel^{\prime}}^{(t)}}_{1} ,\!~\!\!\!\!\!\!
\end{align}
with the different norms having different operational meanings. Indeed, additive error under the $\diamond$-norm at the basis of the inequality chain has a clear interpretation in terms of how easy it is to distinguish states at the output of channels drawn from the ensembles. Suppose two $t$-th order moment operators have additive error of $\varepsilon$ under the $\diamond$-norm. Then for all quantum algorithms that make only one query of a channel from $\ensemblechannel$ or from $\ensemblechannel^{\prime}$, the output state of the channel is at most $\varepsilon$-away under the $1$-norm \cite{schuster2024random}. More generally still, we can also consider relative errors, defined as
\begin{equation}
	(1-\varepsilon){\Tau}_{\ensemblechannel^{\prime}}^{(t)}\preceq {\Tau}_{\ensemblechannel}^{(t)}\preceq(1+\varepsilon){\Tau}_{\ensemblechannel^{\prime}}^{(t)}~,
\end{equation}
where ${\Tau}_{\ensemblechannel^{\prime}}^{(t)}\preceq {\Tau}_{\ensemblechannel}^{(t)}$ denotes that ${\Tau}_{\ensemblechannel}^{(t)}-{\Tau}_{\ensemblechannel^{\prime}}^{(t)}$ is a completely-positive map. Now, suppose two $t$-th order moment operators have a relative error $\varepsilon$. Then for all quantum algorithms that make $k$ queries of channels from $\ensemblechannel$ or from $\ensemblechannel^{\prime}$, the output state of the channels is at most $2\varepsilon$-away under the state $1$-norm \cite{schuster2024random}.

At this point, we find it important to note that a moment operator can be thought of as a Markov transfer matrix like process \cite{hayden2007black,sekino2008fast,brown2012scrambling,lashkari2013towards,hosur2016chaos,nahum2018operator,von2018operator,hunter2019unitary,barak2020spoofing,napp2022quantifying,harrow2018approximate,letcher2023tight,hayden2016holographic,nahum2018operator,hunter2019unitary,harrow2018approximate,belkin2023approximate,mittal2023local,schuster2024random,braccia2024computing,deneris2024exact,garcia2024architectures,west2024random,schatzki2024random,garcia2023deep}. In particular, given some basis of operator space, the moment operator says how the channels maps a given basis element onto a linear combination of other basis elements due to the randomness in the ensemble. When concatenating multiple ensembles, each channel is typically sampled independently, thus leading to a memory-less process.  As an aside, other than explicit relevance to statistics, we note an immediate interpretation of moment operators that is relevant to randomized benchmarking \cite{wood2011tensor}, and follows from the relationship between superoperators and their associated Choi states. In particular for pairs of channels, inner products between their superoperators correspond to inner products between their Choi states. It follows that the norm and trace of moment operators correspond respectively to the average overlap between the Choi states, and the average entanglement fidelity of the Choi states \cite{wood2011tensor}, of the channels from its ensemble. Concurrently, design properties of ensembles of unitaries have been attributed to their expressive power within variational quantum algorithms \cite{du2020expressive,sim2019expressibility,yu2023expressibility,holmes2021connecting}. However it remains to be seen whether similar notions of capability or utility of ensembles of quantum channels directly follow from their design properties. Other questions regarding operational interpretations will arise within this work once exact forms for moment operators over specific ensembles of quantum channels are derived.

\subsection{Special Case: Ensembles of Unitaries}

Since unitaries represent the evolution of closed quantum systems, the study of moments of ensembles of unitaries, has received considerable attention within the field of quantum information \cite{laracuente2024approximate,grevink2025will,zhu2016clifford,collins2006integration,petz2004asymptotics}. Here we review some well-known results concerning ensembles of unitaries.

First, we recall that the $t$-th order moment operator over unitary ensembles $\ensembleunitary$ can now be expressed as
\begin{align}
		\widehat{\Tau}_{\ensembleunitary}^{(t)} =\int_{\ensembleunitary} dU~U^{\otimes t} \otimes U^{\otimes t ~\! *}~,\label{eq:twirl-t-th-vect}
\end{align}
where $dU$ denotes the probability distribution over $\ensembleunitary$. For the simple case when $\ensembleunitary$ contains a single unitary $U$ (which is sampled with probability one), we readily find
\begin{equation}
		\widehat{\Tau}_{U}^{(t)}= U^{\otimes t} \otimes U^{\otimes t ~\! *}~, \quad \text{and} \quad \norm{\widehat{\Tau}_{U}^{(t)}}^{2}=\dsys^{2t}~.
\end{equation}
For the case when $\ensembleunitary = G$ is a group, and $dU=d\mu(U)$ its associated Haar measure, one can use Weingarten calculus \cite{mele2023introduction,ragone2022representation,collins2022weingarten} to evaluate moment operators. Now, $\widehat{\Tau}_{G}^{(t)}$ is a projector onto the $t$-th order commutant, defined as
\begin{align}
{\rm com}^{(t)}(G)=\{X \in \mathcal{B}[\spacesys^{\otimes t}] \!:\! [U^{\otimes t},X] = 0 ,\forall~ U \in G\}~,
\end{align}
and hence satisfies
\begin{equation}\label{eq:proj-group}
\widehat{\Tau}_{G}^{(t)}\widehat{\Tau}_{G}^{(t)}=\widehat{\Tau}_{G}^{(t)} ~~~ \text{and} ~~~ \norm{\widehat{\Tau}_{G}^{(t)}}^{2}={\rm dim}({\rm com}^{(t)}(G))~.
\end{equation}
Equation~\eqref{eq:proj-group} is extremely useful for bounding the difference between moment operators with respect to some ensemble $\ensembleunitary$, and with respect to a reference ensemble of the Haar-distributed group $G$, whenever $\ensembleunitary \subseteq G$. Note that $\ensembleunitary$ does not need to form a group itself. Such comparisons are central when studying approximate $\varepsilon$-$t$-designs. Indeed, we say that $\ensembleunitary$ is an $\varepsilon$-$t$-design over $G$ if and only if $\tnorm{\widehat{\Tau}_{\ensembleunitary}^{(t)} - \widehat{\Tau}_{G}^{(t)}} \leq \varepsilon$, where the choice of norm depends on the task at hand \cite{hearth2025unitary,hunter2019unitary}. In the $2$-norm case, combining Eq.~\eqref{eq:proj-group} along with the left- and right- invariance of the Haar measure, we find that for all ensembles $\ensembleunitary \subseteq G$,
\begin{equation}
\widehat{\Tau}_{G}^{(t)}\widehat{\Tau}_{\ensembleunitary}^{(t)}=\widehat{\Tau}_{\ensembleunitary}^{(t)}\widehat{\Tau}_{G}^{(t)}=\widehat{\Tau}_{G}^{(t)}~, \label{eq:haar-invariance}
\end{equation}
from where it then follows that
\begin{align}
	\norm{\widehat{\Tau}_{\ensembleunitary}^{(t)} - \widehat{\Tau}_{G}^{(t)}}^{2} = \norm{\widehat{\Tau}_{\ensembleunitary}^{(t)}}^{2}-\norm{\widehat{\Tau}_{G}^{(t)}}^{2}~,
\end{align}
and hence,
\begin{equation}\label{eq:ineq}
		{\rm dim}({\rm com}^{(t)}(G))\leq \norm{\widehat{\Tau}_{\ensembleunitary}^{(t)}}^{2}~.
\end{equation}
In addition, one can also readily find that any $k$ concatenations of the Haar ensemble are invariant,
\begin{align}
	\widehat{\Tau}_{\ensembleunitary}^{(t)k} =&~ \widehat{\Tau}_{\ensembleunitary}^{(t)}~.
\end{align}

Notably, Eq.~\eqref{eq:ineq} can be used to establish a partial ordering between the norms of the moment operators for the Haar measure over groups. In particular, given groups $G_1,G_2,\ldots$ such that $\{U\}\subseteq G_1\subseteq G_2 \subseteq \cdots \subseteq\ensemblehaar[\dsys]$ then
\begin{equation}\label{eq:chain-groups}
	\!\norm{\widehat{\Tau}_{\ensemblehaar[\dsys]}^{(t)}}^{2}\!\leq\! \cdots \!\leq\! \norm{\widehat{\Tau}_{G_2}^{(t)}}^{2}\!\leq\! \norm{\widehat{\Tau}_{G_1}^{(t)}}^{2}\!\leq \!\norm{\widehat{\Tau}_{U}^{(t)}}^{2}=\dsys^{2t}~
\end{equation}
Here, we denote $\ensemblehaar[\dsys]$ as the unitary group of all unitaries acting on $\spacesys$ with uniform Haar measure, henceforth referred to as the Haar ensemble. Equation~\eqref{eq:chain-groups} implies that the more "unstructured" a group $G$ is (i.e., the less symmetries it has) the smaller the norm of its moment operator. Thus, the Haar ensemble $\ensemblehaar[\dsys]$, at the base of such a hierarchy of unitary ensembles, serves as an appropriate reference ensemble for other subsets of unitaries. Moreover, we note that such norm inequalities appear within the context of variational quantum algorithms \cite{cerezo2020variationalreview,bharti2021noisy} and quantum machine learning \cite{gujju2024quantum,wang2024comprehensive} as a measure of the expressiveness of the ensemble of unitaries arising from a parameterized circuit \cite{sim2019expressibility,harrow2018approximate,mcclean2018barren,deneris2024exact,bai2024primitivity,garcia2023deep}. In this context, it has been shown that the smaller the norm, the greater the concentration (and the more difficult the optimization) of the variational landscape \cite{sim2019expressibility,holmes2021connecting,du2020expressive,nakaji2021expressibility,garcia2021quantum,larocca2024review}.

Given the privileged role that the Haar ensemble of unitaries $\ensemblehaar[\dsys]$ plays, we find it convenient to recall some of its properties \cite{mele2023introduction}. To begin, the $t$-th order commutant of $\ensemblehaar[\dsys]$ is given by the system permuting representation $\textnormal{com}^{(t)}(\ensemblehaar[\dsys]) = \representations[t]$ of the symmetric group $\permutations[t]$, namely all permutations of $t$ copies a space $\spacesys^{\otimes t}$. Moreover, given a permutation $\sigma \in \permutations[t]$, the system permuting representation $\represent~:~\permutations[t]\rightarrow \representations[t] \subset \mathcal{B}[\spacesys^{\otimes t}]$ is,
\begin{align}\label{eq:rep-S_k}
	\!\!\!\!\!\!\!
	\representation{\sigma} =&~ \!\!\!\!\!\!\!\!\!\!\sum_{\alpha_{0},\dots,\alpha_{t-1} \in [\dsys]} \!\!\!\!\!\!\!\! \ket{\alpha_{\sigma^{-1}(0)},\dots,\alpha_{\sigma^{-1}(t-1)}}\bra{\alpha_{0},\dots,\alpha_{t-1}}~, \!\!\!\!
\end{align}
with associated normalized inner products, or characters,
\begin{align}
	\!\!
	\chi_{\dsys}^{(t)}(\sigma,\pi) =&~ \frac{\bbra{\representation{\sigma}}\kket{\representation{\pi}}}{\dsys^{t}}
	~~ : ~~
	\chi_{\dsys}^{(t)}(\sigma) = \chi_{\dsys}^{(t)}(e,\sigma) ~,\!\!\!
\end{align}
where $e$ denotes the identity permutation with representation $\representation{e} = \idsys^{\otimes t}$. For $t\leq \dsys$ then ${\rm dim}({\rm com}^{(t)}(\ensemblehaar[\dsys]))=t!$, and for any unitary group $G$, we have the bounds
\begin{equation}
		t! ~\leq~ \norm{\widehat{\Tau}_{G}^{(t)}}^{2} ~\leq~ \dsys^{2t}~.
\end{equation}
To finish, we note that since $\widehat{\Tau}_{\ensemblehaar[\dsys]}^{(t)}$ is a projector \cite{mele2023introduction,garcia2023deep}, we can express the Haar ensemble moment operator as,
\begin{equation}\label{eq:haar-moment}
\widehat{\Tau}_{\ensemblehaar[\dsys]}= \frac{1}{\dsys^{t}} \sum_{\sigma, \pi\in \permutations[t]}W_{\dsys}^{(t)-1}(\sigma,\pi)~\kket{\representation{\sigma}}\bbra{\representation{\pi}}~,
\end{equation}
where the $t! \times t!$ transfer matrix $W_{\dsys}^{(t)-1}$ is known as the Weingarten matrix, the inverse of the Gram matrix $W_{\dsys}^{(t)}$ with entries of the overlaps, $W_{\dsys}^{(t)}(\sigma,\pi) = \chi_{\dsys}^{(t)}(\sigma,\pi)$. Moreover, given the norm hierarchy and invariance properties in Eq.~\eqref{eq:haar-invariance}, the moment operator of any unitary ensemble, $\ensembleunitary\subseteq\ensemblehaar[\dsys]$, has support on the projector Haar ensemble moment operator,
\begin{equation}\label{eq:inclusion}
	\widehat{\Tau}_{\ensembleunitary}^{(t)}=\widehat{\Tau}_{\ensemblehaar[\dsys]}^{(t)}~+~\cdots ~.
\end{equation}

\section{Results}\label{sec:results}
In this section, we present the main results of this work, with additional details in the appendices. Specifically, we define reference ensembles of quantum channels, derive their relationships and properties (Appendix \ref{app:spaces_operators_and_permutations} and Appendix \ref{app:properties_of_moment_operators}), calculate moments of specific ensembles of quantum channels (Appendix \ref{app:moment_operators_of_noisy_ensembles}), and investigate concentration phenomena arising in the context of noisy variational quantum algorithms (Appendix \ref{app:designs_and_trainability}).

\subsection{Reference Ensembles}

As previously discussed, it is convenient to introduce reference ensembles to define concepts such as approximate $t$-designs. Here, we consider three candidate reference ensembles for quantum channels and derive their properties and relationships. Ultimately, such properties allow us to determine whether a hierarchy exists between these ensembles that suggests the most appropriate reference ensemble for quantum channels.

\begin{figure*}[t]
		\centering
		\includegraphics[width=.9\textwidth]{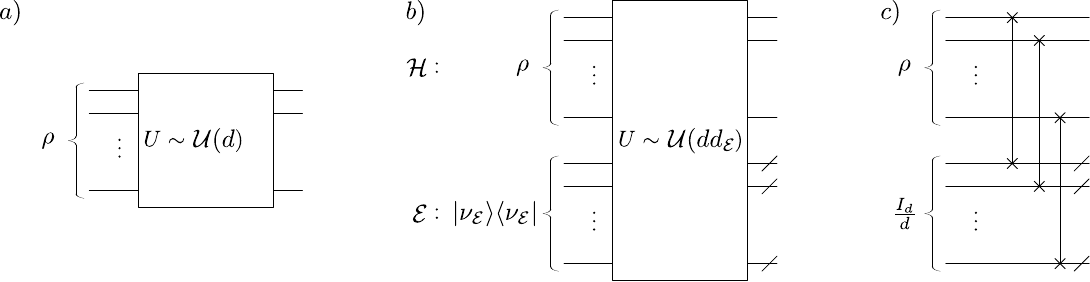}
		\caption{\textbf{Circuits implementing the reference channel ensembles.} (a) The Haar ensemble can be implemented by sending a state $\rho$ through a random unitary uniformly sampled from $\ensemblehaar$. b) The cHaar ensemble can be implemented via its Stinespring dilation: Initialize the joint state of the tensor product Hilbert space $\spacesys \otimes \spaceenv$ in $\rho\otimes\ket{\nu_\mathcal{E}}\bra{\nu_\mathcal{E}}$, apply a joint unitary uniformly sampled from $\ensemblehaar[\dsys,\denv]$, and trace out the environment. c) The Depolarize ensemble can be applied with a circuit that swaps the state $\rho$ with an ancillary maximally mixed state ${\idsys}/{\dsys}$, and then traces out the ancillary registers containing $\rho$. }
		\label{fig:circuits}
\end{figure*}

For our first candidate reference ensemble, we take the Haar ensemble of unitaries, given its place at the base of the unitary ensemble hierarchy. We will denote it as the \emph{Haar} ensemble $\ensemblehaar$, with a $t$-th order twirl of,
\begin{align}\label{eq:Haar}
	\hspace{-0.2cm}
	\Tau_{\ensemblehaar}^{(t)}(X) = \int_{\ensemblehaar} \!dU~ U^{\otimes t}~X~{U}^{\otimes t \dagger}~, \!\!
\end{align}
and an associated moment operator of,
\begin{align}
		\widehat{\Tau}_{\ensemblehaar}^{(t)} =\int_{\ensemblehaar} dU~U^{\otimes t} \otimes U^{\otimes t ~\! *}~.
\end{align}

For our second candidate reference ensemble, we recall the definition of the Stinespring dilation \cite{wilde2013quantum,nielsen2000quantum}, where any quantum channel $\channel : \mathcal{B}[\spacesys]\rightarrow \mathcal{B}[\spacesys]$, can be expressed as a unitary  $\unitary_{\channel}: \spacesys \otimes \spaceenv \rightarrow \spacesys \otimes \spaceenv$ acting on the tensor product of the system Hilbert space $\spacesys$ of dimension $\dsys$ and an environment Hilbert space $\spaceenv$ of dimension $\denv$, with the environment being in a pure state $\nu_{\spaceenv} = \ket{\nu_{\spaceenv}}\bra{\nu_{\spaceenv}} \in \mathcal{B}[\spaceenv]$. Thus the action of any quantum channel is
\begin{align}
	\channel(X) =&~ \Tr_{\spaceenv} \left[\unitary_{\channel} \left(X \otimes \nu_{\spaceenv} \right) \unitary_{\channel}^{\dagger} \right]~,
\end{align}
where the environment $\spaceenv$ is partially traced out. A key advantage of working with the Stinespring dilation is that we can now analyze ensembles of random quantum channels in terms of the random Stinespring unitaries $\unitary_{\channel}$ that generate such channels. Sampling quantum channels can thus be directly related to sampling unitaries in larger composite spaces. Notably, it has been shown that a consistent measure $d\channel$ for quantum channels exists that is equivalent to sampling unitaries $\unitary_{\channel}$ according to the Haar measure $d\unitary$ over $\ensemblehaar[\dsys\denv]$ \cite{kukulski2021generating}. Here, we note that the specific choice of environment state $\nu_{\spaceenv}$ is arbitrary (due to the invariance of the Haar measure) \footnote{While we here consider sampling unitaries from $\ensemblehaar[\dsys\denv]$, one can also consider the case when the unitaries are sampled from some subgroup of $\ensemblehaar[\dsys\denv]$, in which case $\nu_{\spaceenv}$ is no longer arbitrary, but needs to be specified, and the overall ensuing channel properties will depend on these choices. We leave the study of these cases for future work.  }. In fact, when $\denv= \dsys^{2}$, this measure $d\channel$ is the Lebesgue measure over quantum channels.

Given this Stinespring dilation-based approach, we define our second candidate reference ensemble as the environment dimension $\denv$-parameterized family of ensembles of channels, which we denote as the channel-Haar or \emph{cHaar} ensemble $\ensemblechaar$, with a $t$-th order twirl of,
\begin{align}\label{eq:cHaar}
	\hspace{-0.2cm}
	\Tau_{\ensemblechaar}^{(t)}(X) = \int_{\ensemblehaar[\dsys\denv]} \!\!\!\!\!\!\!\!\!\!dU~\Tr_{\spaceenv^{\otimes t}}\left[ U^{\otimes t}(X \otimes \nu_{\spaceenv}^{\otimes t}){U}^{\otimes t \dagger})\right]~, \!\!
\end{align}
and an associated moment operator of,
\begin{align}\label{eq:chaar-moment-op}
	 \hspace{-0.1cm}
\widehat{\Tau}_{\ensemblechaar}^{(t)}
&= \left( \idsys^{\otimes 2t}\!\otimes \bbra{\idenv^{\otimes t}} \right) ~\widehat{\Tau}_{\ensembleunitary(\dsys\denv)}^{(t)}~ \left( \idsys^{\otimes 2t}\!\otimes\kket{\nu_{\spaceenv}^{\otimes t}} \right)~.
\end{align}
Finally, we can readily use our expansions for Haar moment operators in terms of the Haar ensemble commutant of permutations in Eq.~\eqref{eq:haar-moment} to find,
\begin{align}\label{eq:chaar-moment-explicit}
	\!\!\!\!
	\widehat{\Tau}_{\ensemblechaar}^{(t)}= \frac{1}{\dsys^{t}} \!\!\sum_{\sigma, \pi\in \permutations[t]}\!\!\! \chi_{\denv}^{(t)}(\sigma)~W_{\dsys\denv}^{-1}(\sigma,\pi)~\kket{\representation{\sigma}}\bbra{\representation{\pi}}~\!.\!\!\!\!
\end{align}
Above we have used the fact that under a simple reordering of the Hilbert space $(\spacesys\otimes \spaceenv)^{\otimes t}\rightarrow \spacesys^{\otimes t}\otimes\spaceenv^{\otimes t}$ \cite{braccia2024computing} we can express $\kket{\representation[\dsys\denv]{\sigma}}\rightarrow \kket{\representation{\sigma}}\otimes \kket{\representation[\denv]{\sigma}}$, as well as the fact that $\bbra{\representation[\denv]{\pi}}\kket{\nu_{\spaceenv}^{\otimes t}}=1$, $\forall~ \pi\in S_t$ and $\forall~ \ket{\nu_{\spaceenv}}\in \spaceenv$.

For our third candidate reference ensemble, we recall the depolarizing channel
\begin{equation}\label{eq:depol-channel}
		\channeldepolarize{\dsys}(X)=\frac{\Tr\left[X\right]}{\dsys}\idsys~,
\end{equation}
that has been shown to describe the trace-preservation component in all ensembles, and will be key to our analysis. We thus define our third candidate reference ensemble as the ensemble containing only the depolarizing channel with unit probability, which we denote as the \emph{Depolarize} ensemble $\ensembledep$, with a $t$-th order twirl of,
\begin{align}
	\Tau_{\ensembledep}^{(t)}(X) =&~ \channeldepolarize{\dsys}^{\otimes t}(X) = \frac{\Tr\left[X\right]}{\dsys^{t}}\idsys^{\otimes t}~,
\end{align}
and an associated moment operator of,
\begin{align}\label{eq:channel-depol-vect}
	\widehat{\Tau}_{\ensembledep}^{(t)} = \channeldepolarize[\widehat]{\dsys}^{\otimes t} =
		\frac{1}{\dsys^{t}}\kket{\idsys^{\otimes t}}\bbra{\idsys^{\otimes t}}~.
\end{align}
Quantum circuits to experimentally implement each of the three candidate reference ensembles are shown in Fig. \ref{fig:circuits}.

In the following sections, after deriving explicit moments of each ensemble, we will compare the Haar, cHaar and Depolarize reference ensembles, and deduce their hierarchy.

\subsection{Properties of Moment Operators}
Having defined the Haar, cHaar, and Depolarize ensembles, we now proceed to analyze their moment operators. First, simply from their definitions, we can immediately prove the following properties regarding their relationships:
\begin{proposition}[Properties of Reference Ensembles]\label{prop:cHaar}
Let $\widehat{\Tau}_{\ensemblehaar}^{(t)}$, $\widehat{\Tau}_{\ensemblechaar}^{(t)}$ and $\widehat{\Tau}_{\ensembledep}^{(t)}$ respectively denote the moment operators for the Haar, cHaar and Depolarize ensembles. Then, for $t=1$ and any $\denv$, the ensembles are identical,
\begin{equation}
	\widehat{\Tau}_{\ensemblehaar}^{(1)}=\widehat{\Tau}_{\ensemblechaar}^{(1)} =\widehat{\Tau}_{\ensembledep}^{(1)} ~,
\end{equation}
with such equalities not holding for $t>1$. While $\widehat{\Tau}_{\ensemblehaar}^{(t)}$ and $\widehat{\Tau}_{\ensembledep}^{(t)}$ are self-adjoint projectors for all $t$ and $\denv$,
\begin{align}\label{eq:depolarize-proj} \widehat{\Tau}_{\ensemblehaar}^{(t)}\widehat{\Tau}_{\ensemblehaar}^{(t)}=\widehat{\Tau}_{\ensemblehaar}^{(t)}~,&~\quad \widehat{\Tau}_{\ensembledep}^{(t)}\widehat{\Tau}_{\ensembledep}^{(t)}=\widehat{\Tau}_{\ensembledep}^{(t)} ~,
\end{align}
$\widehat{\Tau}_{\ensemblechaar}^{(t)}$ is not a self-adjoint projector for $t>1$ and $\denv>1$,
\begin{align}\label{eq:no-proj-cHaar}
	\widehat{\Tau}_{\ensemblechaar}^{(t)}\widehat{\Tau}_{\ensemblechaar}^{(t)}\neq \widehat{\Tau}_{\ensemblechaar}^{(t)}
	\quad ,&~ \quad
	\widehat{\Tau}_{\ensemblechaar}^{(t) \dagger}\neq \widehat{\Tau}_{\ensemblechaar}^{(t)}~.
\end{align}
Moreover, for any $t$ and any $\denv$, the cHaar ensemble is invariant under any ensemble of unitaries $\ensembleunitary$,
\begin{align}\label{eq:general-unitary}
\widehat{\Tau}_{\ensemblechaar}^{(t)} \widehat{\Tau}_{\ensembleunitary}^{(t)}=\widehat{\Tau}_{\ensembleunitary}^{(t)}\widehat{\Tau}_{\ensemblechaar}^{(t)} =\widehat{\Tau}_{\ensemblechaar}^{(t)} ~.
\end{align}
The Depolarize ensemble is right-invariant under any ensemble of channels $\ensemblechannel$ and is left-invariant under adjoints,
\begin{align}\label{eq:general-channel}
\widehat{\Tau}_{\ensembledep}^{(t)}\widehat{\Tau}_{\ensemblechannel}^{(t)}=\widehat{\Tau}_{\ensembledep}^{(t)}
\quad ,&~ \quad
\widehat{\Tau}_{\ensemblechannel}^{(t)\dagger}\widehat{\Tau}_{\ensembledep}^{(t)}=\widehat{\Tau}_{\ensembledep}^{(t)}~,
\end{align}
and is left-invariant under unital ensembles of channels $\ensembleunital$,
\begin{align}\label{eq:unital-channel}
\widehat{\Tau}_{\ensembleunital}^{(t)}\widehat{\Tau}_{\ensembledep}^{(t)}=\widehat{\Tau}_{\ensembledep}^{(t)}~.
\end{align}
Therefore Hilbert-Schmidt norms of differences of moment operators relative to the Depolarize ensemble are,
\begin{align}
	\norm{\widehat{\Tau}_{\ensemblechannel}^{(t)} - \widehat{\Tau}_{\ensembledep}^{(t)}}^{2} = \norm{\widehat{\Tau}_{\ensemblechannel}^{(t)}}^{2} - \norm{\widehat{\Tau}_{\ensembledep}^{(t)}}^{2}~.
\end{align}
\end{proposition}

These relationships between ensembles already suggest that the Depolarize ensemble, although not corresponding to uniformly sampling all quantum channels, is potentially an appropriate reference ensemble. The relevance of this Depolarize ensemble for quantum channels follows partially from its right, and often left invariance under concatenation with other ensembles. It is also important to note that any moment operator of any trace-preserving ensemble involves a Depolarize moment operator term (which is reminiscent of Eq.~\eqref{eq:inclusion} and how the moment operator for any unitary ensembles contain the term $\widehat{\Tau}_{\ensemblehaar[\dsys]}^{(t)}$).

It is perhaps surprising that the cHaar ensemble, given its lack of invariance properties, does not appear to be a natural generalization of the Haar ensemble, and does not immediately appear to be the most appropriate reference ensemble for quantum channels. After all, the Depolarize ensemble only contains a single channel! To fully support this proposal, we must derive exact properties of the cHaar ensemble and deduce its place in the ensemble hierarchy, as a function of the variables $\denv,t,k$

\begin{theorem}[Hierarchy of moment operator norms]\label{theorem:cHaar-explicit}
The following dimension-dependent relationships between the Depolarize, cHaar and Haar ensembles hold: \\[2pt]
\noindent 1) The cHaar ensemble, and its $k$-concatenations, interpolates with system and environment dimension between the Haar and Depolarize ensembles. That is, for all $k \geq 0$:
\begin{align}
	\lim_{\dsys\denv \to \dsys} \widehat{\Tau}_{\ensemblechaar}^{(t)k} =& \widehat{\Tau}_{\ensemblehaar}^{(t)}~, \quad
	\lim_{\dsys\denv\to \infty} \widehat{\Tau}_{\ensemblechaar}^{(t)k} = \widehat{\Tau}_{\ensembledep}^{(t)}~.\label{eq:limitscHaar}
\end{align}
In fact, the cHaar ensemble moment operator has an exact closed form expression, consisting of trace-preserving, non-unital, and unital terms, expressed for simplicity in terms of some traceless operators $T^{(t,k)}_{\dsys},T^{(t,k)\prime}_{\dsys} \in \mathcal{B}[\spacesys^{\otimes t}]$, and asymptotic scalings of transfer matrix elements,
\begin{align}
		\lim_{\substack{\dsys\denv \to \infty}} \widehat{\Tau}_{\ensemblechaar}^{(t)k} =&~ \frac{1}{\dsys^{t}} \kket{\idsys^{\otimes t}}\bbra{\idsys^{\otimes t}} \\
		&~~+~ \frac{1}{\dsys^{t}} \mathcal{O}\!\left(\frac{1}{\dsys\denv}\right) ~ \kket{T^{(t,k)}_{\dsys}}\bbra{\idsys^{\otimes t}} \nonumber \\
		&~~+~ \frac{1}{\dsys^{t}} \mathcal{O}\!\left(\frac{1}{\dsys^{2}\denv^{k}}\right)\kket{T^{(t,k)}_{\dsys}}\bbra{T^{(t,k)\prime}_{\dsys}}~. \nonumber
\end{align}

\noindent 2) The cHaar ensemble forms an $\varepsilon$-approximate Depolarize $t$-design (as introduced in Eq.~\eqref{eq:channeldesign}), for $\varepsilon \in \mathcal{O}\left(\frac{1}{\denv}\right)$, in the fixed $t$ and the asymptotic $\dsys\denv \to \infty$ limit.

\noindent 3) In the asymptotic $\dsys\denv \to \infty$ limit, there exists a strict hierarchy between the $k$-concatenated $t$-th order ensembles, via their moment operator norms, with $k \geq k^{\prime}$,
\normalsize
\begin{align}\label{eq:chain-channels}
	\norm{\widehat{\Tau}_{\ensembledep}^{(t)}}^{2} \leq \norm{\vphantom{\widehat{\Tau}_{\ensembledep}^{(t)}}{}\widehat{\Tau}_{\ensemblechaar}^{(t)k}}^{2} \leq \norm{\vphantom{\widehat{\Tau}_{\ensembledep}^{(t)}}{}\widehat{\Tau}_{\ensemblechaar}^{(t)k^{\prime}}}^{2} \leq \norm{\vphantom{\widehat{\Tau}_{\ensembledep}^{(t)}}{}\widehat{\Tau}_{\ensemblehaar}^{(t)}}^{2}~.
\end{align}
\end{theorem}

As per part 1) and 2) of Theorem \ref{theorem:cHaar-explicit}, in the appropriate limits, the cHaar ensemble reduces to either the Depolarize ensemble or the Haar ensemble, and can be thought as interpolating with dimension between these ensembles.

More concretely, for large enough systems and environments such that $\dsys\denv \to \infty$ (for example if $\denv$ is a function of $\dsys$, i.e., the maximal $\denv=\dsys^{2}$ necessary to implement any quantum channel) the cHaar ensemble approaches the Depolarize ensemble, and evenly dissipates quantum information in all directions of the Hilbert space. Moreover, even with finite-sized systems and environments, the perturbative next-leading-order non-unital contributions to the moment operator vanish inversely with system and environment dimensions, and are intriguingly independent of concatenations. The exact forms for cHaar moment operators generalize previous derivations of random quantum channels being asymptotically depolarizing  \cite{bai2024primitivity}, and offer a complete depiction of their average behavior.

On the other hand, for trivial environments such that $\denv \to 1$, the cHaar ensemble reduces to the Haar ensemble. The hierarchy between ensembles in part 3) of Theorem \ref{theorem:cHaar-explicit} then implies that concatenating cHaar ensembles decreases their norm, tending towards the base of the hierarchy set by the Depolarize ensemble. Thus, as a consequence of the dissipative nature of tracing out the environment, whereas cHaar ensembles with a finite environment are not maximally dissipative, their behavior can be made more so via concatenation. (We note that while the hierarchy between ensembles in Theorem \ref{theorem:cHaar-explicit} holds asymptotically when $\dsys\denv \to \infty$, we have explicitly verified it for $t=2,3,4$, $k = 1,3$, and various $\dsys,\denv$ in Fig. \ref{fig:moment_operator_norm} of Appendix \ref{app:properties_of_moment_operators}, and we conjecture that it holds for all $t$, $k$, and $\dsys,\denv$.)

Theorem \ref{theorem:cHaar-explicit} is derived explicitly in Appendix \ref{app:properties_of_moment_operators} by studying the spectrum of the cHaar moment operator. Namely, since the cHaar moment operator fails to be a projector, unlike the Haar and Depolarize moment operators, its eigenvalues indicate the inherent average dissipation of information of random quantum channels. This realization, as well as the derived concatenation-independent behavior of its next-leading-order non-unital contributions, prompted us to study the spectral properties of $\widehat{\Tau}_{\ensemblechaar}^{(t)}$.
\begin{theorem}[Spectrum of cHaar Moment Operator]\label{theorem:spectrum_of_chaar_moment_operator}
The $t$-th order cHaar moment operator has the exact leading term of a projector, \!\!
\begin{align}\label{eq:cHaar-leading}
	\widehat{\Tau}_{\ensemblechaar}^{(t)} =& \frac{\kket{\psi}\bbra{\varphi}}{\bbra{\varphi}\kket{\psi}} ~+~ \cdots~,
\end{align}
which implies that it has a leading eigenvalue $\lambda=1$ with associated right- and left-eigenvectors, given by,
\begin{align}\label{eq:right-eigen}
	\kket{\psi} = \sum_{\sigma \in \permutations[t]}
	\chi_{\dsys\denv}^{(t)}(\sigma)
	~\kket{\representation{\sigma}}
	\quad , \quad
	\kket{\varphi} = \kket{\representation{e}}~.
\end{align}
In the fixed $t$, asymptotic $\dsys\denv \to \infty$, the $t$-th order cHaar moment operator has a single leading eigenvalue $\lambda = 1$ from the above contribution, and all non-leading eigenvalues are, up-to $t$-dependent factors, $\lambda \in \mathcal{O}\left(\frac{1}{\denv^{}}\right) < 1$.
\end{theorem}

We note that a direct consequence of the previous theorem is that $k$-concatenations of the cHaar ensemble will have moment operators with the fixed points,
\begin{align}\label{eq:cHaar-leading-k}
	\widehat{\Tau}_{\ensemblechaar}^{(t)k} =& \frac{\kket{\psi}\bbra{\varphi}}{\bbra{\varphi}\kket{\psi}} + \cdots~.
\end{align}
Moreover, Theorem \ref{theorem:spectrum_of_chaar_moment_operator} implies that the cHaar ensemble moment operator's only fixed point is its eigenvector $\kket{\psi}$ of Eq.~\eqref{eq:right-eigen}. In turn, this leading eigenvector converges with dimension to the Depolarize channel's single fixed point
\begin{align}
	\lim_{\dsys\denv \to \infty} \kket{\psi} \to \kket{\varphi}
	\quad \leftrightarrow \quad
	\lim_{\dsys\denv \to \infty} \frac{\kket{\psi}\bbra{\varphi}}{\bbra{\varphi}\kket{\psi}} \to \channeldepolarize[\widehat]{\dsys}^{\otimes t}~.
\end{align}
This observation regarding the eigenvectors and the single, depolarization fixed point of the cHaar ensemble further underlies the hierarchy between ensembles in Theorem \ref{theorem:cHaar-explicit}.

Proposition \ref{prop:cHaar}, and Theorems \ref{theorem:cHaar-explicit} and \ref{theorem:spectrum_of_chaar_moment_operator} have several important implications, that culminate in the Depolarize ensemble being found to be a natural reference ensemble for ensembles of quantum channels.

\begin{claim}\label{claim:depolarize_ensemble_is_natural_generalization_quantum_channels}
The Depolarize ensemble $\ensembledep$ is a natural generalization of the Haar ensemble. The invariance properties of $\ensembledep$ impose that its moment operator is present in the moment operators of all quantum channel ensembles, and place the Depolarize ensemble at the bottom of the moment operator norm hierarchy of all quantum channel ensembles. Therefore, any ensemble of channels $\ensemblechannel$ satisfies
\begin{equation}\label{eq:chain-2} \norm{\widehat{\Tau}_{\ensembledep}^{(t)}}^{2}\leq \norm{\widehat{\Tau}_{\ensemblechannel\vphantom{\ensemblehaar}{}}^{(t)}}^{2}~.
\end{equation}
\end{claim}
\begin{claim}
The Depolarize ensemble therefore serves as a natural reference ensemble for quantum channels for defining $t$-designs.
Concretely, an ensemble $\ensemblechannel$ is defined to be an approximate Depolarize $t$-design with $\varepsilon$ additive error, if and only if
\begin{equation}
	\norm{{\Tau}_{\ensemblechannel}^{(t)} - {\Tau}_{\ensembledep}^{(t)}}_{\diamond} \leq \varepsilon~,
\end{equation}
and an approximate Depolarize $t$-design with $\varepsilon$ relative error if
\begin{equation}\label{eq:relative-error-def}
	(1-\varepsilon){\Tau}_{\ensembledep}^{(t)}~\preceq~ {\Tau}_{\ensemblechannel}^{(t)}~\preceq~(1+\varepsilon){\Tau}_{\ensembledep}^{(t)}~.
\end{equation}
\end{claim}

Importantly, in Appendix \ref{app:proof-lemma-errors} we prove the following lemma, which connects additive and relative errors, and provides operational meaning to these moment operators.
\begin{lemma}[Additive versus relative error]\label{lemma-errors}
 Any approximate Depolarize $t$-design with relative error
$\varepsilon_{\textnormal{rel}}$ is a $\varepsilon_{\textnormal{add}}$-approximate Depolarize $t$-design with an additive error $\varepsilon_{\textnormal{add}}=2\varepsilon_{\textnormal{rel}}$. Conversely, any approximate Depolarize $t$-design with additive error
$\varepsilon_{\textnormal{add}}$ is a $\varepsilon_{\textnormal{rel}}$-approximate Depolarize $t$-design with a relative error $\varepsilon_{\textnormal{rel}}=\dsys^{2t} \varepsilon_{\textnormal{add}}$.
\end{lemma}
Notably, the previous bounds connecting relative and absolute errors for quantum channels are as tight as those known for unitaries \cite{schuster2024random}.

It admittedly remains surprising that from our criteria of positions within hierarchies of moment operator norms, the most appropriate reference ensemble for quantum channels is ultimately found to be the Depolarize ensemble and not the cHaar ensemble. To explain this outcome, we recall that only when $\denv = \dsys^{2}$, does the cHaar ensemble have a uniform Lebesgue measure over quantum channels, and only when $\denv = d$, is there a resulting uniform distribution of quantum states output by the cHaar ensemble \cite{kukulski2021generating}. However, even at such special dimensions, the cHaar ensemble still does not share the defining characteristics of the Haar or Depolarize ensembles that make them the most appropriate reference ensembles, namely invariance under concatenations. In fact, cHaar ensembles for sufficiently large environment dimension $\denv$, with non-leading eigenvalues $\abs{\lambda} < 1$, will dissipate information upon concatenation. It is therefore proposed that criteria of ensemble hierarchies and ensemble invariance, as opposed to (the admittedly desirable property of) uniformity of measure, are used when considering future reference ensembles.

A possible explanation for why the Depolarize ensemble sits at the bottom of the norm hierarchy for quantum channels can be offered by the distribution of its Kraus operators. In particular, the squared Hilbert-Schmidt norm of the $t$-th order moment operator can be interpreted as the $t$-th moment of the overlaps between the Kraus operators in the ensemble, or frame potentials,
\begin{equation}
	\norm{\widehat{\Tau}_{\ensemblechannel}^{(t)}}^2 = \int_{\ensemblechannel \times \ensemblechannel} d\channel~d\channelother \sum_{K_{\channel},K_{\channelother}} \abs{\bbra{K_{\channelother}}\kket{K_{\channel}}}^{2t}~,
\end{equation}
where we used Eq.~\eqref{eq:kraus_superoperators}. In the unitary case, the Haar reference ensemble minimizes the frame potentials, and hence maximizes the spread of the corresponding (unitary) Kraus operators. In the case of more general channels, it is precisely the Depolarize ensemble that maximizes the spread of the Kraus operators, offering a geometric interpretation of its privileged role as a reference ensemble.

\begin{figure}[t!]
		\centering
		\includegraphics[width=0.925\linewidth]{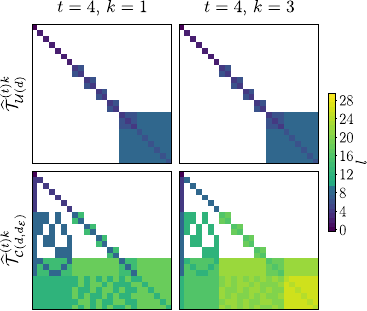}
		\caption{\textbf{Moment operators in a basis that block-diagonalizes the Haar moment operator as a permutation transfer matrix.} We show the scaling of the $k$-concatenated $t$-th order moment operators for the Haar (top) and cHaar (bottom) ensembles for $\denv = \dsys^{2}$, $k=1$ (left) , $k=3$ (right), and $t=4$. In this basis, the $t!$ elements are sorted by their support, from the smallest identity (top-left corner), to the largest cycle (bottom-right corner). Colors represent the leading-order scaling $l$ with $1/\dsys$ of elements $\tau_{\mathcal{C}}^{(t,k)} \sim \mathcal{O}(1/\dsys^{l})$.}
		\label{fig:haar_chaar_coefficients}
\end{figure}

Our derivations of the properties of the cHaar moment operator are made possible via a novel analysis of the transfer matrices $\tau_{\ensemblechannel}^{(t)}$ of the Haar and cHaar moment operator in the permutation basis, namely the $t!\times t!$ matrices that map permutations to permutations. In Appendices \ref{app:spaces_operators_and_permutations} and \ref{app:properties_of_moment_operators} we derive a crucial change of basis, from non-orthogonal permutations \cite{harrow2024approximate}, to so-called localized permutations. This novel localized basis consists of operators, each with support over a strict subset of copies of the Hilbert space, that are orthogonal to any operators with non-identical support. Due to its well-defined support and block-orthogonality, this basis block diagonalizes the Haar transfer matrix and block lower-triangularizes the cHaar transfer matrix \footnote{We note that this change of basis, in terms of $t! \times t!$ matrices, is not a $\dsys^t\times \dsys^t$ Schur transformation matrix \cite{harrow2005applications}.}. Indeed,  this localized basis is not merely a technical convenience, but the main structural ingredient that makes Theorems \ref{theorem:cHaar-explicit} and \ref{theorem:spectrum_of_chaar_moment_operator} accessible, and we expect it to be useful more broadly in permutation-based moment calculations.

The resulting transfer matrix elements are depicted in Fig. \ref{fig:haar_chaar_coefficients}, for $t=4$ and $k=1,3$ concatenations, obtained from closed-form expressions and symbolic methods \cite{cardin2026haarpy}. Shown is the leading-order scaling $l$ of matrix elements with $\dsys$. The Haar ensemble transfer matrix is block-diagonal, with elements that are invariant under concatenations. Such structure reflects that its moment operator is a projector and is support-preserving. The cHaar ensemble transfer matrix is block-lower-triangular, with elements that decay with concatenations, except for the first column which corresponds to its fixed-point leading eigenvector of Eq.~\eqref{eq:cHaar-leading}. Such structure reflects that its moment operator is non-unital and is support-non-decreasing, given it is generated by a support-preserving Haar ensemble, with its output environment traced over. Given the exact forms of the transfer matrices, in particular the $k$-concatenated transfer matrices $\tau_{\ensemblechannel}^{(t,k)}$, the spectral properties, norms and traces of the associated moment operators may be derived in Theorem  \ref{theorem:spectrum_of_chaar_moment_operator}.

\subsection{Moment Operators of Noisy Quantum Circuits}\label{sec:moment_operator_norms_of_noisy_ensemble}

\begin{figure}[t]
		\centering
		\includegraphics[width=0.95\linewidth]{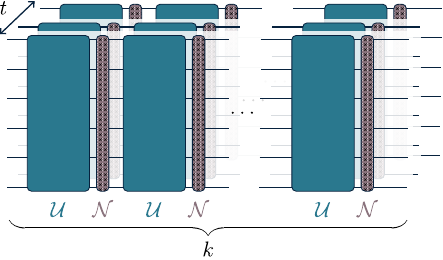}
		\caption{\textbf{Concatenation of random unitaries and noisy channels.} We study the $t$-th moment operator of a noisy circuit where we concatenate $k$ layers of random unitaries sampled from $\ensembleunitary$ (blue) with a fixed noise channel $\ensemblenoise$ (brown).}
		\label{fig:noisy_quantum_circuit}
		\vspace{-0.5cm}
\end{figure}

Here, we use our formalisms to study the moment operators of noisy quantum circuits. As per Fig. \ref{fig:noisy_quantum_circuit}, we will consider $k$-layers of random unitaries sampled from a unitary ensemble $\mathcal{U}$, followed by a fixed noise channel $\channelnoise$. The $t$-th order moment operator for this evolution is $(\widehat{\channelnoise}^{\otimes t} \widehat{\Tau}_{\ensembleunitary}^{(t)})^{k}$.

In what follows, we choose an orthogonal basis for $\mathcal{B}[\mathcal{H}]$, denoted as $\mathcal{P}_{\dsys} = \{\idsys,P,\dots,S\}$. This induces a basis for $\mathcal{B}[\mathcal{H}^{\otimes t}]$ given by $\mathcal{P}_{\dsys}^{(t)}=(\mathcal{P}_{\dsys})^{\otimes t}$. Finally, we define the operator support $\Gamma_{P} \subseteq [t]$ and locality $\abs{P} = \abs{\Gamma_{P}}$.

Next, we consider a constant noise model acting on $\spacesys$ in the basis $P,S \in \mathcal{S}_{\ensemblenoise}$  of,
\footnotesize
\begin{align}\label{eq:noise-channel}
	\!\!\!\widehat{\channelnoise}_{\gamma\eta} \!=\!\frac{1}{\dsys}\!
	\left[
	\begin{array}{cccc}
		1 & 0 & \cdots & 0 \\
		\eta_{\dsys}(P) & 1-\gamma_{\dsys}(P) & \cdots & 0 \\
		\vdots & \vdots & \ddots & \vdots \\
		\eta_{\dsys}(S) & 0 & \cdots & 1-\gamma_{\dsys}(S) \\
	\end{array}\right]\!,\!\!\!\!\!
\end{align}
\normalsize
where $\gamma_{\dsys}(P) \geq \gamma~,~\eta_{\dsys}(P) \leq \eta$, $\forall~ P \in \mathcal{P}_{\dsys}$, with appropriate constraints on $\gamma_{\dsys},\eta_{\dsys}$ such that $\channelnoise_{\gamma\eta}$ is completely positive \cite{greenbaum2015introduction}. This noise model is composed of a strictly diagonal part which reduces the magnitude of operators, and a non-unital part that maps identity to non-identity operators, encompasses both unital Pauli-like and non-unital noise \cite{wilde2013quantum,sharma2019noise}, and allows us to understand how noise propagates across copies of Hilbert spaces. In particular, we note that when $d=q^n$ (e.g., for $q=2$, an $n$ qubit system), Eq.~\eqref{eq:noise-channel} should be viewed as an effective model for constant-rate local noise on $n$ qubits, whereby tensor-product local unital channels scale an operator of locality $l$ by a factor $(1-\gamma)^l$, and non-unital channels generate identity-to-non-identity contributions of order $\eta$. The $t$-th fold tensor product superoperator for the noise component acting on $\spacesys^{\otimes t}$ in the basis $P,S \in \mathcal{S}_{\ensemblenoise_{\gamma\eta}}^{(t)} = \mathcal{P}_{\dsys}^{t}$ is thus,
\begin{equation}
		\widehat{\channelnoise}_{\gamma\eta}^{\otimes t} =~ \frac{1}{\dsys^{t}}\sum_{P,S \in \mathcal{S}_{\ensemblenoise_{\gamma\eta}}^{(t)}} \tau_{\ensemblenoise_{\gamma\eta}}^{(t)}(P,S) ~ \kket{P}\bbra{S}~,
\end{equation}
with transfer matrix elements
\begin{align}
	\tau_{\ensemblenoise_{\gamma\eta}}^{(t)}(P,S)
	\in&~ \mathcal{O}\!\left((1 - \gamma)^{\abs{S}}~\eta^{\abs{P}-\abs{S}}\right)~\delta_{\Gamma_{P} \supseteq \Gamma_{S}}~.
\end{align}
Under this framework, the following theorem holds, regarding the effect of noise on channel design properties.

\begin{theorem}[Noise-Induced Depolarize $t$-Designs]\label{theorem:unital_noise_decreases_and_nonunital_noise_increases_moment_operator_norms}
For $k$-concatenated $t$-th order unitary ensembles, diagonal unital noise decreases, and non-unital noise increases moment operator norms. The respective ensembles approach and deviate from being Depolarize $t$-designs.
\end{theorem}

More explicitly, we can compute the scaling of moment operator norms. First, consider a unitary ensemble $\ensembleunitary$ is an $\varepsilon$-$t$-design with respect to a Haar unitary ensemble $\ensembleunitary^{\prime}$, with $t$-order commutant elements $\mathcal{S}_{\ensembleunitary^{\prime}}^{(t)}$ with locality $\geq l$, and rank-$r$ moment operators. Next, let $\channelnoise_{\gamma}$ denote a diagonal unital noise channel with noise coefficients bounded by $\gamma_{\dsys}(P) \geq \gamma$, $\forall~ P \in \mathcal{P}_{\dsys}$. Here we have the exact bound,
\small
\begin{align}
	\norm{\!\Big({\ensemblenoise}^{\otimes t}_{\gamma}{\Tau}_{\ensembleunitary}^{(t)}\Big)^{k} - \channeldepolarize{\dsys}^{\otimes t}}_{\diamond}  \leq ~ \dsys^{t}~(r-1+\varepsilon)^{k}~(1-\gamma)^{k}~.  \label{eq:depolarization_t_design}
\end{align}
\normalsize
If the unitary ensemble $\ensembleunitary$ is an exact Haar-distributed group $t$-design, or consists of randomly parameterized unitaries $\ensemblevariable = \{\unitary_{\theta}^{G} = e^{-i\theta G}, G \in \mathcal{P}_{\dsys}, \theta \sim \Theta\}$ with parameters $\theta \sim \Theta$, and a single involutory $G$ with commutant $\mathcal{S}_{G}$,  and we apply a diagonal unital $\channelnoise_{\gamma}$ and or non-unital noise $\channelnoise_{\gamma\eta}$ with noise coefficients $\gamma_{\dsys}(P) \geq \gamma~,~\eta_{\dsys}(P) \leq \eta$, $\forall~ P \in \mathcal{P}_{\dsys}$, we have the leading-order scaling,
\begin{align}
\!\!\!
	\norm{\Big(\widehat{\ensemblenoise}^{\otimes t}_{\gamma}\widehat{\Tau}_{\ensembleunitary}^{(t)}\Big)^{k} \!\!- \channeldepolarize[\widehat]{\dsys}^{\otimes t}}_{1}
	=&~ (r-1)~\mathcal{O}\!\left((1-\gamma)^{lk}\right) \nonumber\\
	\!\!\!
	\norm{\Big(\widehat{\ensemblenoise}_{\gamma\eta}^{\otimes t}\widehat{\Tau}_{\ensembleunitary}^{(t)}\Big)^{k} \!\!- \channeldepolarize[\widehat]{\dsys}^{\otimes t}}^{2}_{2}
	=&~ t~\abs{\mathcal{P}_{\dsys}\backslash \{\idsys\}}~\mathcal{O}\!\left(\eta^{2}\right) \label{eq:design}\\
	\!\!\!
	\norm{\Big(\widehat{\ensemblenoise}_{\gamma}^{\otimes t}\widehat{\Tau}_{\ensemblevariable}^{(t)}\Big)^{k} \!\!- \channeldepolarize[\widehat]{\dsys}^{\otimes t}}^{2}_{2}
	=&~ t~\abs{\mathcal{S}_{G} \backslash \{\idsys\}}~\mathcal{O}\!\left((1-\gamma)^{2k}\right)~.\nonumber
\end{align}
Here, in the $\gamma,\eta \to 0$ limit, independent of $\dsys,t$, the leading-order $\mathcal{O}(\gamma,\eta)$ behavior is due to $k$ concatenations of the noise component acting on any unitary component operators of minimum locality $l$. The leading-order scaling is thus only dependent on the locality and rank of the unitary operators, and is otherwise independent of the specific unitary ensemble. Unital noise is inherently depolarizing, decreasing moment operator norms exponentially with concatenations and converges toward a Depolarize $t$-design.

To understand the previous noise-induced behavior, recall that Haar-distributed unitary ensemble moment operators are support-preserving and consist of locality-$l$, rank-$r_{l}\leq r$ projectors. Unital noise predominantly scales such projectors by $\sim (1-\gamma)^{lk}$ throughout the $k$-concatenations, leading to an effect proportional to the total amount of noise, $\sim 1-lk\gamma$, independent of the number of operators scaled. Conversely, non-unital noise predominantly occurs at the last concatenation, independent of $k$, scaling $\mathcal{P}_{\dsys}\backslash \{\idsys\}$ number of non-identity basis operators mapped to from the identity. Any intermediate non-unital noise that increases support at concatenation $s<k$ would contribute next-leading-order scaling $\sim \eta(1-\gamma)^{l(k-s)}$, leading to a mixture of unital and non-unital effects as unitary and noise basis operators are mapped between each other.

\begin{figure*}[t!]
		\centering
		\includegraphics[width=\linewidth]{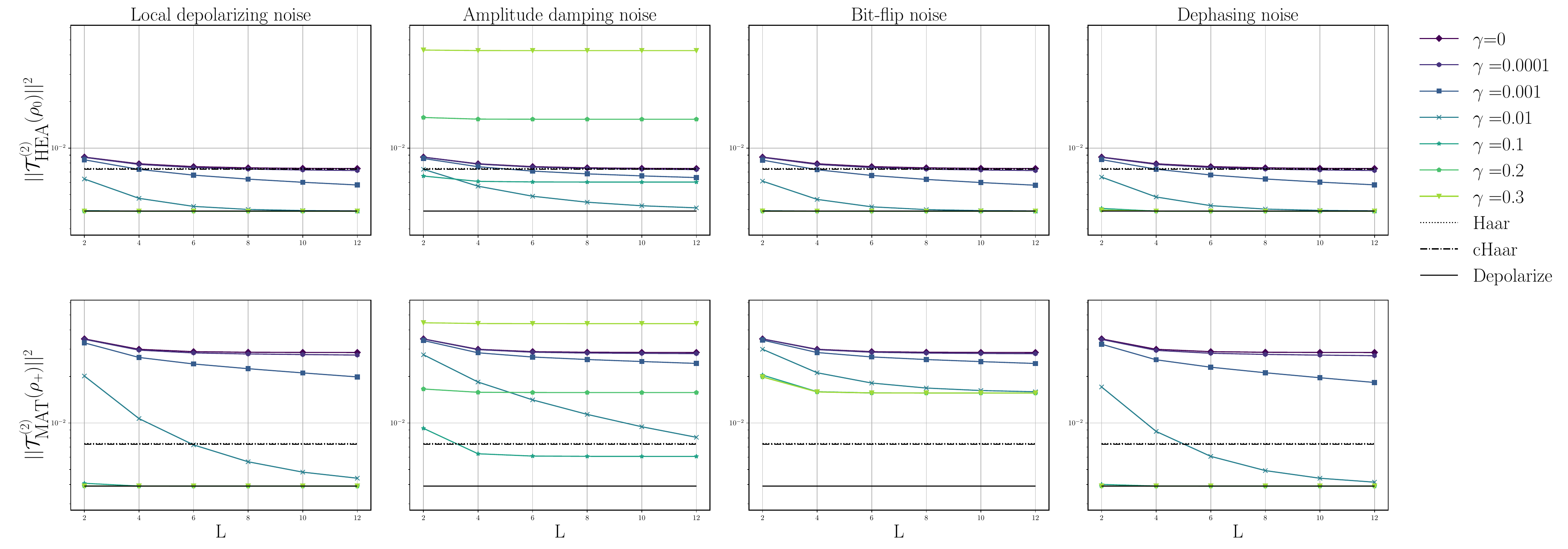}
		\caption{{\bf Effects of noise on $||\Tau_{\mathcal{C}}^{(2)}(\rho)||^2$.} We study two parameterized quantum circuit architectures: a hardware-efficient ans\"atz (HEA) and a matchgate ans\"atz (MAT), on $n=7$ qubits and $L\in[L]$ layers. We take the variational parameters uniformly in $[0,2\pi]$, and we simulate the effect of four different types of single-qubit noise channels, namely, local depolarizing, amplitude damping, bit-flip and dephasing nose, with noise strengths $\gamma\in[0,0.3]$. The input states for the HEA and MAT circuits are $\rho_0$ and $\rho_+$, respectively. We also plot the value for the Haar, cHaar with $\denv=d^2$ (i.e., the Lebesgue measure) and Depolarize reference ensembles. These simulations show that certain types of noise can decrease the Hilbert-Schmidt norm of the $2$-order moment operator (e.g., depolarizing), while others can increase it (e.g., amplitude damping).}
		\label{fig:numerics}
\end{figure*}

Finally, let us consider a direct application of our framework to reinterpret the results of \cite{dalzell2021random,deshpande2021tight}. Therein, it has recently been shown that random measurement probabilities $p_{\gamma}$ from random circuits with local noise converge to measurement probabilities from globally depolarized noisy circuits $p$, in average total variational distance $\tnorm{p_{\gamma}-p}_{1}$ \cite{dalzell2021random,deshpande2021tight}. Using the fact that the moment operator norms serve as bounds for the diamond norm, which in turn bounds the total variational distance, our noisy moment operator norms $\tnorm{\widehat{\Tau}-\widehat{\Tau}_{\gamma}}_{1}$ are shown to be interpretable, scaling similarly, and potentially bounding these variational distance results,
\begin{align}
	\!\!\!
	\expval{\norm{p_{\gamma}-p}_{1}}_{\ensembleunitary\ensemblenoise_{\gamma}} \sim&~ \mathcal{O}\!\left({e^{-2k\gamma}}\right)
	\leq \expval{\norm{\channelnoise_{\gamma}\channelunitary - \channeldepolarize{\dsys}}_{\diamond}}_{\ensembleunitary\ensemblenoise_{\gamma}}~\!.\!\!
\end{align}
Then, since we have derived the bounds,
\begin{align}
	\!\!\!\!
	\norm{\Tau_{\ensembleunitary\ensemblenoise_{\gamma}}^{(t)} - \channeldepolarize{\dsys}^{\otimes t}}_{\diamond}
	\!\!\!\leq&~\!
	\dsys^{t} \norm{\widehat{\Tau}_{\ensembleunitary\ensemblenoise_{\gamma}}^{(t)} - \channeldepolarize[\widehat]{\dsys}^{\otimes t}}_{1}
	\!\!\!\sim\! \mathcal{O}\!\left(\dsys^{t} t!{(1-\gamma)^{2k}}\right)~\!\!\!,\!\!\!\!
\end{align}
our diamond norm bounds imply that for qudit systems, depth that is at least linear in system size, $n = \log(\dsys)$,
\begin{align}
	k = \Omega\left(t~\frac{\log({\dsys})+\log({t})}{\log({\frac{1}{1-\gamma}})}\right)~,
\end{align}
is sufficient for $t$-design Haar unitary ensembles plus unital noise to converge to Depolarize $t$-designs. Such loosely bounded linear scaling depths with $n$ for all $t$-designs are comparable to previous derivations of $k = \Omega(n\log({n}))$ number of channels, or $\Omega(\log({n}))$ depth for $\varepsilon$-$t=2$-design Haar unitary ensembles with local noise \cite{dalzell2021random,deshpande2021tight}. Our bounds can potentially be tightened if focusing specifically on the $t=2$ case, given that such moments operators and their norms can be explicitly calculated. Our explicit expressions for moment operators also suggest their $1$-norm, as opposed to the $2$-norm, can possibly be calculated, leading to tighter bounds on the twirl diamond norm for non-unital noise or parameterized ensembles.

\subsection{Channel Designs and Expectation Value Concentration}\label{subsec:designs_and_trainability}

In this section, we relate concentration of expectation values to channel design properties. As previously discussed, moment operator norms for ensembles of unitaries have been given operational meaning in the context of variational algorithms, as they can be directly related to the concentration phenomena \cite{holmes2021connecting}. Here, we extend such intuition to the case of channels, and show that similar phenomena are exhibited by ensembles of quantum channels.

In particular, let us define the expectation value
\begin{align}\label{eq:loss}
	\mathcal{L}_{\channel}(\rho,O) = \Tr\left[\channel(\rho)O\right]~,
\end{align}
where $\rho$ is a quantum state, $O$ an observable, and $\channel$ is a channel from an ensemble $\ensemblechannel$. We can bound the probability of $\mathcal{L}_{\channel}$ deviating from its mean $\mu_{\ensemblechannel} = \texpval{\Tr\left[\channel(\rho) O \right]^{\vphantom{2}{}}}_{\ensemblechannel}$ by a quantity $\epsilon \geq 0$, in terms of its variance $\sigma^{2}_{\ensemblechannel} = \texpval{\Tr\left[\channel(\rho) O \right]^{2}}_{\ensemblechannel}-\texpval{\Tr\left[\channel(\rho) O \right]^{\vphantom{2}{}}}_{\ensemblechannel}^{2}$, via Chebyshev's inequality
\begin{align}
	\text{Probability}\left[\abs{ \mathcal{L}_{\channel} - \mu_{\ensemblechannel}^{} } \geq \epsilon\right] ~\leq~ \frac{\sigma_{\ensemblechannel}^{2}}{\epsilon^{2}}~.
\end{align}
Since computing variances $\sigma_{\ensemblechannel}^{2}$ for arbitrary ensembles of quantum channels $\ensemblechannel$ may be difficult, we can instead study concentration phenomena by comparing how close $\ensemblechannel$ is to a reference ensemble $\ensemblechannel^{\prime}$. In particular, one can show that the following theorem holds

\begin{theorem}[Variance Bounds in terms of Reference Ensemble]\label{theorem:variance_bounds_in_terms_of_reference_ensembles}
Given an expectation value $\mathcal{L}_{\channel}$ and ensembles of quantum channels $\ensemblechannel,\ensemblechannel^{\prime}$, the variance of $\mathcal{L}_{\channel}$ over $\ensemblechannel$ is upper bounded by, the variance of $\mathcal{L}_{\channel^{\prime}}$ over $\ensemblechannel^{\prime}$, and design terms that depend on $\ensemblechannel$ being an $\varepsilon$-$t$-design with respect to $\ensemblechannel^{\prime}$,
\begin{align}
	\sigma_{\ensemblechannel}^{2}
	~=&~~ \sigma_{\ensemblechannel^{\prime}}^{2} ~+~ \delta_{\ensemblechannel\ensemblechannel^{\prime}}^{(2)}~,
\end{align}
with deviations from the reference ensemble variance, upper bounded by terms that depend on various choices of norm,
\begin{align}
	\delta_{\ensemblechannel\ensemblechannel^{\prime}}^{(2)}
	\leq&~ ~~~~~\min_{1/p + 1/q = 1}\norm{\rho}_{p}^{2}~\norm{O}_{q}^{2}~\tnorm{\Tau_{\ensemblechannel}^{(2)} - \Tau_{\ensemblechannel^{\prime}}^{(2)}}_{p} \\
	 &~+~~ \min_{1/p + 1/q = 1}2~\abs{\mu_{\ensemblechannel^{\prime}}}~\norm{\rho}_{p}\norm{O}_{q}~\tnorm{\Tau_{\ensemblechannel}^{(1)} - \Tau_{\ensemblechannel^{\prime}}^{(1)}}_{p}~. \nonumber
\end{align}

In particular, when $\rho$ is a quantum state, the observable $O$ has trace $\abs{\Tr[O]} \leq 1$ (e.g., a projector or Pauli operator), the reference ensemble is $\ensemblechannel^{\prime} \in \{\ensemblechaar,\ensembledep\}$, and the norms are diamond $\norm{\cdot}_{\diamond}$ or operator $\norm{\cdot}_{\textnormal{op}}$ norms,
\begin{align}
	\hspace{-0.2cm}
	\sigma_{\ensemblechannel^{\prime}}^{2}
	=&~ \left\{ \begin{array}{ll}
		\mathcal{O}\!\left(\frac{1}{\dsys^{2}\denv}\right)~\norm{\rho}_{2}^{2}~\norm{O}_{2}^{2} & \ensemblechannel^{\prime}=\ensemblechaar \\
		0 & \ensemblechannel^{\prime}=\ensembledep
		\end{array} \right. ~. \\
	\delta_{\ensemblechannel\ensemblechannel^{\prime}}^{(2)} \leq&~ \\
	~~~~\norm{O}_{\infty}^{2}&\tnorm{\Tau_{\ensemblechannel}^{(2)} \! - \! \Tau_{\ensemblechannel^{\prime}}^{(2)}}_{\diamond} \!\! ~+~ 2\frac{1}{\dsys}\norm{O}_{\infty}\tnorm{\Tau_{\ensemblechannel}^{(1)} \!\! - \! \Tau_{\ensemblechannel^{\prime}}^{(1)}}_{\diamond} ~,~ \nonumber \\[-6pt] & ~~\quad \quad \quad \quad \quad\quad 	\textnormal{or} ~ \nonumber \\[-6pt]
	\norm{\rho}_{2}^{2}\norm{O}_{2}^{2}&\tnorm{\Tau_{\ensemblechannel}^{(2)} \!\! - \! \Tau_{\ensemblechannel^{\prime}}^{(2)}}_{\textnormal{op}} \!~+~\! 2\frac{1}{\dsys}\norm{\rho}_{2}\norm{O}_{2}\tnorm{\Tau_{\ensemblechannel}^{(1)} \!\! - \! \Tau_{\ensemblechannel^{\prime}}^{(1)}}_{\textnormal{op}}  ~.  \nonumber
\end{align}
\end{theorem}
These bounds must be considered case-by-case, depending on which norms of states, observables, and moment operator differences are optimal. Theorem \ref{theorem:variance_bounds_in_terms_of_reference_ensembles} encompasses the results for unitary ensembles of Ref. \cite{holmes2021connecting}, and allows us to study non-unitary concentration phenomena. For instance, noise-induced barren plateaus in Ref. \cite{wang2020noise} can be understood more generally as arising when the ensemble of noisy channels becomes an approximate $t$-design with respect to the Depolarize ensemble. In particular, within our framework, the key condition underlying the concentration results of Ref. \cite{wang2020noise} translates to unital noise scales $\gamma>0$, in which case the noisy channel ensemble is driven toward the depolarizing channel, as per Eq.~\eqref{eq:design}.

\section{Numerical Experiments}
\label{sec:numerics}

In this section we present numerical simulations for ensembles of noisy parameterized quantum circuits. Specifically, we study the behavior of the norm of $\Tau_{\mathcal{C}}^{(2)}$ twirls applied to an initial pure state $\rho$, for different circuit depths $L$ and noise strengths $\gamma$.

We consider the following (noiseless) unitary ans\"atze for $n$ qubits: hardware efficient ans\"atze (HEA) \cite{larocca2021diagnosing} and matchgate ans\"atze (MAT) \cite{terhal2002classical}, each with a set of generators $\mathcal{G}$ in terms of the $X_i, Y_i$, $Z_i$ local Pauli operators acting on qubit $i \in [n]$,

\begin{align}
	\mathcal{G}_{\rm HEA} =&~ \left\{ X_i \right\}_{i=0}^{n-1} \cup \left\{ Y_i \right\}_{i=0}^{n-1} \cup \left\{ Z_i Z_{i+1} \right\}_{i=0}^{n-2} ~, \label{eq:hea_generators} \\
	\mathcal{G}_{\rm MAT} =&~ \left\{ X_i \right\}_{i=0}^{n-1} \cup \left\{ Z_i Z_{i+1} \right\}_{i=0}^{n-2} ~.\label{eq:tfim_generators}
\end{align}

The noiseless parameterized quantum circuits $\{U(\vec{\theta})\}_{\vec{\theta}}$ have a layered structure, obtained from the generators $\{G_{k}\}$ via exponentiation as
\begin{equation}
	U(\vec{\theta}) = \prod_{l=0}^{L-1}\prod_{k=0}^{K-1} e^{-i\theta_{lk} G_{k}} ~,
\end{equation}
where $\vec{\theta}=(\theta_0,\dots,\theta_{LK-1})$ are variational parameters, distributed uniformly in $[0,2\pi]$, and $G_{k} \in \mathcal{G}_{HEA} ~\textnormal{or}~ \mathcal{G}_{MAT}$. As initial states, for HEA we choose $\rho_0 \equiv(\ketbra{0}{0})^{\otimes n}$, and for MAT we choose $\rho_+\equiv\left(\frac{(\ket{0}+\ket{1})(\bra{0}+\bra{1}])}{2}\right)^{\otimes n}$.

We also consider the following noise models consisting of local noise channels acting on qubit $i \in [n]$ after each unitary gate $e^{-i\theta_{lk}G_{k}}$. In particular, we consider bit-flip ($\channel_{BF}^{i}$), dephasing ($\channel_{D}^{i}$), local-depolarizing ($\channel_{LD}^{i}$) and amplitude damping ($\channel_{AD}^{i}$) noise. Given a probability $\gamma$ of noise occurring, also referred to as the noise strength, the action of these local noise channels is,
\begin{align}
	\channel_{BF}^{i}(\rho) &= (1-\gamma) \rho + \gamma ~X_i \rho X_i ~, \nonumber\\
		\channel_{D}^{i}(\rho) &= (1-\gamma) \rho + \gamma ~Z_i \rho Z_i ~, \nonumber\\
	\channel_{LD}^{i}(\rho) &= (1-\gamma) \rho + \frac{\gamma}{3} \left(X_i \rho X_i + Y_i \rho Y_i + Z_i \rho Z_i\right) ~, \nonumber\\
	\channel_{AD}^{i}(\rho) &= K_{AD_{1}}^{i} \rho K_{AD_{1}}^{i \dagger} + K_{AD_{2}}^{i} \rho K_{AD_{2}}^{i \dagger} ~,
\end{align}
where the amplitude damping channel Kraus operators are,
\begin{equation}
	\!\!\!\!
	K_{AD_{1}}^{i}\equiv \sqrt{\gamma} ~\ket{0}_i\bra{1}~, K_{AD_{2}}^{i}\equiv \ket{0}_i\bra{0} + \sqrt{1-\gamma}~ \ket{1}_i\bra{1}. \!\!\!\!
\end{equation}

In Fig. \ref{fig:numerics}, we plot the (squared) Hilbert-Schmidt norm, or the purity, of the average of $t=2$ copies of states acted on by channels from the ensemble generated by noisy circuits, as a function of circuit depth $L$ and noise strength $\gamma$. From these results, we can compare the behavior of the HEA and MAT ensembles, in both the noiseless and noisy cases.

First, we consider the noiseless case. We observe that $||\Tau_{\rm HEA}^{(2)}(\rho_0)||^2$ converges with depth towards the Haar random value, meaning the circuit becomes an $\varepsilon$-approximate state $2$-design \cite{mcclean2018barren}. In contrast, $||\Tau_{\rm MAT}^{(2)}(\rho_+)||^2$ does not converge towards the Haar random value, but instead plateaus at a higher value and does not become a $\varepsilon$-approximate state $2$-design. Intuitively, this is a consequence of the fact that the HEA generators in Eq.~\eqref{eq:hea_generators} produce controllable circuits (meaning that any unitary from $\mathbb{SU}(2^n)$ can be implemented given sufficient depth and an appropriate choice of parameters). Conversely, the MAT generators in Eq.~\eqref{eq:tfim_generators} produce free-fermionic evolutions, which are a small subgroup of $\mathbb{SU}(2^n)$ that do not form state designs.

Next, we consider the noisy case. Under local depolarizing noise channels, both HEA and MAT norms converge with depth to the absolute minimum norm, namely, the one given by the Depolarize ensemble (as is expected from previous studies of noise-induced phenomena in Ref. \cite{wang2020noise}). Similarly, under local dephasing noise, both HEA and MAT norms also converge with depth to the absolute minimum Depolarize ensemble norm. However, under local amplitude damping noise, interestingly both HEA and MAT norms decrease when the noise strength is small, and increases with increasing noise strength. Finally, under local bit-flip errors, both HEA and MAT norms are decreased, although the HEA norm is driven towards the minimum Depolarize ensemble norm, whereas the MAT norm converges to a larger value.

These numerical experiments showcase that certain types of noise can decrease or increase the norm of the moment operator, and reveal a subtle interplay between the exact circuit that is run on a quantum computer, and the type and strength of noise present in the device.

The simulations have been performed with the open-source library \texttt{Qibo} \cite{efthymiou2020qibo,efthymiou2022quantum}.
One of the main technical issues in numerically evaluating twirl norms, via sampling from an ensemble, is that the error in such estimations is often large compared to the norm itself, unless a prohibitively large number of samples are simulated. Hence, we employ a method to evaluate twirl norms for $t = 2$ exactly, avoiding any sampling error. The method is similar to that in Ref. \cite{heyraud2023efficient} and is based on the following proposition, whose proof can be found in Appendix \ref{subsec:unital_noise_and_parameterized_random_unitary_channels}.

\begin{proposition} \label{prop2}
	The average adjoint action of $\unitary^{\otimes 2}$, with $U \equiv e^{i \theta G} \sim \ensemblevariable$, for fixed generators $G \in \mathcal{G}$, $G^{2}=\idsys$, and a uniform distribution of parameters $\theta \in \Theta = [0,2\pi]$, is
		\small
	\begin{equation} \label{eq:prop2}
		\Tau_{\ensemblevariable}^{(2)}\!\left(\cdot\right) =~ \frac{3}{8} \left(\cdot + G^{\otimes 2}\cdot G^{\otimes 2}\right) -\frac{1}{8} \left\{\cdot~, G^{\otimes 2} \right\} ~+\frac{1}{2}G^{(2)} \cdot G^{(2)}~ ,\nonumber
	\end{equation}
 \normalsize
 where $G^{(2)}= G\otimes \idsys +\idsys\otimes G$.
\end{proposition}

\section{Conclusions}\label{sec:conclusions}

In this work, we introduced a framework for studying moment operators of ensembles of random quantum channels. A key contribution is the proposal of three reference ensembles: Haar, cHaar, and Depolarize, for which we derived exact expressions for their spectra, traces, and norms, and obtained a hierarchy between these ensembles. In particular, we showed that the cHaar ensemble interpolates (via the environment dimension) between the unitary and depolarizing limits. Indeed, we found that its dominant fixed point is predominantly depolarizing with a perturbatively non-unital component, highlighting persistent non-unital effects and the intrinsic dissipative nature of quantum channels \cite{fefferman2024effect,mele2024noise,singkanipa2024beyond,hirche2020contraction}. Finally, instead of uniformity of measure, derived hierarchies between ensembles, generally arising from invariance properties of ensembles, are shown to be more versatile criteria to compare ensembles in these more general settings. Based on these criteria, we have shown that the Depolarize ensemble is the most natural reference ensemble for ensembles of quantum channels.

The norm of the ensemble moment operator plays a central role in describing relationships between ensembles. It is natural to ask for further operational interpretation of this quantity, particularly to describe the capabilities of random quantum channels. In the case of ensembles of random unitaries, this quantity has been interpreted as the expressive power of the ensemble \cite{sim2019expressibility}. However, in the case of ensembles of random quantum channels, given their inherent depolarizing and therefore not necessarily useful behavior, this feels less applicable to describe channel-centric algorithms \cite{lohani2021improving}. The Depolarize or asymptotic dimension cHaar reference ensembles do capture in some sense the "uniformity" or "symmetry" of an ensemble, in terms of its ability to dissipate information evenly in all directions, however neither term is quite satisfactory. Perhaps, as channel-centric algorithms become more prevalent, similarly to how the laser was presented originally as a "solution looking for a problem", here we have a "property looking for a name". We leave finding a nice, operationally meaningful, name for this important quantity as future work.

Applying the formalism to noisy parameterized circuits with diagonal unital and non-unital Pauli-like noise, we observed opposite trends for moment-operator norms. Unital noise typically decreases such norms, while non-unital amplitude damping increases them. Circuit depth drives ensembles toward the appropriate reference designs, and concentration of observables emerges. Such analysis ultimately offers a channel-design-centric view on noisy barren-plateau–type phenomena \cite{larocca2024review,holmes2021connecting,wang2020noise,fefferman2024effect,mele2023introduction,singkanipa2024beyond}. More broadly, this perspective suggests that noise need not be viewed solely as a passive perturbation. Instead, by driving channel ensembles toward or away from the Depolarize reference ensemble, it can itself become the mechanism governing concentration phenomena.

By establishing the previous clear connection to quantum machine learning, we highlight that our framework can also be directly relevant to more general hybrid classical–quantum protocols that rely on sampling and inversion (e.g., randomized benchmarking, classical shadows, and error-mitigation). Known design parameters and exact moments can potentially tighten sample-complexities and guide experimental choices, particularly when noise models are difficult to characterize within devices \cite{elben2022randomized,magesan2011scalable,kunjummen2023shadow,hama2023quantum,ijaz2024more,berg2022probabilistic}.

Moreover, we note that recent work has shown that design-like randomness can emerge dynamically in chaotic many-body systems, even without explicit sampling from a Haar ensemble. In particular, projected or subsystem ensembles generated by chaotic dynamics can realize exact or approximate quantum state-design behavior \cite{ho2022exact,cotler2023emergent,claeys2022emergent,ippoliti2023dynamical}, and this perspective has also been explored experimentally \cite{choi2023preparing}. Our framework suggests a natural channel-level extension for these ideas, whereby one studies the subsystem quantum channels induced by chaotic or noisy dynamics on a system plus environment. Here, one may ask whether the associated channel moments approach those of a reference channel ensemble, and if any design properties emerge. This viewpoint is also closely related to recent work on push-forward designs, including channel-design constructions arising from Stinespring-type maps \cite{czartowski2025quantum}.

Likewise, recent results on noisy random circuits indicate that sufficiently random dynamics can scramble local unital noise into an effectively global white-noise contribution, whereas non-unital noise can lead to qualitatively different behavior \cite{dalzell2021random,fefferman2024effect}. Establishing general conditions under which chaotic open-system dynamics, monitored dynamics, or noisy many-body evolution give rise to approximate channel designs is an interesting direction for future work.

To finish, we note that our work paves the way for several future research directions that merit follow-up: (i) non-asymptotic spectral analyses and representation-theoretic proofs for the cHaar–Haar hierarchy and related block structures, with links to Weingarten calculus and partition algebras \cite{collins2017weingarten,harrow2024approximate,bai2024primitivity}; (ii) characterization of more specific, physically motivated, and experimentally relevant ensembles of channels (e.g., Stinespring unitaries beyond Haar, structured entanglement, composite ans\"atze), and principled criteria for channel designs \cite{garcia2023deep,garcia2024architectures,ge2016area,czartowski2025quantum,webb2016clifford,diaz2023showcasing}; and (iii) targeted experiments that probe higher-order moments and benchmark predicted concentration trends, alongside implications for simulability and noise-assisted or active-error-mitigation strategies \cite{guimaraes2023noise,lidar2014review,so2025symmetry,mele2024noise,strydom2022implementation}.

\section*{Data Availability}
The data that support the findings of this article are not publicly available upon publication because it is not technically feasible and/or the cost of preparing, depositing, and hosting the data would be prohibitive within the terms of this research project. The data are available from the authors upon reasonable request.

\begin{acknowledgments}
The authors would like to thank Martin Larocca, Zachary Mann, and Mark M. Wilde, for their valuable insights into noisy quantum processes. M.D. would like to acknowledge the support of the Natural Sciences and Engineering Research Council of Canada (NSERC), and of the U.S. Department of Energy (DOE) through a quantum computing program sponsored by the Los Alamos National Laboratory (LANL) Information Science \& Technology Institute. Research at the Perimeter Institute is supported in part by the Government of Canada through the Department of Innovation, Science and Economic Development Canada and by the Province of Ontario through the Ministry of Economic Development, Job Creation and Trade. DGM acknowledges support by the Laboratory Directed Research and Development (LDRD) program of LANL under Project Number 20260043DR, as well as by LANL's ASC Beyond Moore’s Law project. Z.H. acknowledges support from the Sandoz Family Foundation-Monique de Meuron program for Academic Promotion. M.C. was supported by the U.S. DOE, Office of Science, Office of Advanced Scientific Computing Research through the Accelerated Research in Quantum Computing Program MACH-Q project. This research used resources provided by the LANL Institutional Computing Program, which is supported by the U.S. DOE National Nuclear Security Administration under Contract No. 89233218CNA000001
\end{acknowledgments}

\bibliography{main}

\begin{thebibliography}{139}
\bibitem [{1()}]{1}%
  \BibitemOpen
  \href@noop {} {}\bibinfo {note} {While we here consider sampling unitaries
  from $\ensemblehaar [\dsys \denv ]$, one can also consider the case when the
  unitaries are sampled from some subgroup of $\ensemblehaar [\dsys \denv ]$,
  in which case $\nu _{\spaceenv }$ is no longer arbitrary, but needs to be
  specified, and the overall ensuing channel properties will depend on these
  choices. We leave the study of these cases for future work.}\BibitemShut
  {Stop}%
\end{thebibliography}

\newpage
\clearpage
\onecolumngrid
\appendix
\section{Spaces, Operators, and Permutations} \label{app:spaces_operators_and_permutations}
In this appendix, we present preliminary definitions, including relevant spaces, operators, norms, and permutation operators, that will be useful throughout the rest of this work.\\

In what follows, we will denote the $\dsys$-dimensional Hilbert space as $\spacesys=\mathbb{C}^{\dsys}$, the set of bounded linear operators acting on $\spacesys$ as $\mathcal{B}[\spacesys]$, and the set of bounded linear superoperators acting on operators in $\mathcal{B}[\spacesys]$ as $\mathcal{B}[\mathcal{B}[\spacesys]]$. For convenience, we also recall the vectorization map, where given an operator $X \in \mathcal{B}[\spacesys]$, its vectorization is $X = \sum_{\alpha,\beta \in [\dsys]} X_{\alpha\beta} \ket{\alpha}\bra{\beta} ~\xrightarrow[{\rm vec}]{}~ \kket{X} = \sum_{\alpha,\beta \in [\dsys]} X_{\alpha\beta} \ket{\alpha\beta}$. Given two operators $X,Y \in \mathcal{B}[\spacesys]$, their normalized inner product, or overlap is
\begin{align}
	\chi_{\dsys}^{}(X,Y) = \frac{1}{\dsys}\Tr\!\left[X^{\dagger} Y\right] = \frac{\bbra{X}\kket{Y}}{\dsys}~
	\quad \quad ~,~ \quad \quad
	\chi_{\dsys}^{}(X) = \frac{1}{\dsys}\Tr\!\left[X\right] = \frac{\bbra{\idsys}\kket{X}}{\dsys} ~,
\end{align}
We will also make use of an orthonormal basis for the space of operators, denoted as $\mathcal{P}_{\dsys} = \{P,S \in \mathcal{B}[\spacesys] ~:~ \chi_{\dsys}^{}(P,S) = \delta_{PS}\}$, and which satisfies $\mathcal{B}[\spacesys] = \textnormal{span}_{\mathbb{C}}\{\mathcal{P}_{\dsys}\}$. To simplify analyses, we assume $\idsys \in \mathcal{P}_{\dsys}$, where $\idsys$ denotes the $\dsys \times \dsys$ dimensional identity, such that non-identity basis operators $P \in \mathcal{P}_{\dsys} \backslash \{\idsys\}$ are traceless, $\chi_{\dsys}(P) = 0$.\\

The $t$-fold tensor product of the Hilbert space is denoted as $\spacesys^{\otimes t}$, and we define the set $[t] = \{0,1,\dots,t-1\}$. Moreover, from $\mathcal{P}_{\dsys}$ we define a basis for the space of linear operators $\mathcal{B}[\spacesys^{\otimes t}]$, which is simply given by $\mathcal{P}_{\dsys}^{t} = \mathcal{P}_{\dsys}^{\otimes t} $. Given two operators $X,Y \in \mathcal{B}[\spacesys^{\otimes t}]$, their normalized inner product is denoted as $\chi_{\dsys}^{(t)}(X,Y) = {\bbra{X}\kket{Y}}/\dsys^t$. Note that in what follows, we indicate $t$-fold tensor product generalizations with $t$ or $(t)$ superscripts, and particular copies are denoted with indices $i \in [t]$. In addition, given an operator $X = \otimes_{i \in [t]}X_{i} \in \mathcal{B}[\spacesys^{\otimes t}]$ with tensor product structure, we define its support $\Gamma_{X}$ and locality as the indices of non-identity operators over the $t$ copies of the space,
\begin{align}
	\Gamma_{X} = \{i \in [t] ~:~ X_{i} \notin \{\idsys\}\}
	\quad \quad ,&~ \quad \quad
	\abs{X} = \abs{\Gamma_{X}} \in [t+1] ~.
\end{align}
Operators $X = \otimes_{i \in \Gamma_{X}}X_{i}$, restricted to a subset of their support $\Gamma \subseteq \Gamma_{X}$, are denoted as $X_{\Gamma} = \otimes_{i \in \Gamma}X_{i}$. For general operators $X = \sum_{x} X_{x}$ with support $\Gamma_{X} = \cup_{x}\Gamma_{X_{x}}$, we refer to such operators as localized with definitive support, if all operators within their expansion have identical support $\Gamma_{X_{x}} = \Gamma_{X}$. We also refer to operators $X,Y \in \mathcal{B}[\spacesys^{\otimes t}]$ as being orthogonal with respect to support if they are orthogonal due to having non-identical support, $\chi_{\dsys}^{(t)}(X,Y) \propto \delta_{\Gamma_{X}\Gamma_{Y}}$.\\

Next, we will denote the symmetric group as $\permutations[t] = \{\sigma : [t] \to [t]\}$ \cite{sagan2001symmetric}. We will denote the identity permutation as $e=()$, and transpositions acting on indices $i \neq j \in [t]$ as $\tau =(ij)$. Here we find it important to recall that any permutation $\sigma \in \permutations[t]$ can be described by the support $\Gamma_{\sigma} \subseteq [t]$ and locality $l_{\sigma}=\abs{\Gamma_{\sigma}} \leq t$ where they act non-trivially; by their size $\abs{\sigma}$ of the number of transpositions $\sigma = \prod_{\tau \in \sigma}\tau$; or by their cycle structure of disjoint cycles $\sigma = \prod_{\lambda \in \sigma}\lambda$. In addition, permutations can be related in terms of the size of their overlap $\abs{\pi^{-1}\sigma}$, or in terms of their sub-permutations $\pi \subseteq \sigma$, a partial ordering between permutations \cite{brady2001partial}, that corresponds to factorizations $\sigma = (\pi)(\pi^{-1}\sigma)$ such that,
\begin{align}
	\pi \subseteq \sigma \quad\leftrightarrow\quad \abs{\pi^{-1}\sigma} = \abs{\sigma} - \abs{\pi}~.
\end{align}
Further, the size of permutations satisfies a triangle inequality, with upper and lower bounds satisfied by sub-permutations. Indeed, for $\sigma,\pi \in \permutations[t]$, one has
\begin{align}
	\big|\abs{\sigma} - \abs{\pi} \big| ~~\leq~~ \abs{\pi^{-1}\sigma} ~~\leq~~ \big|\abs{\sigma} + \abs{\pi} \big|~.
\end{align}

The representation of the symmetric group which permutes copies of the Hilbert space is denoted as $\represent ~:~ \sigma \in \permutations[t] \to \representation{\sigma} \in \representations[t] \subset \mathcal{B}[\spacesys^{\otimes t}]$. From here, we can define the overlap between the representations of permutations $\sigma,\pi \in \permutations[t]$ as
\begin{align}
	\chi_{\dsys}^{(t)}(\sigma,\pi) = \frac{1}{\dsys^{\abs{\sigma^{-1}\pi}}}~.
\end{align}

These overlaps form a $t! \times t!$ matrix over the permutations, $W_{\dsys}^{(t)}$, referred to as a Gram matrix, with elements $W_{\dsys}^{(t)}(\sigma,\pi) = \chi_{\dsys}^{(t)}(\sigma,\pi)$. The inverse of this Gram matrix exists over the permutations, $W_{\dsys}^{(t)} \to W_{\dsys}^{(t)-1}$, referred to as the Weingarten matrix, with elements, whose closed-form dependencies on $d,t$ are only only known to leading-order \cite{collins2006integration},
\begin{equation}
	W_{\dsys}^{(t)-1}(\sigma,\pi)
	= \mobius[t]{\sigma^{-1}\pi}~\frac{1}{\dsys^{\abs{\sigma^{-1}\pi}}}\left[1 + \mathcal{O}\!\left(\frac{1}{\dsys}\right) \right] \leq \weingarten{t}~\mobius[t]{\sigma^{-1}\pi}~\frac{1}{\dsys^{\abs{\sigma^{-1}\pi}}} ~.
\end{equation}
Here for permutations $\sigma$, the Möbius function, $\mobius[t]{\sigma} = \prod_{\lambda \in \sigma} \mobius[l_{\sigma}]{\lambda}$, is the product of the signed Catalan numbers, $\mobius[l_{\sigma}]{\lambda} = (-1)^{\abs{\lambda}}{c}_{\abs{\lambda}}$ for each of its cycles $\lambda \in \sigma$, with Catalan numbers of ${c}_{l} = \tfrac{1}{l+1}\tbinom{2l}{l}$, and $\weingarten{t}$ is a $t$-dependent upper bound \cite{collins2017weingarten}.

To simplify our analysis, we will define the set of character-normalized permutations $\normalizations[t]$ given by,
\begin{align}
	\normalizations[t] = \{\normalization{\sigma} = \representation{\sigma}/\chi_{\dsys}^{(t)}(\sigma)\}_{\sigma \in \permutations[t]}
	\quad \quad : \quad \quad
	\chi_{\dsys}^{(t)}(\normalization{\sigma},\normalization{\pi}) = \frac{\chi_{\dsys}^{(t)}(\sigma,\pi)}{\chi_{\dsys}^{(t)}(\sigma)~\chi_{\dsys}^{(t)}(\pi)}
	~.
\end{align}
Here, it is important to note that while the elements of $\permutations[t]$ are linearly independent for $\dsys \geq t$ \cite{mele2023introduction}, they are not orthogonal \cite{harrow2024approximate}. Hence, we find it convenient to derive an alternative basis denoted as the localized permutations. The elements of this basis are in one-to-one correspondence with the original permutations, are localized with definitive support, and crucially are orthogonal with respect to support.

We can motivate our localized basis by considering a simple example on $t=3$. First, let us first recall that the normalized operator for a transposition $\tau = (ij)$ acting on indices $i \neq j \in [t]$ is
\begin{align}
		\normalization{\tau} =&~ \sum_{P \in \mathcal{P}_{\dsys}} P_{i} \otimes P_{j}^{\dagger}~.
\end{align}
Then, given the $l=3$-cycle $\lambda = (012)$, we can decompose it as $(012)=(01)(02)=(12)(01)$. As such, we can express
\begin{align}
	\normalization{\lambda}=&\sum_{P \in \mathcal{P}_{\dsys}} P_0 \otimes P_1^{\dagger}\sum_{S \in \mathcal{P}_{\dsys}} S_0 \otimes S_2^{\dagger}
	~=~\sum_{P,S \in \mathcal{P}_{\dsys}} P_0S_0 \otimes P_1^{\dagger} \otimes S_2^{\dagger}\nonumber\\
	=&~ \underbrace{\idsys \otimes \idsys \otimes \idsys}_{()}
	~+~ \!\!\!\!\!\!\sum_{P \in \mathcal{P}_{\dsys} \backslash \{\idsys\}}\!\!\!\!\!\! \underbrace{P \otimes P^{\dagger} \otimes \idsys}_{(01)}
	~+~ \!\!\!\!\!\sum_{P \in \mathcal{P}_{\dsys} \backslash \{\idsys\}} \!\!\!\!\!\!\! \underbrace{P \otimes \idsys \otimes P^{\dagger}}_{(02)}
	~+~ \!\!\!\!\!\sum_{P \in \mathcal{P}_{\dsys} \backslash \{\idsys\}} \!\!\!\!\!\!\! \underbrace{\idsys \otimes P \otimes P^{\dagger}}_{(12)}
	~+~ \!\!\!\!\!\sum_{P \neq S \in \mathcal{P}_{\dsys} \backslash \{\idsys\}} \!\!\!\!\!\!\!\!\!\underbrace{PS \otimes P^{\dagger} \otimes S^{\dagger}}_{(012)} ~. \nonumber
\end{align}
The brackets in the previous equation showcase that the final decomposition contains grouping of tensor products of operators that match the expansion of $\lambda$ into its sub-permutations, i.e., $\{\pi = (),(01),(02),(12),(012) \subseteq (012) = \lambda\}$. Importantly, since the decomposition of a permutation into transpositions is not unique, the previous decomposition is also not unique. However, such expansions suggest there exist general relationships between representations of permutations, and their sub-permutations. We seek to exploit such relationships to define the basis of localized permutations.

\begin{theorem}[Localized Basis for Permutations]\label{theorem:app:localized_basis_for_permutations}
Let $\normalizations[t]$ be the set of character normalized permutations. There exists a change of basis, $\phi_{t} ~:~ \normalizations[t] \to \localizations[t]$, to a localized permutation basis $\localizations[t]$,
\begin{align}
	\normalization{\sigma} ~=&~~ \sum_{\pi \subseteq \sigma}~ \localization{\pi}
	\quad \leftrightarrow \quad
	\localization{\sigma} ~=~ \sum_{\pi \subseteq \sigma} ~\phi_{t}(\sigma,\pi) ~\normalization{\pi}~,
\end{align}
where the elements of $\localizations[t]= \{\localization{\sigma}\}_{\sigma \in \permutations[t]}$ depend on sub-permutation relationships $\pi \subseteq \sigma$, and obey the following properties:
	\begin{enumerate}
		\item Change of basis transformation: $\phi_{t} ~:~ \normalizations[t] \to \localizations[t]$ is invertible, lower-triangular as per the sub-permutation ordering, is dimension-$\dsys$-independent, and is solely dependent on the structure of the $t$-th order permutations:
		\begin{align}
				\phi_{t}(\sigma,\pi) ~=&~~ \mobius[t]{\pi^{-1}\sigma}~\delta_{\sigma \supseteq \pi}
				\quad\quad , \quad\quad
				\mobius[t]{\sigma} = \prod_{\lambda \in \sigma}\! \mobius[l_{\sigma}]{\lambda}
				\quad , \quad
				\mobius[l_{\sigma}]{\lambda}= (-1)^{\abs{\lambda}}~\catalan[t]{\abs{\lambda}}
				\quad\quad , \quad \quad
				\phi_{t}^{-1}(\sigma,\pi) ~=~ \delta_{\sigma \supseteq \pi}~.
		\end{align}~\\[-4pt]
		\item Linearly independent for $\dsys \geq t$, and in one-to-one correspondence with permutations: $\localization{\sigma} \in \localizations[t] \leftrightarrow \sigma \in \permutations[t]$~\\[12pt]
		\item Orthogonal with respect to support, and localized with definitive support $\Gamma_{\localization{\sigma}} = \Gamma_{\sigma}$:
		\begin{align}
			\!\!\!\!\!\!\!\!\!\!
			\chi_{\dsys}^{(\localizations[t])}
			~=~ \bigoplus_{\Gamma \subseteq [t]} \chi_{\dsys}^{(\Gamma)}
			~:~
			\chi_{\dsys}^{(\Gamma)} =
			\left[\begin{array}{ccc}
			\chi_{\dsys}^{(t)}(\localization{\sigma},\localization{\sigma}) & \cdots & \chi_{\dsys}^{(t)}(\localization{\sigma},\localization{\pi}) \\
			\vdots & \ddots & \vdots \\
			\chi_{\dsys}^{(t)}(\localization{\pi},\localization{\sigma}) & \cdots & \chi_{\dsys}^{(t)}(\localization{\pi},\localization{\pi})
			\end{array}	\right]_{\substack{\sigma,\pi \in \permutations[t]\\\Gamma_{\sigma}=\Gamma_{\pi} = \Gamma}}
			\!\!\!	\!\!\!\!\!\! ~:~~~
			\abs{\chi_{\dsys}^{(t)}(\localization{\sigma},\localization{\pi})} ~\!\leq~\! \mobiuscatalan[l]{\sigma}^{2}~\!\mobiuscatalan[l]{\pi}^{2}~\!\dsys^{\abs{\sigma} + \abs{\pi}}~\delta_{\Gamma_{\sigma}\Gamma_{\pi}} \!\!\!\!\!\!\!\!\!\!\!\!\!\!\!
		\end{align}
	\end{enumerate}
\end{theorem}

\begin{figure}[ht]
	\centering
	\includegraphics[width=1\linewidth]{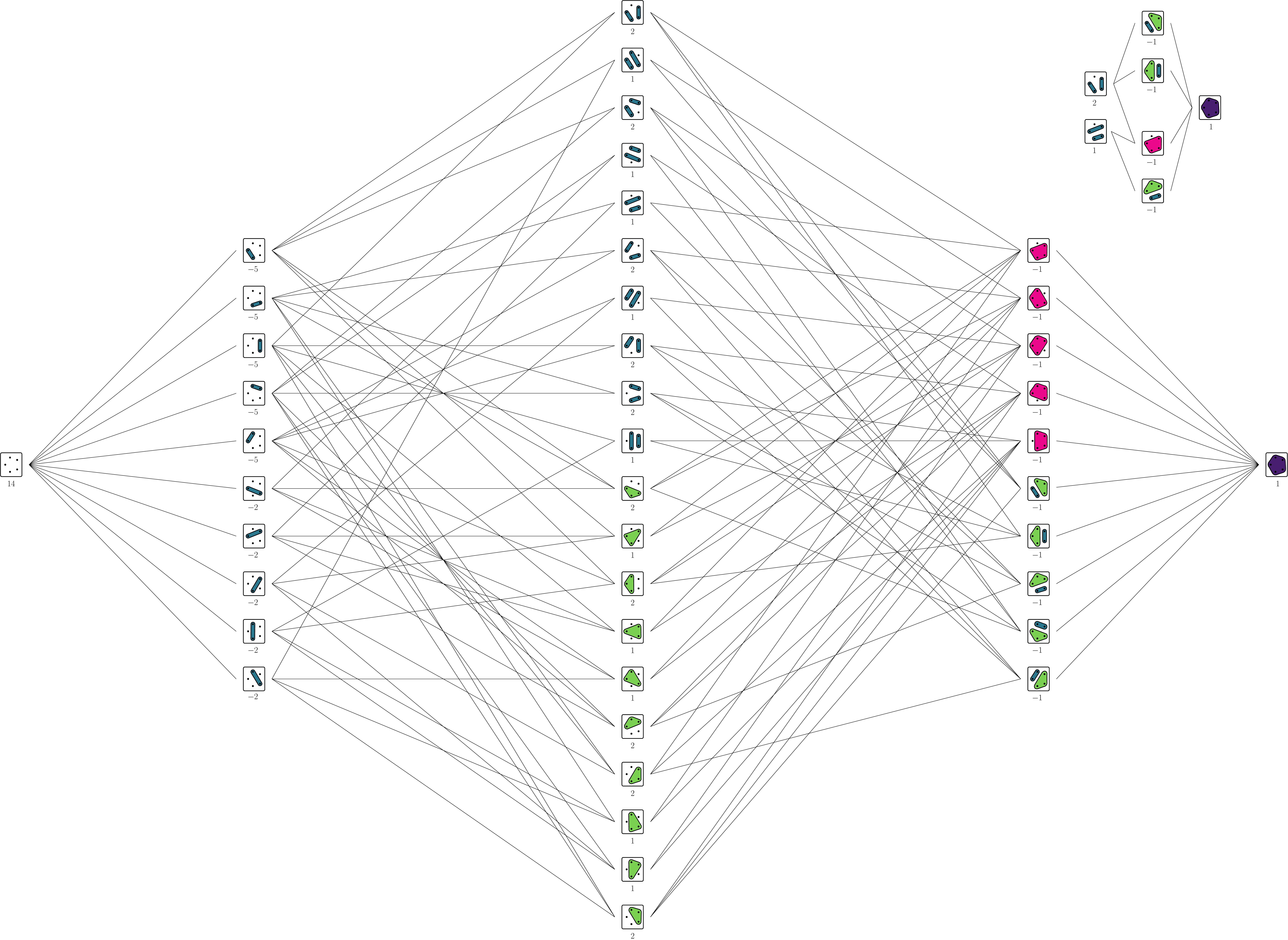}
	\caption{\textbf{Sub-permutation lattice for an $l=5$ cycle $\sigma=(01234)$.} All $\catalan[t]{5}=42$ sub-permutations $e \subseteq \xi \subseteq \sigma$ are shown as nodes on a lattice, each as non-crossing partitions of points, ordered counter-clockwise from the leftmost point on an $l=5$-polygon. Nodes of the lattice depict sub-permutations $e \subseteq \xi \subseteq \sigma$ between minimal $e$ and maximal $\sigma$, ordered in $l$ columns, by size $0 \leq \abs{\xi} \leq l-1$, from left to right. Edges of the lattice depict sub-permutation relationships, from left to right. At a given node $\xi$, edges and paths to rightward nodes $\xi^{\prime}$ represent larger sup-permutations $\xi \subset \xi^{\prime} \subseteq \sigma$, such that $1 \leq \abs{\xi^{-1}\xi^{\prime}} \leq \abs{\xi^{-1}\sigma} \leq l-1$. Such sup-permutations $\{\xi \subseteq \xi^{\prime} \subseteq \sigma \}$ of $\xi$ form a sub-lattice with minimal $\xi$ and maximal $\sigma$. Permutations $\xi$ are labeled with their localized permutation change of basis coefficient $\phi_{l}(\sigma,\xi) = \mobius[l]{\xi^{-1}\sigma}$, such that $\sum_{\pi \subseteq \xi \subseteq \sigma}\phi_{l}(\sigma,\xi) = \delta_{\sigma\pi}$. Inset (top right) is the sub-lattices forming a set $S_{\sigma}$, with minimal permutations $\pi \in T_{\sigma}=\{\pi = (01)(23),(03)(12) \subset \sigma = (01234)\}$, with support $\Gamma = \Gamma_{\pi}=\{0,1,2,3\} \subset \Gamma_{\sigma} = \{0,1,2,3,4\}$, and a smallest common sup-permutation $\pi_{\sigma}=(0123) \subset \sigma = (01234)$, such that $\sum_{\xi \in S_{\sigma}}\phi_{l}(\sigma,\xi) = \sum^{\pi \in T_{\sigma}}_{\pi \subseteq \xi \subseteq \sigma}\phi_{l}(\sigma,\xi) - \sum^{\pi=\pi_{\sigma}}_{\pi \subseteq \xi \subseteq \sigma} \phi_{l}(\sigma,\xi) = 0$.}
	\label{fig:app:permutations}
	\vspace{-12pt}
\end{figure}

\newpage
\begin{proof}\label{proof:app:localized_basis_for_permutations}
		To prove properties of the localized permutation basis, we use properties of sub-permutations $\pi \subseteq \sigma$ such that $\abs{\pi^{-1}\sigma} = \abs{\sigma}-\abs{\pi}$, which represent a partial ordering between permutations \cite{brady2001partial,kreweras1972sur}, and is transitive, such that if $\pi \subseteq \xi$, and $\xi \subseteq \sigma$, then $\pi \subseteq \sigma$, and if $\pi \subseteq \xi \subseteq \sigma$, then $\pi^{-1}\xi \subseteq \pi^{-1}\sigma$. Concurrently, the number of sub-permutations of an $l$-length cycle also corresponds to the number of non-crossing partitions of the set $[l]$ that respect the order of the permuted indices \cite{brady2001partial}. As depicted in Fig. \ref{fig:app:permutations}, the set of sub-permutations $\{\pi \subseteq \xi \subseteq \sigma\} \cong \{e \subseteq \xi \subseteq \pi^{-1}\sigma\}$ forms a lattice containing all paths of sub-permutations $\xi$ between some minimum $\pi$ and maximum $\sigma$ permutations. We also note that any sub-permutations $\{\pi \subseteq \lambda\}$ of a cycle $\lambda$, with identical support $\Gamma_{\pi}=\Gamma \subseteq \Gamma_{\lambda}$ and locality $l_{\pi} = l \leq l_{\lambda}$, are within a set of permutations that is isomorphic to the $l$-order permutations $\permutations[l]$. Further, any subset $T \subseteq \{\pi \subseteq \lambda\}$ of such $\Gamma$-support sub-permutations of $\lambda$ have a unique smallest common sup-permutation $\pi^{\prime}$ (which is referred to as the join of $T$ in the lattice literature \cite{nica2006lectures}). Further, given the isomorphism to $\permutations[l]$, this join $\pi^{\prime}$ has $\Gamma$-support and is a sub-permutation of the unique $\Gamma$-support, $l$-length cycle $\pi_{\lambda} \subseteq \lambda$ (joins of $T$ can be seen to preserve common singletons, the elements $[t]\backslash \Gamma$ not in the support of, and acted on trivially by all $\pi \in T$ \cite{ebrahimifard2024noncrossing}).

		Further, permutation-dependent quantities can often be partitioned by the support $\Gamma_{\sigma} = \Gamma \subseteq [t]$ and locality $l_{\sigma}=l = \abs{\Gamma} \leq t$ of permutations $\sigma$. Properties of sub-permutations also hold independently for disjoint cycles, given permutations $\pi = \prod_{\lambda \in \sigma}\!\pi_{\lambda}$ can be decomposed into disjoint sub-permutations $\{\pi_{\lambda} \subseteq \lambda\}_{\lambda \in \sigma}$ for any $\sigma = \prod_{\lambda \in \sigma}\lambda \supseteq \pi$. It follows that sup-permutation dependent quantities can often be factorized, such as $\localization{\pi} = \prod_{\lambda \in \sigma}~\localization[_{\lambda}]{\pi}$, and $\phi_{t}(\sigma,\pi) = \prod_{\lambda \in \sigma}\phi_{l_{\lambda}}(\lambda,\pi_{\lambda})$.

		Finally, we can count the number of sub-permutations and localized permutations. We note that the number of sub-permutations, which equals the number of non-crossing partitions, equals the Catalan numbers, $\catalan[l]{l} = \tfrac{1}{l+1}\tbinom{2l}{l}$ \cite{linton2010permutation}. The number of sub-permutations between permutations $\pi \subseteq \sigma$ subsequently are,
		\begin{align}
			\mobiuscatalan[t]{\sigma,\pi} = \abs{\{\xi \in \permutations[t] ~:~ \pi \subseteq \xi \subseteq \sigma\}} = \mobiuscatalan[t]{\pi^{-1}\sigma}~\delta_{\sigma \supseteq \pi}
			\quad : \quad
			\mobiuscatalan[t]{\sigma} = \abs{\{\xi \in \permutations[t] ~:~ \xi \subseteq \sigma\}}
			= \prod_{\lambda \in \sigma}\catalan[t]{l_{\lambda}} ~.
		\end{align}
		The number of locality-$l$ blocks of permutations is $\tbinom{t}{l}$, and the number of permutations non-trivially permuting their $l$-locality support $\Gamma$, is by definition, $!l = \floor{\frac{l!}{e} + \frac{1}{2}}$, the number of length-$l$ derangements \cite{schroeder2009permutations}.\\

		\noindent We will prove each property $i)$ to $iii)$ of the mapping between the permutation and localized permutation bases separately.\\

		\noindent \emph{$i)$} To prove the existence of the localized permutation basis, we will show that an invertible relationship exists, also known as Möbius inversion \cite{nica2006lectures}, between $\normalizations[t]$ and $\localizations[t]$. Let $\phi_{t}^{-1} ~:~ \localizations[t] \to \normalizations[t]$ be such a change of basis, with matrix elements $\phi_{t}^{-1}(\sigma,\pi) = \delta_{\sigma\supseteq \pi}$, which is lower-triangular, with unit diagonal. Its sub-permutation dependent structure further impose that its strictly lower-triangular component is $t$-nilpotent, given the $(\sigma,\pi)$ element of its $k$ powers is the number of paths, of sub-permutations in the lattice formed by $\pi \subset \sigma$, of length $k$, and each path has at most $\abs{\pi^{-1}\sigma} < t$ elements. The change of basis matrix inverse thus exists, $\phi_{t} ~:~ \normalizations[t] \to \localizations[t]$, in fact with a closed form Möbius inverse \cite{nica2006lectures}, with identical sub-permutation dependent structure to its inverse, $\phi_{t}(\sigma,\pi),\phi_{t}^{-1}(\sigma,\pi) \propto \delta_{\sigma \supseteq \pi}$. Such relationships only depend on the overlaps and structure of permutations, and are independent of the representation dimension $\dsys$,
		\begin{align}
			\!\!\!\!
			\phi_{t} = \sum_{k \in [t]}~(\idpermutations-\phi^{-1}_{t})^{k}
			~~ , ~~
			\phi_{t}(\sigma,\pi) = \phi_{t}(\pi^{-1}\sigma)~\delta_{\sigma \supseteq \pi}
			~~ : ~~
			\phi_{t}(\sigma) = \mobius[t]{\sigma} = \prod_{\lambda \in \sigma} \mobius[l_{\sigma}]{\lambda}
			~~ , ~~
			\sum_{\pi \subseteq \sigma}\phi_{t}(\pi) = \delta_{\sigma e}~\!.\!\!\!
		\end{align}

		\noindent \emph{$ii)$} From the invertible relationship between the bases, both the permutations and the localized permutations are linearly independent for $\dsys \geq t$ \cite{mele2023introduction}, and the localized permutations are in one-to-one correspondence with the permutations. \\

		\noindent \emph{$iii)$} To prove that localized permutations are localized, we consider the effect of the support of a basis operator string, $P = \otimes_{i \in \Gamma_{P}}P_{i} \in \mathcal{P}_{\dsys}^{t}$, with support $\Gamma_{P} \subseteq \Gamma_{\sigma} \subseteq [t]$, on its overlap with a localized permutation $\localization{\sigma} \in \localizations[t]$, with potentially non-localized support $\Gamma_{\localization{\sigma}} = \cup_{\pi \subseteq \sigma}\Gamma_{\pi} = \Gamma_{\sigma}$. As per Fig. \ref{fig:app:diagrams}, the overlap between a permutation and an operator string is the tracefullness of products of operators, in the order of the permuted indices within each cycle. Such products of operators are invariant under any sup-permutations of the permutation, meaning non-zero overlaps of operator strings with all permutations $\pi,\pi^{\prime}$ with a common sup-permutation $\sigma \supseteq \pi,\pi^{\prime}$, must be identical,
		\begin{align}
			\!\!\!\!\!\!\!\!
			\chi_{\dsys}^{(t)}(\normalization{\pi},P)
			~=~ \chi_{\dsys}^{(t)}(\normalization{\pi}^{\prime},P)
			~=~ \chi_{\dsys}^{(t)}(\normalization{\sigma},P)
			~=~ \frac{1}{\chi_{\dsys}^{(t)}(\sigma)}~\prod_{\lambda \in \sigma}~\chi_{\dsys}^{(t)}(\prod_{i \in \Gamma_{\lambda}}P_{i})
			~\propto~ \delta_{\Gamma_{\sigma} \supseteq \Gamma_{P}}~\delta_{\Gamma_{\pi}\supseteq \Gamma_{P}}~\delta_{\Gamma_{\pi^{\prime}} \supseteq \Gamma_{P}}
			~~~~
			\forall~ \sigma \supseteq \pi,\pi^{\prime} ~. \!\!\!\!\!\!\!\!
		\end{align}
		Therefore overlaps of localized permutations with an operator string reduce to summations of coefficients $\phi_{t}$ over the subset of permutations, $S_{\sigma} = \{\xi \subseteq \sigma ~:~ \Gamma_{\xi} \supseteq \Gamma_{P} ~,~ \chi_{\dsys}^{(t)}(\xi,P) \neq 0 \}$ that have non-zero overlap with the operator string,
		\begin{align}
			\frac{\chi_{\dsys}^{(t)}(\localization{\sigma},P)}{\chi_{\dsys}^{(t)}(\normalization{\sigma},P)}
			~=&~~ \sum_{\substack{\xi \subseteq \sigma}}\!\!~\phi_{t}(\sigma,\xi) ~\frac{\chi_{\dsys}^{(t)}(\normalization{\xi},P)}{\chi_{\dsys}^{(t)}(\normalization{\sigma},P)}
			~=~ \sum_{\substack{\xi \in S_{\sigma}}}\!\!~\phi_{t}(\sigma,\xi) ~\frac{\chi_{\dsys}^{(t)}(\normalization{\xi},P)}{\chi_{\dsys}^{(t)}(\normalization{\sigma},P)}
			~=~ \sum_{\substack{\xi \in S_{\sigma}}}\!\!~\phi_{t}(\sigma,\xi)
			~~\overset{?}{\propto}~~ \delta_{\Gamma_{P}\Gamma_{\sigma}} ~.
		\end{align}

		\begin{figure}[ht]
			\centering
			\includegraphics[width=0.7\linewidth]{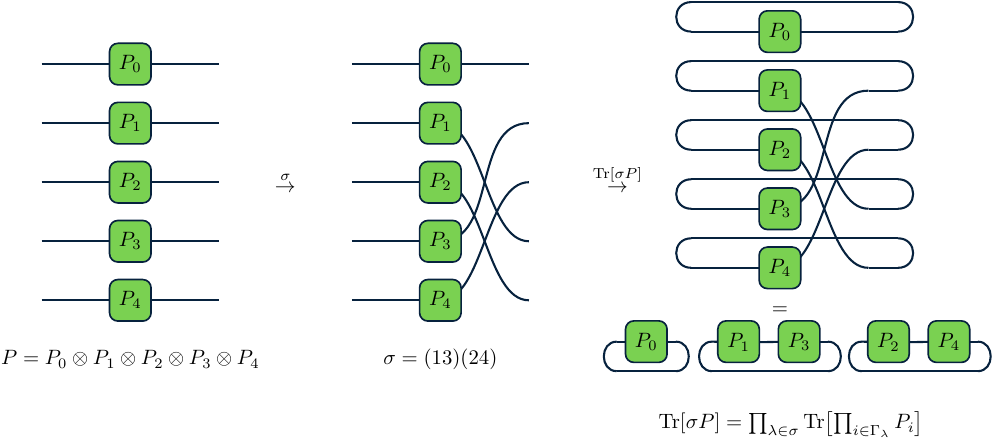}
			\caption{\textbf{Overlap between an operator string $P$, and a permutation $\sigma$.} Overlaps are equal to the tracefullness of the products of an operator's components $\otimes_{\lambda \in \sigma} \prod_{i \in \Gamma_{\lambda}} P_{i}$, in the order of indices $\cup_{\lambda \in \sigma}\Gamma_{\lambda}$ permuted by a permutation's disjoint cycles.}
			\label{fig:app:diagrams}
			\vspace{-12pt}
		\end{figure}

		\noindent Given summations of the coefficients $\phi_{t}$ over lattices $\{\pi \subseteq \xi \subseteq \sigma\}$ are non-zero if and only if $\pi=\sigma$, we seek to express the unknown summations over $S_{\sigma}$ in terms of these known summations over lattices, potentially leading to cancellations. The set $S_{\sigma} = \cup_{\pi \in T_{\sigma}}S_{\sigma\pi}$, although not a lattice, can be partitioned into lattices, $S_{\sigma\pi} = \{\xi \in S_{\sigma} ~:~ \pi \subseteq \xi \subseteq \sigma\}$, each defined by a unique minimal permutation, $\pi \in T_{\sigma} = \{\pi \in S_{\sigma} ~:~ \pi \not\supset \xi~~\forall~ \xi \in S_{\sigma}\}$, as in the inset of Fig. \ref{fig:app:permutations}. As will be crucial, all minimal $\pi \in T_{\sigma}$ have identical support $\Gamma_{\pi} = \Gamma_{P}$, otherwise they would not be minimal permutations in $S_{\sigma}$. As such, as above, any subset $T \subseteq T_{\sigma}$ of permutations always has smallest common sup-permutations $\pi^{\prime}$ with $\Gamma_{P}$-support, and which is itself a sub-permutation $\pi^{\prime} \subseteq \pi_{\sigma}$, of the unique $\Gamma_{P}$-support permutation $\pi_{\sigma} = \prod_{\lambda \in \sigma}\pi_{\sigma\lambda} \subseteq \sigma$, comprised of disjoint cycles $\{\pi_{\sigma\lambda} : \pi_{\sigma\lambda}\subseteq \lambda\}_{\lambda \in \sigma}$. Finally, in the equal support case, $\Gamma_{P} = \Gamma_{\sigma}  \to  \pi_{\sigma} \subseteq \sigma$, whereas in the strict subset of support case, $\Gamma_{P} \subset \Gamma_{\sigma} \to \pi_{\sigma} \subset \sigma$, and any smallest common sup-permutations of $T$ satisfy $\pi^{\prime} \subseteq \pi_{\sigma}$.\\

		\noindent Given this decomposition of a set of permutations into a union of lattices, we use the inclusion-exclusion principle \cite{rama2025inclusion}, that relates counting of unions as counting of intersections, less multiply counted elements in multiple intersections,
		\begin{align}
			\sum_{\xi \in \cup_{\pi \in T_{\sigma}}S_{\sigma\pi}} ~=&~ \sum_{\{\} \subset T \subseteq T_{\sigma}} (-1)^{\abs{T}-1}\sum_{\xi \in \cap_{\pi \in T} S_{\sigma\pi}}~.
		\end{align}~\\~\\
		\noindent As one example, in the equal support case, if there is a single minimal permutation $\pi = \sigma$, forming a single lattice, $S_{\sigma\pi} = S_{\sigma} = \{\sigma\}$, then trivially $\sum_{\substack{\xi \in S_{\sigma}}}\!\!~\phi_{t}(\sigma,\xi) = 1$. As another example, in the strict subset of support case, if there is a single minimal permutation $\pi \subset \sigma$, forming a single lattice, $S_{\sigma\pi} = S_{\sigma} = \{\pi \subseteq \xi \subseteq \sigma\}$, then trivially $\sum_{\substack{\xi \in S_{\sigma}}}\!\!~\phi_{t}(\sigma,\xi) = 0$. It is therefore essential to understand any cases of $\sigma,P$ that yield non-trivial intersections of lattices $S_{\sigma\pi}$, in order to understand the summations over $S_{\sigma}$ using the inclusion-exclusion principle.\\

		\noindent Although intersections of lattices, $S_{\sigma}^{\prime} = \cap_{\pi \in T}S_{\sigma\pi}$ for some $T \subseteq T_{\sigma}$, are not necessarily lattices \cite{topkis1976structure}, any sup-permutation $\xi^{\prime\prime} \supseteq \xi^{\prime}$ of any permutation in the intersection $\xi^{\prime} \in S_{\sigma}^{\prime}$, is necessarily also in this intersection, $\xi^{\prime\prime} \in S_{\sigma}^{\prime}$. Therefore, intersections of lattices can be themselves decomposed into new lattices $S_{\sigma\pi}^{\prime}$, such that $S_{\sigma}^{\prime} = \cup_{\pi^{\prime} \in T_{\sigma}^{\prime}}S_{\sigma\pi^{\prime}}^{\prime}$, each defined by a unique minimal permutation, $\pi^{\prime} \in T_{\sigma}^{\prime} = \{\pi \in S_{\sigma}^{\prime} ~:~ \pi \not\supset \xi ~~\forall~ \xi \in S_{\sigma}^{\prime}\}$, and crucially $\pi^{\prime} \subseteq \pi_{\sigma}$.\\

		\noindent Intersections of lattices can thus be recursively defined as unions of sub-lattices, $S_{\sigma} \to S_{\sigma}^{\prime} \to \cdots \to S_{\sigma}^{*}$. Such intersecting sub-lattices are depicted in Fig. \ref{fig:app:intersections}, where colored patches and nodes represent their non-intersecting components, with highlighted minimal elements in one of $T_{\sigma} \to T_{\sigma}^{\prime} \to \cdots \to T_{\sigma}^{*}$. Given the unique common sup-permutation $\pi_{\sigma}$ of any subsets of $T_{\sigma},T_{\sigma}^{\prime},\dots,T_{\sigma}^{*}$, and that each lattice has paths of at most length $\abs{\sigma}+1$ such a sequence of unions of sublattices terminates after at most $\abs{\sigma}$ iterations, yielding a single lattice $S_{\sigma}^{*} = \{\pi_{\sigma} \subseteq \xi \subseteq \sigma\}$, with minimal element $T_{\sigma}^{*} = \{\pi_{\sigma}\}$.

		\vspace{-0.3cm}
		\begin{figure}[ht]
			\centering
			\includegraphics[width=0.36\linewidth]{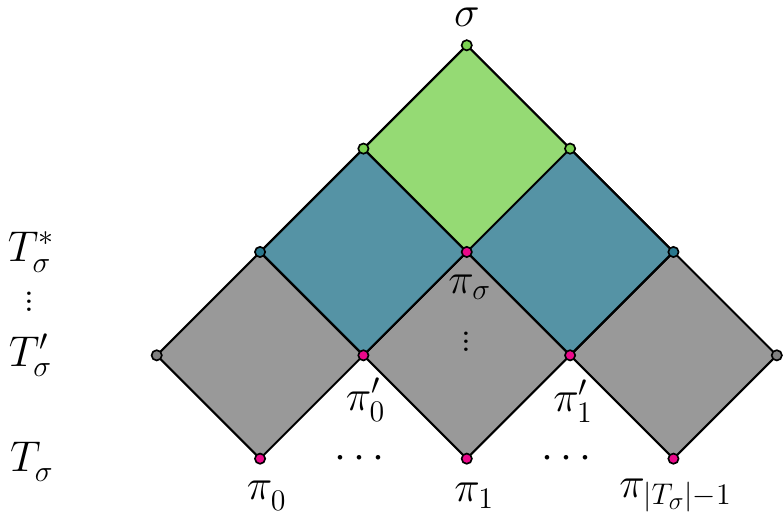}
			\caption{\textbf{Decomposition of $S_{\sigma} = \cup_{\pi \in \bar{T}_{\sigma}}S_{\sigma\pi}$ into intersecting sub-lattices $S_{\sigma\pi} = \{\pi \subseteq \xi \subseteq \sigma\}$.} Gray, blue, green patches and nodes depict disjoint subsets of each sub-lattice, with pink nodes for minimal permutations $\pi \!\in\! \bar{T}_{\sigma} = \cup\{T_{\sigma},T_{\sigma}^{\prime},\dots,T_{\sigma}^{*}\} \!~:~\! \pi \!\subseteq\! \pi_{\sigma} \!\subseteq \sigma$.}
			\label{fig:app:intersections}
		\end{figure}

		\noindent Summations over $S_{\sigma}$ therefore take the form of summations over lattices, defined by $\bar{T}_{\sigma} = \{\xi : ~\exists~ \pi \in T_{\sigma} :\pi \subseteq \xi \subseteq \pi_{\sigma}\}$, weighted by some non-negative multiplicities and signs $\{n_{\pi},k_{\pi}\}_{\pi \in \bar{T}_{\sigma}}$. Finally, given summations over $\phi_{t}(\sigma,\pi)$ are only non-zero when $\pi=\sigma$, summations over $S_{\sigma}$ reduce to summations over the lattice $\{\pi_{\sigma} \subseteq \xi \subseteq \sigma\}$,
		\begin{align}
			\sum_{\xi \in S_{\sigma}}\phi_{t}(\sigma,\xi) ~=&~ \sum_{\pi \in \bar{T}_{\sigma}}(-1)^{k_{\pi}}~n_{\pi}\sum_{\pi \subseteq \xi\subseteq \sigma}\phi_{t}(\sigma,\xi) ~=~ (-1)^{k_{\sigma}}~n_{\sigma}~\delta_{\pi_{\sigma}\sigma} ~\propto~\delta_{\Gamma_{P}\Gamma_{\sigma}}~.
		\end{align}
		\noindent Localized permutations $\localization{\sigma}$ thus have zero overlap with operator strings $P$ with $\Gamma_{P} \subset \Gamma_{\sigma}$, meaning localized permutations are localized, with definitive support, $\Gamma_{\localization{\sigma}}=\Gamma_{\sigma}$,
		\begin{align}
			\localization{\sigma} ~=&~~ \sum_{\substack{P \in \mathcal{P}_{\dsys}^{t}}}\chi_{\dsys}^{(t)}(\localization{\sigma},P)~P
			\quad : \quad
			\chi_{\dsys}^{(t)}(\localization{\sigma},P) ~=~ (-1)^{k_{\sigma}}~n_{\sigma}~\frac{\chi_{\dsys}^{(t)}(\sigma,P)}{\chi_{\dsys}^{(t)}(\sigma)}\delta_{\Gamma_{P}\Gamma_{\sigma}}
			~~\quad \to \quad~~
			\chi_{\dsys}^{(t)}(\localization{\sigma},\localization{\pi}) ~\propto~ \delta_{\Gamma_{\sigma}\Gamma_{\pi}}~.
		\end{align}

		\noindent Intriguingly, such arguments could potentially hold for the localization of other change of bases $\{\normalization{X}\} \to \{\localization{X}\}$ that are in terms of other partial orderings $Y \subseteq X$ between operators, their Möbius inverses $\phi_{t}$, and their summations $\sum_{\substack{Y \in S_{X}}}\!\!~\phi_{t}(X,Y)$, given symmetries of overlaps with sub and sup-operators $\chi_{\dsys}^{(t)}(\normalization{Y},P) = \chi_{\dsys}^{(t)}(\normalization{X},P)$ for $Y \subseteq X$, $P \in \mathcal{P}_{\dsys}^{t}$.\\

		\noindent To prove the bounds on the overlap between localized permutations, we recall their definition,
		\begin{align}
			\chi_{\dsys}^{(t)}(\localization{\sigma},\localization{\pi})
			~=~ \sum_{\substack{\eta \subseteq \sigma\\\kappa \subseteq \pi}} ~\phi_{t}(\sigma,\eta)~\phi_{t}(\pi,\kappa)\frac{\chi_{\dsys}^{(t)}(\eta,\kappa)}{\chi_{\dsys}^{(t)}(\eta)\chi_{\dsys}^{(t)}(\kappa)}
			\quad \quad \leftrightarrow \quad \quad
			\frac{\chi_{\dsys}^{(t)}(\sigma,\pi)}{\chi_{\dsys}^{(t)}(\sigma)\chi_{\dsys}^{(t)}(\pi)}
			~=~ \sum_{\substack{\eta \subseteq \sigma\\\kappa \subseteq \pi}} ~\chi_{\dsys}^{(t)}(\localization{\eta},\localization{\kappa})~.
		\end{align}
		Given bounds on permutation sizes, $\abs{\sigma} \!+\! \abs{\pi} \!-\! \abs{\sigma^{-1}\pi} \leq \abs{\sigma} \!+\! \abs{\pi}$, their overlaps are, ${\chi_{\dsys}^{(t)}(\sigma,\pi)}/{\chi_{\dsys}^{(t)}(\sigma)\chi_{\dsys}^{(t)}(\pi)} \!\leq\! \dsys^{\abs{\sigma}+\abs{\pi}}$, given the monotonicity of the Catalan numbers $\catalan[l]{l} \leq \catalan[l]{l+1}$, the Möbius functions are, $\abs{\mobius[t]{\pi^{-1}\sigma}} \!\leq\! \abs{\mobius[t]{\sigma}} \!\leq\! \mobiuscatalan[t]{\sigma}$, and given summations are bounded by the largest summand times the number of summands, $|{\chi_{\dsys}^{(t)}(\localization{\sigma},\localization{\pi})}| \!\leq\! \mobiuscatalan[t]{\sigma}^{2}~\!\mobiuscatalan[t]{\pi}^{2}~\!\dsys^{\abs{\sigma} + \abs{\pi}} ~\!\delta_{\Gamma_{\sigma}\Gamma_{\pi}}$.\\

		\noindent Finally, we note there is a trade-off in this choice of localized basis. The mapping from the localized permutations to the permutations has a simple, attractive closed form, with known enumerations of the expansions in terms of sub-permutations, with unit expansion coefficients. Permutations can equally be expanded in terms of a basis of strings of orthogonal operators, grouped by their support. However such groupings are not enumerated by the sub-permutations, and the associated expansion coefficients have no closed form, requiring the algorithms used in this proof.

\end{proof}

\section{Properties of Moment Operators}\label{app:properties_of_moment_operators}
In these appendices, we consider properties of $t$-th order moment operators for various ensembles, generalizing previous expressions \cite{holmes2021connecting}, and adapt approaches for integrating over ensembles of operators \cite{garcia2023deep,collins2022weingarten,collins2006integration}.

\subsection{Twirls and Moment Operators}\label{subsec:moment_operators}
Here, we denote the $t$-th order moment operator for an ensemble of channels $ \ensemblechannel \subseteq \mathcal{B}[\mathcal{B}[\spacesys]]$ as $\widehat{\Tau}_{\ensemblechannel}^{(t)} \!=\! \int_{\ensemblechannel}d\channel~\widehat{\channel^{\otimes t}}$. Then, the moment operator for $k$-concatenated independent ensembles $\{\ensemblechannel_{i}\}_{i\in [k]}$ is the product of the moment operators,
\begin{align}\label{eq:op-twirl-ap}
	\vspace{-0.1cm}
	\hspace{-0.3cm}
	\widehat{\Tau}_{\{\ensemblechannel_{i}\}}^{(t)}
	=&~ \prod_{i \in [k]}\widehat{\Tau}_{\ensemblechannel_{i}}^{(t)}
	~=~ \prod_{i \in [k]}\int_{\ensemblechannel_{i}}d\channel_{i}~\widehat{\channel_{i}^{\otimes t}}
	~=~ \channeldepolarize[\widehat]{\dsys}^{\otimes t} ~+~ \widehat{\Delta}_{\{\ensemblechannel_{i}\}}^{(t)}~,
\end{align}
where we defined
\begin{equation}
	\channeldepolarize[\widehat]{\dsys}^{\otimes t} = \displaystyle{\frac{1}{\dsys^{t}}\kket{\idsys^{\otimes t}}\bbra{\idsys^{\otimes t}}}
	\quad \text{and}\quad
	\widehat{\Delta}_{\{\ensemblechannel_{i}\}}^{(t)} = \displaystyle{\frac{1}{\dsys^{t}} \sum_{\substack{P \in \mathcal{S}_{\ensemblechannel_{0}}^{(t)} \backslash \{\idsys^{\otimes t}\},~S \in \mathcal{S}_{\ensemblechannel_{k-1}}^{(t)}}} \tau_{\{\ensemblechannel_{i}\}}^{(t)}(P,S) \kket{P}\bbra{S}}~.
\end{equation}
Here, $\tau_{\{\ensemblechannel_{i}\}}^{(t)}$ denote the composite expansion coefficients, also known as the transfer matrix elements. These composite transfer matrices can be expanded in terms of the individual transfer matrices $\{\tau_{\channel_{i}}^{(t)}\}_{i \in [k]}$ for each moment operator, and overlaps $\chi_{\dsys}^{(t)}$ between the elements of some ensemble-dependent bases $\{\mathcal{S}_{\ensemblechannel_{i}}^{(t)} \subseteq \mathcal{B}[\spacesys^{\otimes t}]\}$ as,
\begin{align}
	\tau_{\{\ensemblechannel_{i}\}}^{(t)}(P,S) =&~ \sum_{\substack{\{P_{i},S_{i} \in \mathcal{S}_{\ensemblechannel_{i}}^{(t)}\}_{i \in [k]}\\P_{0}=P,~S_{k-1}=S}} \prod_{i \in [k]}\tau_{\ensemblechannel_{i}}^{(t)}(P_{i},S_{i}) \prod_{i \in [k-1]}\chi_{\dsys}^{(t)}(S_{i},P_{i+1}) ~.
\end{align}
In the case of concatenating $k$ identical ensembles, i.e., when $\ensemblechannel_{i} = \ensemblechannel$ $\forall i\in[k]$, our notation is replaced with $k$-powers, or $k$-superscripts, namely $\widehat{\Tau}_{\{\ensemblechannel_{i}\}}^{(t)} \to \widehat{\Tau}_{\ensemblechannel}^{(t)k}$,~ $\widehat{\Delta}_{\{\ensemblechannel_{i}\}}^{(t)} \to \widehat{\Delta}_{\ensemblechannel}^{(t,k)}$,~ and~ $\tau_{\{\ensemblechannel_{i}\}}^{(t)} \to \tau_{\ensemblechannel}^{(t,k)}$.\\

To compare two ensembles $\ensemblechannel,\ensemblechannel^{\prime}$, let us define the difference between their $t$-th order moment operators as
\begin{align}
	\widehat{\Delta}_{\ensemblechannel\ensemblechannel^{\prime}}^{(t)} = \widehat{\Tau}_{\ensemblechannel}^{(t)} - \widehat{\Tau}_{\ensemblechannel^{\prime}}^{(t)}
	\quad \quad \to \quad \quad
	\Epsilon_{\ensemblechannel\ensemblechannel^{\prime}}^{(t)} = \norm{\widehat{\Delta}_{\ensemblechannel\ensemblechannel^{\prime}}^{(t)}} \quad ~,~ \quad \Epsilon_{\ensemblechannel}^{(t)} = \norm{\widehat{\Delta}_{\ensemblechannel}^{(t)}}~.
\end{align}
Various norms $\norm{\cdot}_{*}$ may be chosen, denoted with $\Epsilon^{(t|*)}$ superscripts, and unless noted, we use the Hilbert-Schmidt norm. \\

Our choice of Hilbert-Schmidt norm $\norm{\cdot} = \norm{\cdot}_{2}$ for moment operators can be motivated, by its intuitive form that lends itself to simplified calculations, but also due to it bounding the more commonly used diamond norm $\norm{\cdot}_{\diamond}$.

\begin{lemma}[Moment Operator Hilbert-Schmidt Norms Bound Twirl Operator Diamond Norms]\label{lemma:app:moment_operator_hilbert_schmidt_norms_bound_twirl_operator_diamond_norms}
Given ensembles of quantum channels $\ensemblechannel,\ensemblechannel^{\prime}$, the completely bounded diamond and Hilbert-Schmidt norm of their twirl operators are bounded by the Hilbert-Schmidt norm of their moment operators, up to a dimension-dependent scaling,
\begin{align}
	\norm{\Tau_{\ensemblechannel}^{(t)} - \Tau_{\ensemblechannel^{\prime}}^{(t)}}_{\diamond} ~\leq&~~ \dsys^{t}~\norm{\widehat{\Tau}_{\ensemblechannel}^{(t)} - \widehat{\Tau}_{\ensemblechannel^{\prime}}^{(t)}}_{2} ~\leq~ \dsys^{t}~\norm{\widehat{\Tau}_{\ensemblechannel}^{(t)} - \widehat{\Tau}_{\ensemblechannel^{\prime}}^{(t)}}_{1} ~.
\end{align}
\end{lemma}

\begin{proof}\label{proof:app:moment_operator_hilbert_schmidt_norms_bound_twirl_operator_diamond_norms}
Given vectorizations, quantum channels $\channel \in \mathcal{B}[\mathcal{B}[\spacesys]]$ can also be represented via channel-operator or even channel-state dualities, as superoperators $\hat{\channel}$, or Choi states $\Phi_{\channel}$,
\begin{align}
	\hat{\channel} = \sum_{K_{\channel}} K_{\channel} \otimes K_{\channel}^{*}
	~~ \quad \leftrightarrow \quad ~~
	\Phi_{\channel} = \sum_{K_{\channel}} \kket{K_{\channel}}\bbra{K_{\channel}^{*}}~,\!\!\!\!
\end{align}
In particular, superoperator norms and traces correspond to the Choi state's purity and entanglement fidelity with respect to unnormalized maximally entangled states $\ket{\Omega} = \sum_{\alpha \in [\dsys]}\ket{\alpha\alpha} \in \spacesys \otimes \spacesys \to \Omega = \ket{\Omega}\bra{\Omega} \in \mathcal{B}[\spacesys \otimes \spacesys]$ \cite{wood2011tensor},
\begin{align}
	\big\lVert{\hat{\channel}}\big\rVert^{2} = \Tr\!\left[{\Phi_{\channel}^{\dagger}\Phi_{\channel}}\right]
	\quad ,&~ \quad
	\Tr\!\left[{\hat{\channel}}\right] = \Tr\!\left[{\vphantom{\Phi_{\channel}^{2}}{} \Omega \Phi_{\channel}}\right] ~.
\end{align}
More generally, for quantum channels $\channel,\channel^{\prime} \in \mathcal{B}[\mathcal{B}[\spacesys]]$, inner products between their superoperators equal inner products between their Choi states,
\begin{align}
	\expval{\hat{\channel}^{\prime},\hat{\channel}} = \Tr\!\left[{\hat{\channel}^{\prime\dagger}\hat{\channel}}\right] = \Tr\!\left[{\vphantom{\hat{\channel}}{}\Phi_{\channel^{\prime}}^{\dagger}\Phi_{\channel}}\right] = \expval{\vphantom{\hat{\channel}^{\prime}}{}\Phi_{\channel^{\prime}},\Phi_{\channel}}~.
\end{align}
Finally, given its convexity, the completely-bounded $1$-norm, or diamond $\diamond$ norm of a map $\channel$,
\begin{align}
	\norm{\channel}_{\diamond} = \max_{\ket{\psi} \in \spacesys \otimes \spacesys} \norm{(\channel \otimes \idsys)(\psi)}_{1}~,
\end{align}
is the $p=1$-norm of the operator acting on some normalized pure state $\ket{\psi} = (1/\dsys)U \otimes \idsys\ket{\Omega}$ on the doubled space $\spacesys \otimes \spacesys$ for some unitary $U \in \ensemblehaar[\dsys]$ \cite{watrous2018thetheory}. Further, the $p$-norm obeys Holder's inequality of $\norm{\cdot\cdot}_{r} \leq \norm{\cdot}_{p}\norm{\cdot}_{q}$ for $1/p + 1/q = 1/r$, the $2$-norm obeys $\norm{\cdot}_{1} \leq \sqrt{\dsys}\norm{\cdot}_{2}$, and unitaries have $\infty$-norms of the $1$-norm one of their columns $u$, bounded by $\norm{U}_{\infty} = \norm{u}_{1} \leq \sqrt{\dsys}$. The completely bounded $1$-norm of the difference of quantum channels $\channel,\channel^{\prime}$ can thus be bounded by the $2$-norm of their associated Choi states \cite{watrous2018thetheory}, and thus by the $2$ and $1$-norms of their superoperators,
\begin{align}
	\norm{\channel - \channel^{\prime}}_{\diamond}
	=&~ \norm{(\channel \otimes \idsys)(\psi) - (\channel^{\prime} \otimes \idsys)(\psi)}_{1} \\
	=&~ \frac{1}{\dsys}~\norm{(\channel \otimes \idsys)((\channelunitary \otimes \idsys)(\Omega)) - (\channel^{\prime} \otimes \idsys)((\channelunitary \otimes \idsys)(\Omega))}_{1} \\
	=&~ \frac{1}{\dsys}~\norm{(\idsys \otimes U^{*})\left((\channel \otimes \idsys)(\Omega) - (\channel^{\prime} \otimes \idsys)(\Omega)\right)(\idsys \otimes U^{T})}_{1} \\
	=&~ \frac{1}{\dsys}~\norm{(\idsys \otimes U^{*})\left(\Phi_{\channel} - \Phi_{\channel^{\prime}}\right)(\idsys \otimes U^{T})}_{1} \\
	\leq&~ \frac{1}{\dsys}~\norm{\idsys \otimes U^{*}}_{\infty}~\norm{\idsys \otimes U^{T}}_{\infty}~\norm{\Phi_{\channel} - \Phi_{\channel^{\prime}}}_{1} \\
	\leq&~ \frac{1}{\dsys}~\sqrt{\dsys}~\sqrt{\dsys}~\norm{\Phi_{\channel} - \Phi_{\channel^{\prime}}}_{1} \\
	=&~ \norm{\Phi_{\channel} - \Phi_{\channel^{\prime}}}_{1} \\
	\leq&~ \dsys~\norm{\Phi_{\channel} - \Phi_{\channel^{\prime}}}_{2} \\
	=&~ \dsys~\norm{\widehat{\channel} - \widehat{\channel}^{\prime}}_{2} \\
	\leq&~ \dsys~\norm{\widehat{\channel} - \widehat{\channel}^{\prime}}_{1} ~.
\end{align}
\end{proof}

For the special cases when the ensembles contain strictly unital operators or have projector moment operators, we can readily prove how the associated moment operators preserve the support of their inputs.

\begin{definition}[Projector Ensembles]\label{def:app:projector_ensembles}
Projector ensembles $\ensemblechannel$ have self-adjoint projector moment operators:
\begin{align}
	\widehat{\Tau}_{\ensemblechannel}^{(t)}\widehat{\Tau}_{\ensemblechannel}^{(t)} = \widehat{\Tau}_{\ensemblechannel}^{(t)}
	\quad \quad \quad , \quad \quad \quad
	\widehat{\Tau}_{\ensemblechannel}^{(t)\dagger} = \widehat{\Tau}_{\ensemblechannel}^{(t)}~.
\end{align}
\end{definition}

\begin{lemma}[Support Preservation of Unital Ensemble Twirls]\label{lemma:app:support_preservation_of_unital_ensemble_twirls}
	Given an orthogonal with respect to support basis $\mathcal{P}_{\dsys}^{t}$ for $\mathcal{B}[\spacesys^{\otimes t}]$, unital ensemble twirls $\ensembleunital$ are support-non-increasing and projector ensemble twirls $\ensembleunital$ are support-preserving,
	\begin{align}
			\widehat{\Tau}_{\ensembleunital}^{(t)} = \frac{1}{\dsys^{t}}\sum_{\substack{P,S \in \mathcal{P}_{\dsys}^{t}\\\Gamma_{P} \subseteq \Gamma_{S}}}\tau_{\ensembleunital}^{(\abs{S})}(P,S)~\kket{P}\bbra{S}
			\quad \quad \leftrightarrow& \quad \quad
			\tau_{\ensembleunital}^{(t)}(P,S) = \tau_{\ensembleunital}^{(\abs{S})}(P,S)~\delta_{\Gamma_{P} \subseteq \Gamma_{S}} \\
			\widehat{\Tau}_{\ensembleunital}^{(t)}\widehat{\Tau}_{\ensembleunital}^{(t)} = \widehat{\Tau}_{\ensembleunital}^{(t)} \quad , \quad \widehat{\Tau}_{\ensembleunital}^{(t)\dagger} = \widehat{\Tau}_{\ensembleunital}^{(t)} \quad \to \quad \widehat{\Tau}_{\ensembleunital}^{(t)} = \sum_{\Gamma \subseteq [t]} \widehat{\Tau}_{\ensembleunital}^{(\Gamma)} \quad :& \quad \widehat{\Tau}_{\ensembleunital}^{(\Gamma)} = \frac{1}{\dsys^{t}}\sum_{\substack{P,S \in \mathcal{P}_{\dsys}^{t}\\\Gamma_{P} = \Gamma_{S} = \Gamma}}\tau_{\ensembleunital}^{(\abs{S})}(P,S)~\kket{P}\bbra{S}~.
	\end{align}
\end{lemma}
\begin{proof}\label{proof:app:support_preservation_of_unital_ensemble_twirls}
Let us consider consider the action of unital ensemble $\ensembleunital$ twirls on $\abs{X}$-local operators $X = \otimes_{i \in \Gamma_{X}}X_{i}$, given identity operators are invariant under unital operators,
\begin{align}
		\Tau_{\ensembleunital}^{(t)}(X)
		= \int_{\ensembleunital}d\channel~ \channel^{\otimes t}(X)~
		= \int_{\ensembleunital}d\channel~ \otimes_{i \in \Gamma_{X}}\channel_{i}(X)~
		= \hspace{-0.3cm} \sum_{\substack{P,S \in \mathcal{P}_{\dsys}^{\abs{X}}\\\Gamma_{P} \subseteq \Gamma_{S} = \Gamma_{X}}}\tau_{\ensembleunital}^{(\abs{X})}(P,S)~\chi_{\dsys}^{(\abs{X})}(S,X)~P~
		= \Tau_{\ensembleunital}^{(\abs{X})}(X) ~.
\end{align}
It follows that unital $t$-order transfer matrices reduce to $\abs{S}$-order, that are non-zero $\tau_{\ensembleunital}^{(t)}(P,S) = \tau_{\ensembleunital}^{(\abs{S})}(P,S) \sim \delta_{\Gamma_{P} \subseteq \Gamma_{S}}$ for $P,S \in \mathcal{P}_{\dsys}^{t}$ when mapping between basis operators of strictly common support. In the case of projector ensembles $\ensembleunital$, moment operators are self-adjoint, $\widehat{\Tau}_{\ensembleunital}^{(t)\dagger} = \widehat{\Tau}_{\ensembleunital}^{(t)}$, constraining such twirls to map between operators of identical support. Projector ensembles can thus be written as a sum of projectors onto $\Gamma \subseteq [t]$-support spaces.
\end{proof}

Further, we note that via Stinespring dilations, an ensemble of unitaries induces an ensemble of quantum channels.

\begin{definition}[Unitary Induced Quantum Channel Ensembles]\label{def:app:unitary_induced_quantum_channel_ensembles}
	Unitary ensembles $\ensembleunitary$ over a composite space $\spacesys \otimes \spaceenv$, and an environment state $\nu_{\spaceenv} \in \mathcal{B}[\spaceenv]$, induce a quantum channel ensemble $\ensemblechannel$ over the system space $\spacesys$, with moment operators,
\begin{align}
	\widehat{\Tau}_{\ensemblechannel}^{(t)}
	=&~ \idsys^{\otimes t} \otimes \bbra{\idenv^{\otimes t}} ~~ \widehat{\Tau}_{\ensembleunitary}^{(t)} ~~ \idsys^{\otimes t} \otimes \kket{\nu_{\spaceenv}^{\otimes t}} ~.
\end{align}
\end{definition}

Finally, moment operators of channel ensembles induced by projector unitary ensembles are support-non-decreasing.

\begin{theorem}[Support Preservation of Quantum Channel Ensembles Induced by Unitary Ensembles]\label{theorem:app:support_preservation_of_quantum_channel_ensembles_induced_by_unitary_ensembles}
Moment operators of channel ensembles $\ensemblechannel$ over a $\dsys$-dimensional space $\spacesys$, induced via Stinespring dilations by projector unitary ensembles $\ensembleunitary$ over a composite $\dsys\denv$-dimensional space $\spacesys \otimes \spaceenv$, with an environment state $\nu_{\spaceenv} \in \mathcal{B}[\spaceenv]$, are support-non-decreasing,
\begin{align}
	\widehat{\Tau}_{\ensemblechannel}^{(t)} =&~ \frac{1}{\dsys^{t}}\sum_{\substack{P,S \in \mathcal{P}_{\dsys}^{t}\\\Gamma_{P} \supseteq \Gamma_{S}}}\tau_{\ensemblechannel}^{(t)}(P,S)~\kket{P}\bbra{S}
	\quad \quad \quad : \quad \quad \quad
	\tau_{\ensemblechannel}^{(t)}(P,S) = \!\!\!\!\!\! \sum_{\substack{T_{\spaceenv} \in \mathcal{P}_{\denv}^{t}\\\Gamma_{P_{\spacesys}} = \Gamma_{S_{\spacesys}} \cup \Gamma_{T_{\spaceenv}}}} \!\!\!\!\!\!  \bbra{T_{\spaceenv}}\kket{\nu_{\spaceenv}^{\otimes t}} ~\tau_{\ensembleunitary}^{(t)}(P_{\spacesys} \otimes \idenv^{\otimes t},S_{\spacesys} \otimes T_{\spaceenv})~,
\end{align}
given orthogonal with respect to support bases $\mathcal{P}_{\dsys}^{t},\mathcal{P}_{\denv}^{t}$ for $\spacesys$ and $\spaceenv$.
\end{theorem}

\begin{proof}\label{proof:app:support_preservation_of_quantum_channel_ensembles_induced_by_unitary_ensembles}
Given composite orthogonal with respect to support basis $\mathcal{P}_{\dsys}^{t} \otimes \mathcal{P}_{\denv}^{t}$ for $\spacesys \otimes \spaceenv$, from Lemma \ref{lemma:app:support_preservation_of_unital_ensemble_twirls}, projector unitary ensemble moment operators are support-preserving,
\begin{align}
	\widehat{\Tau}_{\ensembleunitary}^{(t)} =&~ \frac{1}{\dsys^{t}}\!\!\!\!\!\!\sum_{\substack{P_{\spacesys},S_{\spacesys} \in \mathcal{P}_{\dsys}^{t}\\P_{\spaceenv},S_{\spaceenv} \in \mathcal{P}_{\denv}^{t}\\\Gamma_{P_{\spacesys}} \cup \Gamma_{P_{\spaceenv}} = \Gamma_{S_{\spacesys}} \cup \Gamma_{S_{\spaceenv}}}}\!\!\!\!\!\!\tau_{\ensembleunitary}^{(t)}(P_{\spacesys} \otimes P_{\spaceenv},S_{\spacesys} \otimes S_{\spaceenv})~\kket{P_{\spacesys}P_{\spaceenv}}\bbra{S_{\spacesys}S_{\spaceenv}}~,
\end{align}
and the form of the induced channel ensemble moment operators follows directly from their definition.
\end{proof}

\subsection{Properties of Haar and cHaar Twirls}\label{subsec:properties_of_chaar_twirls}
Here, we derive expressions and properties for moment operators for the Haar, cHaar, and Depolarize ensembles. Throughout this section, we will use the fact that the $t$-th moment operator over the representation of a group projects onto its $t$-th order commutant. We begin with the following definitions of the relevant ensembles.

~\\~\\
\begin{definition}[Haar ensemble]\label{def:haar_ensemble} The $t$-th order twirl over the Haar ensemble $\ensemblehaar$ is defined as the twirl over $t$ copies of unitaries sampled from the group $\ensemblehaar$ according to the Haar measure. The associated twirl and moment operator are
	\begin{align}
			{\Tau}_{\ensemblehaar}^{(t)}(\cdot)
			~=~ \int_{\ensembleunitary(\dsys)}dU~U^{\otimes t}~\cdot~U^{\otimes t \dagger}
			\quad\quad \to&~ \quad\quad
			\widehat{\Tau}_{\ensemblehaar}^{(t)}
			~=~ \int_{\ensembleunitary(\dsys)}dU~U^{\otimes t} \otimes U^{\otimes t *}~.
	\end{align}
	Since the $t$-th order commutant of the unitary group corresponds to the system-permuting representation of the symmetric group we can always express the moment operator as
	\begin{align}
		\widehat{\Tau}_{\ensemblehaar}^{(t)}
		=&~ \frac{1}{\dsys^{t}}\sum_{\sigma,\pi \in \permutations[t]} \tau_{\ensemblehaar}^{(t)}(\sigma,\pi) ~ \kket{\sigma}\bbra{\pi}
		\quad\quad \text{with} \quad\quad
		\tau_{\ensemblehaar}^{(t)}(\sigma,\pi) = W_{\dsys}^{(t)-1}(\sigma,\pi) ~.
	\end{align}
\end{definition}
~\\
\begin{definition}[cHaar ensemble]\label{def:chaar_ensemble}
	The $t$-th order twirl over the channel-Haar (cHaar) ensemble $\ensemblechaar$ is defined as a measure over $t$ copies of random channels, and is a uniform, Lebesgue measure when $\denv = \dsys^{2}$. Using the Stinespring formalism, we can re-write this average as a twirl with respect to the Haar ensemble of unitaries acting on $t$ copies of a composite space, of a system and environment $\spacesys \otimes \spaceenv$ of dimensions $\dsys, \denv$, and a local pure environment state $\nu_{\spaceenv} \in \mathcal{B}[\spaceenv]$,
	\begin{align}
			{\Tau}_{\ensemblechaar}^{(t)}(\cdot) ~=~ \Tr_{\spaceenv^{\otimes t}}\left[{\Tau}_{\ensembleunitary(\dsys\denv)}^{(t)}(\cdot~ \otimes~ \nu_{\spaceenv}^{\otimes t}) \right]
			\quad\quad \to&~ \quad\quad
			\widehat{\Tau}_{\ensemblechaar}^{(t)} ~=~
			\idsys^{\otimes t} \otimes \bbra{\idenv^{\otimes t}} ~~ \widehat{\Tau}_{\ensembleunitary(\dsys\denv)}^{(t)} ~~ \idsys^{\otimes t} \otimes \kket{\nu_{\spaceenv}^{\otimes t}}~.
	\end{align}
	Since the cHaar ensemble is generated by the unitary Haar ensemble over $\ensembleunitary(\dsys\denv)$, $\widehat{\Tau}_{\ensemblechaar}^{(t)}$ can be obtained by projecting onto the commutant of $\ensembleunitary(\dsys\denv)$, acting on the environment state $\nu_{\spaceenv}$, and then tracing out the environment. This leads to
	\begin{align}
		\widehat{\Tau}_{\ensemblechaar}^{(t)}
		=&~ \frac{1}{\dsys^{t}}\sum_{\sigma,\pi \in \permutations[t]} \tau_{\ensemblechaar}^{(t)}(\sigma,\pi) ~\kket{\sigma}\bbra{\pi}
		\quad\quad \text{with} \quad\quad
		\tau_{\ensemblechaar}^{(t)}(\sigma,\pi) = \chi_{\denv}^{(t)}(\sigma)~\tau_{\ensembleunitary(\dsys\denv)}^{(t)}(\sigma,\pi)~.
	\end{align}
\end{definition}
~\\
\begin{definition}[Depolarize ensemble]\label{def:depolarize_ensemble}
	The $t$-th order twirl over the depolarizing (Depolarize) ensemble $\ensembledep$ is defined via a single maximally depolarizing channel $\channeldepolarize{\dsys}(\cdot)$, with unit probability. The associated twirl and moment operator are
	\begin{align}
		{\Tau}_{\ensembledep}^{(t)}(\cdot) = \channeldepolarize{\dsys}^{\otimes t}(\cdot) = \frac{\Tr\!\left[\cdot\right]}{\dsys^{t}}\idsys^{\otimes t}
		\quad\quad \to&~ \quad\quad
		\widehat{\Tau}_{\ensembledep}^{(t)} = \channeldepolarize[\widehat]{\dsys}^{\otimes t} = \frac{1}{\dsys^{t}}\kket{\idsys^{\otimes t}}\bbra{\idsys^{\otimes t}}~.
	\end{align}
	The Depolarize moment operator projects onto an invariant, trace-preserving basis, of solely the identity operator $\{\idsys^{\otimes t}\}$.
\end{definition}~\\

We now derive explicit forms of the Haar, cHaar, and Depolarize moment operators.

\begin{theorem}[Haar Moment Operators are Support-Preserving and Block-Diagonal in Localized Basis]\label{theorem:app:haar_moment_operators_are_support_preserving_and_block_diagonal_in_localized_basis}
	The Haar moment operator is block-diagonal in the localized permutation basis, as per its support,
	\begin{align}
		\widehat{\Tau}_{\ensemblehaar}^{(t)}
		=&~ \bigoplus_{\Gamma \subseteq [t]} \widehat{\Tau}_{\ensemblehaar}^{(\Gamma)}
		= \left[\begin{array}{ccccc}
		\widehat{\Tau}_{\ensemblehaar}^{(\{\})} & & & \\
		& \widehat{\Tau}_{\ensemblehaar}^{(\{i,j\})} & & \\
		& & \ddots & \\
		& & & \widehat{\Tau}_{\ensemblehaar}^{([t])}
		\end{array}\right]
	\end{align}
	where
	\begin{equation}
		\displaystyle{\widehat{\Tau}_{\ensemblehaar}^{(\Gamma)}
		= \hspace{-0cm}\frac{1}{\dsys^{t}}\sum_{\substack{\sigma,\pi \in \permutations[t]\\\Gamma_{\sigma}=\Gamma_{\pi}=\Gamma\\l = \abs{\Gamma}}} \hspace{-0cm} \tau_{\ensemblehaar}^{(l)}(\localization{\sigma},\localization{\pi}) ~\kket{\localization{\sigma}}\bbra{\localization{\pi}}} ~,\quad\text{and}\quad
		\tau_{\ensemblehaar}^{(l)}(\localization{\sigma},\localization{\pi}) = \!\!\!\displaystyle\sum_{\substack{\sigma \subseteq \eta \in \permutations[l]\\\pi \subseteq \kappa \in \permutations[l]}}\!\!\! ~\chi_{\dsys}^{(l)}(\eta)~W_{\dsys}^{(l)-1}(\eta,\kappa)~\chi_{\dsys}^{(l)}(\kappa)~,
	\end{equation}
	with orthogonal support-dependent block moment operators $\widehat{\Tau}_{\ensemblehaar}^{(\Gamma)}$, spanned by localized permutations $\localization{\sigma},\localization{\pi} \in \localizations[t]$ with support $\Gamma = \Gamma_{\localization{\sigma}} = \Gamma_{\localization{\pi}} ~:~ \Gamma \subseteq [t]$, and exact identity and transposition coefficients of,
	\begin{align}
		\tau_{\ensemblehaar}^{(t)}(\localization{\sigma},\localization{e}) = \delta_{\sigma e}
		\quad \quad ,&~ \quad \quad
		\tau_{\ensemblehaar}^{(t)}(\localization{\tau},\localization[^{\prime}]{\tau}) = \frac{1}{\dsys^{2}-1}\delta_{\tau\tau^{\prime}}~.
	\end{align}
\end{theorem}

\begin{theorem}[cHaar Moment Operators are Support-Non-Decreasing and Block-Lower-Triangular in Localized Basis]\label{theorem:app:chaar_moment_operators_are_support_non_decreasing_and_block_diagonal_in_localized_basis}
	The cHaar moment operator is block-lower-triangular in the localized permutation basis, as per its support,
	\begin{align}
		\hspace{-0.5cm}
		\widehat{\Tau}_{\ensemblechaar}^{(t)}
		=&~ \left[\begin{array}{ccccc}
		\widehat{\Tau}_{\ensemblechaar}^{(\{\})} & & & & \\
		\widehat{\Tau}_{\ensemblechaar}^{(\{\},\{i,j\})} & \widehat{\Tau}_{\ensemblechaar}^{(\{i,j\})} & & &\\
		\widehat{\Tau}_{\ensemblechaar}^{(\{\},\{i,j,k\})} & \cdots & \widehat{\Tau}_{\ensemblechaar}^{(\{i,j,k\})} & &\\
		\vdots & \ddots & \widehat{\Tau}_{\ensemblechaar}^{(\Gamma,\Gamma^{\prime})} & \ddots & \\
		\widehat{\Tau}_{\ensemblechaar}^{(\{\},[t])} & \cdots & \cdots & \cdots & \widehat{\Tau}_{\ensemblechaar}^{([t])}
		\end{array}\right]
	\end{align}
	where
	\begin{equation}
		\displaystyle{\widehat{\Tau}_{\ensemblechaar}^{(\Gamma,\Gamma^{\prime})}
		= \frac{1}{\dsys^{t}}\hspace{-0.5cm}\sum_{\substack{\sigma,\pi \in \permutations[t]\\\Gamma_{\sigma}=\Gamma \supseteq \Gamma_{\pi}=\Gamma^{\prime}\\l = \abs{\Gamma} \geq l^{\prime} = \abs{\Gamma^{\prime}}}} \hspace{-0.5cm} \tau_{\ensemblechaar}^{(l)}(\localization{\sigma},\localization{\pi}) ~\kket{\localization{\sigma}}\bbra{\localization{\pi}} }~,\quad \text{and}\quad
		\tau_{\ensemblechaar}^{(l)}(\localization{\sigma},\localization{\pi}) = \!\!\!\displaystyle\sum_{\substack{\sigma \subseteq \eta \in \permutations[l]\\\pi \subseteq \kappa \in \permutations[l]}}\!\!\! ~\chi_{\dsys\denv}^{(l)}(\eta)~W_{\dsys\denv}^{(l)-1}(\eta,\kappa)~\chi_{\dsys}^{(l)}(\kappa)~,
	\end{equation}
	with orthogonal support-dependent block moment operators $\widehat{\Tau}_{\ensemblechaar}^{(\Gamma,\Gamma^{\prime})}$, spanned by localized permutations $\localization{\sigma},\localization{\pi} \in \localizations[t]$ with support $\Gamma_{\localization{\sigma}} = \Gamma \supseteq \Gamma^{\prime} =\Gamma_{\localization{\pi}} ~:~ \Gamma,\Gamma^{\prime} \subseteq [t]$, and exact identity and transposition coefficients of,
	\begin{align}
		\tau_{\ensemblechaar}^{(t)}(\localization{e},\localization{e}) = 1
		\quad ,&~ \quad
		\tau_{\ensemblechaar}^{(t)}(\localization{\tau},\localization{e}) = \frac{\denv-1}{\dsys^{2}\denv^{2}-1}
		\quad ,&~ \quad
		\tau_{\ensemblechaar}^{(t)}(\localization{\tau},\localization[^{\prime}]{\tau}) = \frac{\denv}{\dsys^{2}\denv^{2}-1}\delta_{\tau\tau^{\prime}}~.
	\end{align}
\end{theorem}

\begin{proof}\label{proof:app:chaar_moment_operators_are_support_non_decreasing_and_block_diagonal_in_localized_basis}
Let us express the moment operators in terms of the orthogonal with respect to support localized permutation basis $\localizations[t]$. In particular, cHaar ensemble moment operators are generated by support-preserving Haar ensemble moment operators in the composite system and environment space,
\begin{align}
		\widehat{\Tau}_{\ensembleunitary(\dsys\denv)}^{(t)}
		=&~ \frac{1}{\dsys^{t}\denv^{t}}\sum_{\sigma,\pi \in \permutations[t]} \tau_{\ensembleunitary(\dsys\denv)}^{(t)}(\sigma,\pi)~\kket{\sigma_{\spacesys}\sigma_{\spaceenv}}\bbra{\pi_{\spacesys}\pi_{\spaceenv}} \\
		=&~ \frac{1}{\dsys^{t}\denv^{t}}\sum_{\substack{\eta,\eta^{\prime} \in \permutations[t] \\\kappa,\kappa^{\prime} \in \permutations[t]\\\Gamma_{\eta}\cup \Gamma_{\eta^{\prime}} = \Gamma_{\kappa} \cup \Gamma_{\kappa^{\prime}}\\\Gamma = \Gamma_{\eta}\cup \Gamma_{\eta^{\prime}} ~\!,~\! l = \abs{\Gamma}}}\sum_{\substack{\eta,\eta^{\prime} \subseteq \sigma \in \permutations[l] \\\kappa,\kappa^{\prime} \subseteq \pi \in \permutations[l] }} ~\chi_{\dsys\denv}^{(l)}(\sigma) ~\tau_{\ensembleunitary(\dsys\denv)}^{(l)}(\sigma,\pi)~\chi_{\dsys\denv}^{(l)}(\pi)~\kket{\localization[_{\spacesys}]{\eta}\localization[^{\prime}_{\spaceenv}]{\eta}}\bbra{\localization[_{\spacesys}]{\kappa}\localization[^{\prime}_{\spaceenv}]{\kappa}} ~.
		\end{align}~\\

		\noindent In particular, we note that the support of the input permutations $\kappa,\kappa^{\prime}$ can be changed from $\Gamma_{\eta}\cup \Gamma_{\eta^{\prime}} = \Gamma_{\kappa} \cup \Gamma_{\kappa^{\prime}} ~\to~ \Gamma_{\eta}\cup \Gamma_{\eta^{\prime}} \supseteq \Gamma_{\kappa} \cup \Gamma_{\kappa^{\prime}}$, as any coefficients arising from summations over $\pi \supseteq \kappa,\kappa^{\prime}$ will cancel to zero for $\Gamma_{\kappa} \cup \Gamma_{\kappa^{\prime}} \subset \Gamma_{\eta}\cup \Gamma_{\eta^{\prime}}$. Hence, we can write the Haar ensemble moment operator in the composite system and environment space as,
		\begin{align}
		\widehat{\Tau}_{\ensembleunitary(\dsys\denv)}^{(t)}
		=&~ \frac{1}{\dsys^{t}\denv^{t}}\sum_{\substack{\eta,\eta^{\prime} \in \permutations[t] \\\eta,\eta^{\prime} \subseteq \sigma \in \permutations[l]\\\Gamma = \Gamma_{\eta}\cup \Gamma_{\eta^{\prime}} ~\!,~\! l = \abs{\Gamma} ~\! ,}}\sum_{\substack{\kappa,\kappa^{\prime} \in \permutations[l] \\\kappa,\kappa^{\prime} \subseteq \pi \in \permutations[l]\\ \Gamma_{\eta}\cup \Gamma_{\eta^{\prime}} \supseteq \Gamma_{\kappa} \cup \Gamma_{\kappa^{\prime}}}} ~\chi_{\dsys\denv}^{(l)}(\sigma) ~\tau_{\ensembleunitary(\dsys\denv)}^{(l)}(\sigma,\pi)~\chi_{\dsys\denv}^{(l)}(\pi)~\kket{\localization[_{\spacesys}]{\eta}\localization[^{\prime}_{\spaceenv}]{\eta}}\bbra{\localization[_{\spacesys}]{\kappa}\localization[^{\prime}_{\spaceenv}]{\kappa}} \\
		=&~ \frac{1}{\dsys^{t}\denv^{t}}\sum_{\substack{\eta,\eta^{\prime} \in \permutations[t] \\\eta,\eta^{\prime} \subseteq \sigma \in \permutations[l]\\\Gamma = \Gamma_{\eta}\cup \Gamma_{\eta^{\prime}} = \Gamma_{\sigma} ~\!,~\! l = \abs{\Gamma}}}\sum_{\substack{\pi \in \permutations[l] \\\Gamma_{\pi} \subseteq \Gamma}} \chi_{\dsys\denv}^{(l)}(\sigma) ~\tau_{\ensembleunitary(\dsys\denv)}^{(l)}(\sigma,\pi)~\kket{\localization[_{\spacesys}]{\eta}\localization[^{\prime}_{\spaceenv}]{\eta}}\bbra{\pi_{\spacesys}\pi_{\spaceenv}} ~.
\end{align}
From here, we can show that the cHaar ensemble moment is support-non-decreasing
\begin{align}
	\!\!\!\!
		\widehat{\Tau}_{\ensemblechaar}^{(t)}
		=&~ \idsys^{\otimes t} \otimes \bbra{\idenv^{\otimes t}} ~~ \widehat{\Tau}_{\ensembleunitary(\dsys\denv)}^{(t)} ~~ \idsys^{\otimes t} \otimes \kket{\nu_{\spaceenv}^{\otimes t}} \\
		=&~ \frac{1}{\dsys^{t}\denv^{t}}\sum_{\substack{\eta,\eta^{\prime} \in \permutations[t] \\\eta,\eta^{\prime} \subseteq \sigma \in \permutations[l]\\\Gamma = \Gamma_{\eta}\cup \Gamma_{\eta^{\prime}} ~\!,~\! l = \abs{\Gamma}}}\sum_{\substack{\pi \in \permutations[l]\\\Gamma_{\pi} \subseteq \Gamma}} \chi_{\dsys\denv}^{(l)}(\sigma) ~\tau_{\ensembleunitary(\dsys\denv)}^{(l)}(\sigma,\pi)~\bbra{\idenv^{\otimes t}}\kket{\localization[^{\prime}_{\spaceenv}]{\eta}}~ \bbra{\pi_{\spaceenv}}\kket{\nu_{\spaceenv}^{\otimes t}} ~\kket{\localization[_{\spacesys}]{\eta}}\bbra{\pi_{\spacesys}}\\
		=&~ \frac{1}{\dsys^{t}}\sum_{\substack{\eta \in \permutations[t] \\\eta \subseteq \sigma \in \permutations[l]\\\Gamma = \Gamma_{\eta} ~\!,~\! l = \abs{\Gamma}}}\sum_{\substack{\pi \in \permutations[l]\\\Gamma_{\pi} \subseteq \Gamma}} \chi_{\dsys\denv}^{(l)}(\sigma) ~\tau_{\ensembleunitary(\dsys\denv)}^{(l)}(\sigma,\pi)~\kket{\localization[_{\spacesys}]{\eta}}\bbra{\pi_{\spacesys}}\\
		=&~ \frac{1}{\dsys^{t}}\sum_{\substack{\eta \in \permutations[t] \\\eta \subseteq \sigma \in \permutations[l]\\\Gamma = \Gamma_{\eta} ~\!,~\! l = \abs{\Gamma}}}\sum_{\substack{\pi \in \permutations[l]\\\kappa \subseteq \pi \in \permutations[l]\\\Gamma_{\eta} \subseteq \Gamma}}\chi_{\dsys\denv}^{(l)}(\sigma) ~\tau_{\ensembleunitary(\dsys\denv)}^{(l)}(\sigma,\pi)~\chi_{\dsys}^{(l)}(\pi)~\kket{\localization[_{\spacesys}]{\eta}}\bbra{\localization[_{\spacesys}]{\kappa}}\\
		=&~ \frac{1}{\dsys^{t}}\!\!\!\!\!\!\!\!\sum_{\substack{\sigma,\pi \in \permutations[t] \\\Gamma = \Gamma_{\sigma} \supseteq \Gamma_{\pi} ~\!,~\! l = \abs{\Gamma}}}\!\!\!\!\!\!\!\!\tau_{\ensemblechaar}^{(l)}(\localization{\sigma},\localization{\pi})~\kket{\localization{\sigma}}\bbra{\localization{\pi}}
		~~~ : ~~~
		\tau_{\ensemblechaar}^{(l)}(\localization{\sigma},\localization{\pi}) = \!\!\!\sum_{\substack{\sigma \subseteq \eta \in \permutations[l]\\\pi \subseteq \kappa \in \permutations[l]}} \!\!\chi_{\dsys\denv}^{(l)}(\eta) ~\tau_{\ensembleunitary(\dsys\denv)}^{(l)}(\eta,\kappa)~\chi_{\dsys}^{(l)}(\kappa)~. \!\!\!
\end{align}
\vspace{-8pt}
\end{proof}

\begin{figure}[t]
		\centering
		\includegraphics[width=0.8\linewidth]{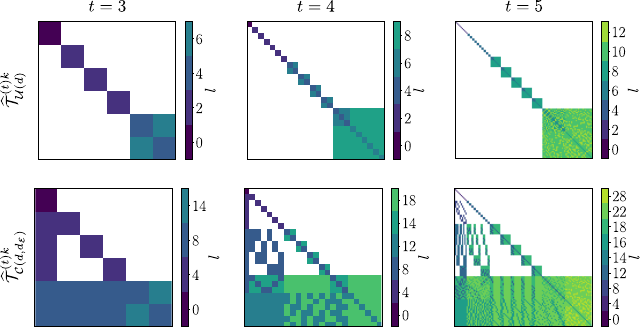}
		\caption{\textbf{Moment operators in a basis that block diagonalizes the Haar moment operator as a permutation transfer matrix.} Scaling of the $t$-th order moment operator transfer matrix elements for the Haar (top) and cHaar (bottom) ensembles in the localized permutation basis, for $t=3,4,5$ and $\denv = \dsys^{2}$. Depicted are matrix elements corresponding to all pairs of $t!$ permutations $\sigma,\pi \in \permutations[t]$, sorted by their support, from the smallest identity (top-left corner), to largest cycles (bottom-right corner), with leading-order scaling $l$ (colored gradient), such that, $\tau_{\ensemblehaar}^{(t)}(\localization{\sigma},\localization{\pi}),\tau_{\ensemblechaar}^{(t)}(\localization{\sigma},\localization{\pi}) \sim \mathcal{O}(1/\dsys^{l})$. A sparse, support-dependent block-diagonal (Haar) or block-lower-triangular (cHaar) structure of coefficients occurs, and transposition coefficients (light-purple top-left diagonal) are uniquely diagonal $1/(\dsys^{2}-1)$ (Haar), or diagonal and non-unital $(\denv - \delta_{\pi e})/(\dsys^{2}\denv^{2}-1)$ (cHaar).}
		\label{fig:app:haar_chaar_coefficients}
\end{figure}

To illustrate these properties, we compute the exact $t=3,4,5$-order Haar and cHaar transfer matrices, for $\denv = \dsys^{2}$~ \cite{cardin2026haarpy}, as shown in Fig. \ref{fig:app:haar_chaar_coefficients}. Here, we can see that a sparse, support-dependent block-diagonal or block-lower-triangular structure occurs. Within the transfer matrices, many of the $\binom{t}{l}$ locality-$l$ blocks, of size $!l$, are identical, and correspond to $l \leq t$-order twirl blocks, due to isomorphisms between permutations. Such structure is primarily attributed to the localized permutation basis being orthogonal with respect to support. For the Haar ensemble, the Haar invariance enforces the support-preserving behavior, and for the cHaar ensemble, the environment space partial trace enforces the support increasing behavior. From these results, and inspired by similar discussions in \cite{collins2017weingarten}, we provide intriguing identities for permutations $\sigma,\pi \in \permutations[t]$ of support $\Gamma = \Gamma_{\sigma} \supseteq \Gamma_{\pi}$ and locality $l = \abs{\Gamma}$, and general dimensions $\dsys,\denv$,
\begin{align}
	\!\!\!
	\sum_{\substack{\eta,\kappa \in \permutations[t] \\ \eta \supseteq \sigma,~ \kappa \supseteq \pi}} \hspace{-0cm} \chi_{\dsys\denv}^{(t)}(\eta)~W_{\dsys\denv}^{(t)-1}(\eta,\kappa) ~ \chi_{\dsys}^{(t)}(\kappa)
	~~=&~ \hspace{-0cm} \sum_{\substack{\eta,\kappa \in \mathcal{S}_{l} \\ \eta \supseteq \sigma,~ \kappa \supseteq \pi}} \hspace{-0cm} \chi_{\dsys\denv}^{(l)}(\eta)~W_{\dsys\denv}^{(l)-1}(\eta,\kappa)~\chi_{\dsys}^{(l)}(\kappa) ~\left(\delta_{\Gamma_{\sigma} \supseteq\Gamma_{\pi}}\delta_{\denv>1} ~+~ \delta_{\Gamma_{\sigma}=\Gamma_{\pi}}\delta_{\denv=1}\right)~\!. \!\!\!\!\!\!
\end{align}

Next, we turn our attention to computing the exact $t=1,2$-order moment operator for the $k$-concatenated cHaar ensemble. In particular, $\widehat{\Tau}_{\ensemblechaar}^{(t)}$ is not a projector (it is not obtained as the average of a group), then $\widehat{\Tau}_{\ensemblechaar}^{(t)k}{}\neq \widehat{\Tau}_{\ensemblechaar}^{(t)}$, and we are interested in its convergence properties with $k$. A straightforward calculation leads to the following lemma.
\begin{lemma}[Exact $t=1,2$-order cHaar Moment Operators]\label{lemma:app:exact_t12_order_chaar_moment_operators}
	$k$-concatenated $t=1,2$-order cHaar moment operators, for all dimensions $\dsys,\denv$, are,
	{
	\setlength{\abovedisplayskip}{-4pt}
	\begin{align}
		\widehat{\Tau}_{\ensemblechaar}^{(1)k}
		=&~ \frac{1}{\dsys}\kket{\localization{e}}\bbra{\localization{e}} \nonumber \\[-6pt]
	\end{align}
	}
	\vspace{-24pt}
	\begin{align}
		\hspace{-0.5cm}
		\widehat{\Tau}_{\ensemblechaar}^{(2)k}
		=&~ \frac{1}{\dsys^{2}}\kket{\localization{e}}\bbra{\localization{e}} + \frac{1}{\dsys^{2}}\frac{\denv-1}{\dsys^{2}\denv^{2}-1} \hspace{-0cm} \left[1 + \sum_{s > 0}^{k-1}\left[\denv\frac{\dsys^{2}-1}{\dsys^{2}\denv^{2}-1}\right]^{s}\right] \hspace{-0.1cm} \kket{\localization{\tau}}\bbra{\localization{e}}
		+ \frac{1}{\dsys^{2}}\frac{1}{\dsys^{2}-1}\left[\denv\frac{\dsys^{2}-1}{\dsys^{2}\denv^{2}-1}\right]^{k}\hspace{-0.2cm} \kket{\localization{\tau}}\bbra{\localization{\tau}} ~, \!\!\!\!\!\!\!\! \nonumber
	\end{align}
	where $\localization{e} = \idsys^{\otimes t}$ is the identity operator, and $\localization{\tau} = \dsys S_{\dsys} - \idsys^{\otimes 2} = \sum_{P \in \mathcal{P}_{\dsys} \backslash \{\idsys\}} \hspace{-0cm} P \otimes P^{\dagger}$ is the localized transposition operator.
\end{lemma}

From the previous results for individual transfer matrix elements, and overlaps between basis operators, we can bound the transfer matrix elements for general $t$-th order moment operators for the $k$-concatenated cHaar ensemble.

\begin{theorem}[Asymptotic $t$-th order cHaar Moment Operators]\label{theorem:app:asymptotic_t_order_chaar_moment_operators}
$k$-concatenated $t$-th order cHaar moment operators, in the asymptotic limit $\dsys\denv \to \infty$, have transfer matrix elements upper bounded by,
	\begin{align}
		\abs{\tau_{\ensemblechaar}^{(l,k)}(\localization{\sigma},\localization{\pi})}
		~\leq&~~ \frac{1}{\dsys^{\abs{\sigma} + \abs{\pi}}\denv^{\abs{\sigma} + (k-1)\abs{\Gamma_{\pi}}/2}}~\constant_{\tau}^{(l,k)} ~\delta_{\Gamma_{\sigma} \supseteq \Gamma_{\pi}}~,
	\end{align}
	for $\sigma,\pi \in \permutations[l]$, $l \in [t+1]$, given $t,k$-dependent constants $\constant_{\tau}^{(l,k)} = \frac{1}{l!^{2}~\catalan[l]{l}^{4}}~\constant_{\lambda}^{(l)k}$, ~and~ $\constant_{\lambda}^{(l)} = l!^{2}~\weingarten[]{l}~\catalan[l]{l}^{7}$.
\end{theorem}

\begin{proof}\label{proof:app:moment_operators}
	By definition, for $\sigma,\pi \in \permutations[l]$ we have bounds for the individual transfer matrix elements and overlaps of,
	\begin{align}
		\tau_{\ensemblechaar}^{(l)}(\localization{\sigma},\localization{\pi})
		~=&~~ \!\!\!\!\sum_{\substack{\sigma \subseteq \eta \!~,\!~\pi \subseteq \kappa}}\!\! \!\!\chi_{\dsys\denv}^{(l)}(\eta) ~W_{\dsys\denv}^{(l)-1}(\eta,\kappa)~\chi_{\dsys}^{(l)}(\kappa)~\delta_{\Gamma_{\sigma} \supseteq \Gamma_{\pi}}
		~~\leq~~ \frac{1}{\dsys^{\abs{\sigma} + \abs{\pi}}\denv^{\abs{\sigma}}}~\constant_{\tau}^{(l)}~\delta_{\Gamma_{\sigma} \supseteq \Gamma_{\pi}}
		\\[10pt]
		\chi_{\dsys}^{(l)}(\localization{\sigma},\localization{\pi})
		~=&~~ \!\!\!\!\sum_{\substack{\eta \subseteq \sigma\!~,\!~\kappa \subseteq \pi}}\!\!\!\!\!\!\! ~\mobius[l]{\eta^{-1}\sigma}~\mobius[l]{\kappa^{-1}\pi}~\frac{\chi_{\dsys}^{(l)}(\eta,\kappa)}{\chi_{\dsys}^{(l)}(\eta)\chi_{\dsys}^{(l)}(\kappa)}~\delta_{\Gamma_{\sigma} \Gamma_{\pi}}
		~~\leq~~ \dsys^{\abs{\sigma} + \abs{\pi}}~\constant_{\chi}^{(l)}~\delta_{\Gamma_{\sigma}\Gamma_{\pi}} \\[10pt]
		\constant_{\tau}^{(l)} ~=&~~ \weingarten{l}~\catalan[l]{l}^{3}
		\quad\quad , \quad \quad
		\constant_{\chi}^{(l)} ~=~ \mobiuscatalan[l]{\sigma}^{2}~\mobiuscatalan[l]{\pi}^{2}
		~~,
	\end{align}
	 up to $l$-dependent constants $\weingarten{l} > 0$ \cite{collins2017weingarten}. We can also bound the Catalan, Möbius, and Weingarten functions as,
	\begin{align}
		\hspace{-0.6cm}
		\mobius[l]{\sigma} ~\leq~ \abs{\mobius[l]{\sigma}} ~=&~~ \prod_{\lambda \in \sigma}\catalan[l]{\abs{\lambda}} ~\leq~ \mobiuscatalan[l]{\sigma} ~=~ \prod_{\lambda \in \sigma}\catalan[l]{\abs{\lambda}+1} ~\leq~ \catalan[l]{l} ~\leq~ e^{2l} ~~:~~ \binom{t}{l} \leq e^{2l}\left(\frac{t}{2l}\right)^{l} \!\!\\[4pt]
		\hspace{-0.6cm}
		W_{\dsys}^{(l)-1}(\sigma) ~\leq&~~ \weingarten{l}~\mobius[l]{\sigma}~\frac{1}{\dsys^{\abs{\sigma}}} ~\leq~ \weingarten{l}~\mobiuscatalan[l]{\sigma}~\frac{1}{\dsys^{\abs{\sigma}}} \\[4pt]
		\hspace{-0.6cm}
		\abs{\sigma^{\vphantom{-1}{}}} + \abs{\pi^{\vphantom{-1}{}}} \pm \abs{\sigma^{-1}\pi} ~~\substack{\displaystyle{\geq}\\\displaystyle{\leq}}&~~ \abs{\sigma^{\vphantom{-1}{}}} + \abs{\pi^{\vphantom{-1}{}}}
		~~ , ~~
		\abs{\sigma^{\vphantom{-1}{}}} + \abs{\sigma^{-1}\pi} ~\geq~ \abs{\sigma^{\vphantom{-1}{}}}
		~~ , ~~
		\ceil{l/2} ~\leq~ \abs{\sigma^{\vphantom{-1}{}}} ~\leq~ l-1
		\!\!\!\!\!\!\!\!\! 		\\[10pt]
		\hspace{-0.6cm}
		\abs{\{\sigma \supseteq \pi \in \permutations[l]\}} ~=&~~ \mobiuscatalan[l]{\sigma} ~\leq~ \catalan[l]{l}
		\quad , \quad
		\abs{\{\sigma \subseteq \pi \in \permutations[l]\}} ~\leq~ \catalan[l]{l-\abs{\sigma}} ~\leq~ \catalan[l]{l} ~.
	\end{align}
	Then, $k$-concatenations of $l \leq t$-order cHaar moment operators have transfer matrix elements for $\sigma,\pi \in \permutations[l]$ of the form,
	\begin{align}
		\tau_{\ensemblechaar}^{(l,k)}(\localization{\sigma},\localization{\pi})
		~=&~~ \!\!\!\!\!\!\!\!\!\!\sum_{\substack{\{\sigma_{i},\pi_{i} \in \permutations[l_{i}]\}_{i \in [k]}\\\Gamma_{\pi_{i-1}} = \Gamma_{\pi_{i}} \subseteq \Gamma_{\sigma_{i}} = \Gamma_{\sigma_{i+1}}\\\sigma_{k-1} = \sigma~,~\pi_{0}=\pi~,~l_{i}=\abs{\Gamma_{\sigma_{i}}}}}\!\!\!\! ~\prod_{i \in [k]}~\tau_{\ensemblechaar}^{(l_{i})}(\localization[_{i}]{\sigma},\localization[_{i}]{\pi})~\prod_{i \in [k-1]}~ W_{\dsys\denv}^{(l_{i})-1}(\localization[_{i+1}]{\pi},\localization[_{i}]{\sigma})~.
	\end{align}
	Given the support constraints $\{\} \subseteq \Gamma_{\pi} \subseteq \cdots \subseteq \Gamma_{\pi_{i}} \subseteq \Gamma_{\sigma_{i}} \subseteq \cdots \subseteq \Gamma_{\sigma} \subseteq [l]$ of the propagated permutations, the telescoping of the system dimension scalings cancelling from identical scaling of the transfer matrices $\sim \dsys^{\abs{\sigma_{i}}+\abs{\pi_{i}}}$ and overlaps $\sim 1/\dsys^{\abs{\sigma_{i}}+\abs{\pi_{i+1}}}$, and the environment dimension scalings $\sim 1/\denv^{\abs{\sigma_{i}}}$ being bounded by the at least $\abs{\Gamma_{\pi}}$-locality minimal permutations $\{\sigma_{i} ~:~ \Gamma_{\sigma_{i}}=\Gamma_{\pi_{i}}=\Gamma_{\pi}~,~\abs{\sigma_{i}} \geq \abs{\Gamma_{\pi}}/2 \geq \abs{\Gamma_{\pi}}/2\}_{i \in [k-1]}$, propagating through until the $k-1$-th moment operator, the $k$-concatenated $l \leq t$-order transfer matrix elements are bounded by,
	\begin{align}
		\abs{\tau_{\ensemblechaar}^{(l,k)}(\localization{\sigma},\localization{\pi})}
		~\leq&~~ \frac{1}{\dsys^{\abs{\sigma} + \abs{\pi}}\denv^{\abs{\sigma} + (k-1)\abs{\Gamma_{\pi}}/2}}~\constant_{\tau}^{(l,k)} ~\delta_{\Gamma_{\sigma} \supseteq \Gamma_{\pi}} \\[12pt]
		\constant_{\tau}^{(l,k)} ~=~ l!^{2(k-1)}~\weingarten[k]{l}~\catalan[l]{l}^{3k + 4(k-1)}
		~=&~~ \frac{1}{l!^{2}~\catalan[l]{l}^{4}}~\left(l!^{2}~\weingarten[]{l}~\catalan[l]{l}^{7}\right)^{k}
		~=~ \frac{1}{l!^{2}~\catalan[l]{l}^{4}}~\constant_{\lambda}^{(l)k}
		\quad \quad , \quad \quad \constant_{\lambda}^{(l)} ~:=~ l!^{2}~\weingarten[]{l}~\catalan[l]{l}^{7}~.
	\end{align}
	Finally, we note that the simplest non-trivial transpositions are orthogonal, with exact transfer matrices and overlaps of,
	\begin{align}
		\hspace{-0.5cm}
		\tau_{\ensemblechaar}^{(l,k)}(\localization{\tau},\localization{e})
		= \frac{\denv-1}{\dsys^{2}\denv^{2}-1} \hspace{-0cm} \left[1 + \sum_{s > 0}^{k-1}\left[\denv\frac{\dsys^{2}-1}{\dsys^{2}\denv^{2}-1}\right]^{s}\right] \hspace{-0.1cm}
		\quad \quad ,&~ \quad \quad
		\tau_{\ensemblechaar}^{(l,k)}(\localization{\tau},\localization[^{\prime}]{\tau})
		= \frac{1}{\dsys^{2}-1}\left[\denv\frac{\dsys^{2}-1}{\dsys^{2}\denv^{2}-1}\right]^{k}~\delta_{\tau\tau^{\prime}} \\[6pt]
		\chi_{\dsys}^{(l)}(\localization{\tau},\localization[^{\prime}]{\tau})
		=&~ \left(\dsys^{2}-1\right)~\delta_{\tau\tau^{\prime}}~.
	\end{align}

\end{proof}

\subsection{Spectral Properties and Hierarchies of Moment Operators}\label{subsec:spectral_properties_and_hierarchies_of_moment_operators}
In this section we derive spectral properties for the moment operators of the reference ensembles. These will allow us to derive a hierarchy between the Haar, cHaar, and Depolarize ensembles via their moment operator norms.\\

First, we investigate the leading eigenvectors of the cHaar moment operator.
\begin{theorem}[Leading Eigenvectors of cHaar Moment Operators]\label{theorem:app:leading_eigenvectors_of_chaar_moment_operators}
The $t$-th order $k$-concatenated cHaar moment operator has a unit leading eigenvalue $\lambda = 1$, a leading right-eigenvector $\kket{\psi}$ of a uniform combination of permutations, weighted by their characters, and a leading left-eigenvector $\kket{\varphi}$ reflecting trace-preservation. That is, the cHaar moment operator
\begin{align}
	\widehat{\Tau}_{\ensemblechaar}^{(t)k} = \frac{\kket{\psi}\bbra{\varphi}}{\bbra{\varphi}\kket{\psi}} ~+~ \cdots~,\quad \quad :&~ \quad \quad
	\kket{\psi} = \sum_{\sigma \in \permutations[t]}\chi_{\dsys\denv}^{(t)}(\sigma)^{}~\kket{\sigma}
	\quad \quad , \quad \quad
	\kket{\varphi} = \kket{e}~
\end{align}
has a leading-order, concatenation-invariant, non-unital, and trace preserving contribution.
\end{theorem}

\begin{proof}\label{proof:app:leading_eigenvectors_of_chaar_moment_operators}
To confirm that such vectors $\kket{\psi},\kket{\varphi}$ are eigenvectors of the cHaar moment operator, our goal is to show that,
	\begin{align}
		\widehat{\Tau}_{\ensemblechaar}^{(t)} \kket{\psi} = \kket{\psi}
		\quad\quad \textnormal{and} \quad\quad
		\bbra{\varphi}\widehat{\Tau}_{\ensemblechaar}^{(t)} = \bbra{\varphi}~.
	\end{align}
	Given the orthogonality of the characters and Weingarten functions, the invariance of the characters under inverses of permutations, $\chi_{\dsys}^{(t)}(\pi,\varsigma)=\chi_{\dsys}^{(t)}(\pi^{-1}\varsigma) = \chi_{\dsys}^{(t)}(\varsigma^{-1}\pi)$, and letting $\varsigma \to \pi^{-1}\varsigma$, we find,
	\begin{align}
		\widehat{\Tau}_{\ensemblechaar}^{(t)} \kket{\psi}
		=&~ \!\!\!\!\!\!\sum_{\sigma,\pi,\varsigma \in \permutations[t]}\!\!\!\! \chi_{\denv}^{(t)}(\sigma)~W_{\dsys\denv}^{(t)-1}(\sigma,\pi)~\chi_{\dsys}^{(t)}(\pi,\varsigma)~\chi_{\dsys\denv}^{(t)}(\varsigma) ~\kket{\sigma}
		= \!\!\!\!\!\!\sum_{\sigma,\pi,\varsigma \in \permutations[t]}\!\!\!\! \chi_{\denv}^{(t)}(\sigma)~W_{\dsys\denv}^{(t)-1}(\sigma,\pi)~\chi_{\dsys\denv}^{(t)}(\pi,\varsigma^{-1})~\chi_{\dsys}^{(t)}(\varsigma) ~\kket{\sigma}
		= \kket{\psi} \\
	\bbra{\varphi}\widehat{\Tau}_{\ensemblechaar}^{(t)}
		=&~ \sum_{\sigma,\pi \in \permutations[t]} \chi_{\dsys}^{(t)}(\sigma)~\chi_{\denv}^{(t)}(\sigma)~W_{\dsys\denv}^{(t)-1}(\sigma,\pi)~\bbra{\pi}
		= \sum_{\sigma,\pi \in \permutations[t]} \chi_{\dsys\denv}^{(t)}(\sigma)~W_{\dsys\denv}^{(t)-1}(\sigma,\pi)~\bbra{\pi}
		= \sum_{\pi \in \permutations[t]}\delta_{\pi e}~\bbra{\pi} = \bbra{\varphi} ~. \nonumber
	\end{align}
	Given the Jucys-Murphy character sums \cite{zinnjustin2009jucys}, $\sum_{\sigma \in \permutations[t]}\chi_{\dsys}^{(t)}(\sigma) = \tbinom{\dsys+t-1}{t}\frac{t!}{\dsys^{t}}$, the eigenvectors have overlaps of,
	\begin{align}
		\chi_{\dsys}^{(t)}(\varphi,\psi) = \chi_{\dsys}^{(t)}(\psi)
		=&~ \frac{1}{\dsys^{t}}\bbra{\varphi}\kket{\psi}
		= \sum_{\sigma \in \permutations[t]}\chi_{\dsys^{2}\denv}^{(t)}(\sigma)
		= \binom{\dsys^{2}\denv + t - 1}{t}\frac{t!}{\dsys^{2t}\denv^{t}}
		= 1 + \binom{t}{2}\frac{1}{\dsys^{2}\denv} + \mathcal{O}\!\left(\frac{1}{\dsys^{4}\denv^{2}}\right) ~.
	\end{align}
\end{proof}

In fact, we can bound all of the eigenvalues, the norm, and the trace of the Haar, cHaar, and Depolarize ensembles, by making use of the following lemma.

\begin{lemma}[Spectrum with respect to Localized Permutation Basis]\label{lemma:app:spectrum_with_respect_to_localized_permutation_basis}
	The spectrum of the cHaar moment operator in the linearly-independent, but not strictly orthogonal localized permutation basis is,
	\begin{align}
	\!\!\!\!
	\textnormal{Spectrum of}~\widehat{\Tau}_{\ensemblechaar}^{(t)k} =~ \textnormal{Spectrum of}~\left[\tilde{\tau}_{\ensemblechaar}^{(t,k)}(\sigma,\pi)\right]_{\sigma,\pi \in \permutations[t]}
	~~ : ~~
	\begin{array}{l}
	\tilde{\tau}_{\ensemblechaar}^{(t,k)}(\sigma,\pi) = \displaystyle\sum_{\varsigma \in \permutations[t]}\tau_{\ensemblechaar}^{(t,k)}(\localization{\sigma},\localization{\varsigma})~\chi_{\dsys}^{(t)}(\localization{\varsigma},\localization{\pi}) \\
	\tilde{\tau}_{\ensemblechaar}^{(t,k)\prime}(\sigma,\pi) = \displaystyle\sum_{\varsigma \in \permutations[t]}\chi_{\dsys}^{(t)}(\localization{\sigma},\localization{\varsigma})~\tau_{\ensemblechaar}^{(t,k)}(\localization{\varsigma},\localization{\pi})
	\end{array}
	~, \!\!
\end{align}
expressed as the spectrum of the modified transfer matrix $\tilde{\tau}_{\ensemblechaar}^{(t,k)}$ in the localized permutation basis, with bounded elements,
\begin{align}
	 \tilde{\tau}_{\ensemblechaar}^{(l,k)}(\sigma,\pi) ~\leq&~~ \frac{1}{\dsys^{\abs{\sigma}-\abs{\pi}}}\frac{1}{\denv^{\abs{\sigma} + (k-1)l^{\prime}/2}}~ \constant_{\tilde{\tau}}^{(l,l^{\prime},k)} ~ \delta_{\Gamma_{\sigma} \supseteq\Gamma_{\pi}} \\
	 \tilde{\tau}_{\ensemblechaar}^{(l,k)\prime}(\sigma,\pi) ~\leq&~~ \dsys^{\abs{\sigma}-\abs{\pi}}\frac{1}{\denv^{l/2 + (k-1)l^{\prime}/2}}~ \constant_{\tilde{\tau}^{\prime}}^{(l,l^{\prime},k)} ~ \delta_{\Gamma_{\sigma} \supseteq\Gamma_{\pi}}~,
\end{align}
given $l,k$-dependent constants $\constant_{\tilde{\tau}}^{(l,l^{\prime},k)} = l!^{2(k-1)}~l^{\prime}!~\weingarten[k]{l}~ \catalan[l]{l}^{3k + 4(k-1)}~\catalan[l^{\prime}]{l^{\prime}}^{4}$ ~and~ $\constant_{\tilde{\tau}^{\prime}}^{(l,l^{\prime},k)} = l!^{2(k-1)}~l!~\weingarten[k]{l}~ \catalan[l]{l}^{3k + 4(k-1)}~\catalan[l^{\prime}]{l}^{4}$.
\end{lemma}

\begin{proof}\label{proof:app:spectrum_with_respect_to_localized_permutation_basis}
Let $\widehat{\Tau} = \sum_{\alpha,\beta \in \mathcal{S}}\tau(\alpha,\beta)\kket{\alpha}\bbra{\beta}$ be a matrix, expressed in terms of a linearly independent basis $\mathcal{S} \ni \alpha,\beta$, with overlaps $\chi(\alpha,\beta) = \bbra{\alpha}\kket{\beta}$. Its eigenvalue equation, given an eigenvector $\kket{\nu} = \sum_{\alpha \in \mathcal{S}}\nu(\alpha)~\kket{\alpha}$, and eigenvalue $\lambda$, is,
\begin{align}
	\widehat{\Tau}\kket{\nu} = \lambda~\kket{\nu}
	~~ \to&~ ~~
	\sum_{\xi,\beta \in \mathcal{S}}\tau(\alpha,\xi)~\chi(\xi,\beta)~\nu(\beta) = \lambda~\nu(\alpha) ~~~\forall~\alpha
	~~ \to ~~
	\tilde{\tau}~\nu = \lambda ~\nu
	~~ : ~~ \tilde{\tau}(\alpha,\beta) = \sum_{\xi \in \mathcal{S}}\tau(\alpha,\xi)~\chi(\xi,\beta)~.
\end{align}
Therefore the eigenvectors and eigenvalues of $\widehat{\Tau}$ are equivalent to the eigenvectors of components $\{\nu(\alpha)\}_{\alpha \in \mathcal{S}}$ and eigenvalues $\lambda$ of the modified matrix $\tilde{\tau}$, with respect to the basis $\mathcal{S}$.\\

To bound the modified cHaar transfer matrix elements, we rely on the block-lower-triangular structure of the cHaar moment operator in this localized basis, The cHaar moment operator can be partitioned into blocks labeled by supports $\Gamma,\Gamma^{\prime}$, with associated localities $l= \abs{\Gamma}, l^{\prime} = \abs{\Gamma^{\prime}}$, and spanned by permutations $\sigma \in \mathcal{S}_{l}$ and $\pi \in \mathcal{S}_{l^{\prime}}$ such that $\Gamma_{\sigma} = \Gamma \supseteq \Gamma^{\prime} = \Gamma_{\pi}$. We also recall that the transfer matrix and overlaps for these blocks have bounds of,
\begin{align}
	\!\!\!\!\!\!\!\!\!\!\!\!
	\abs{\tau_{\ensemblechaar}^{(l,k)}(\localization{\sigma},\localization{\pi})}
		~\leq~ \frac{1}{\dsys^{\abs{\sigma} + \abs{\pi}}\denv^{\abs{\sigma} + (k-1)l/2}}~\constant_{\tau}^{(l,k)} ~\delta_{\Gamma_{\sigma} \supseteq \Gamma_{\pi}}
	\quad ,&~ \quad	 \abs{\chi_{\dsys}^{(l)}(\localization{\sigma},\localization{\pi})}
	~\leq~ \dsys^{\abs{\sigma} + \abs{\pi}}~\constant_{\chi}^{(l)}~\delta_{\Gamma_{\sigma}\Gamma_{\pi}} , \!\!\!\!\!\!\!\! \\
	\constant_{\tau}^{(l,k)} ~=~ l!^{2(k-1)}~\weingarten[k]{l}~\catalan[l]{l}^{3k + 4(k-1)}
	\quad ,&~ \quad
	\constant_{\chi}^{(l)} ~=~ \catalan[l]{l}^{4}~.
\end{align}
\noindent The modified transfer matrix elements, as sums over permutations $\varsigma \in \permutations[l]$ can thus be bounded by,
\begin{align}
	\!\!\!\!\!\!\!\!
	\tilde{\tau}_{\ensemblechaar}^{(l,k)}(\sigma,\pi)
	=&~ \!\!\!\!\!\!\!\!\!\! \sum_{\substack{\varsigma \in \permutations[l]\\\Gamma_{\sigma} = \Gamma \supseteq \Gamma^{\prime} = \Gamma_{\varsigma} = \Gamma_{\pi}}}\!\!\!\!\!\!\!\!\!\!\tau_{\ensemblechaar}^{(l,k)}(\localization{\sigma},\localization{\varsigma})~\chi_{\dsys}^{(l^{\prime})}(\localization{\varsigma},\localization{\pi})
	~\leq~ \!\!\!\!\!\!\!\!\!\! \max_{\substack{\sigma,\varsigma,\pi \in \permutations[l]\\\Gamma_{\sigma} = \Gamma \supseteq \Gamma^{\prime} = \Gamma_{\varsigma} = \Gamma_{\pi}}}\!\!\!\!\!\!\!\!\!\! !l^{\prime}~ \abs{\tau_{\ensemblechaar}^{(l,k)}(\localization{\sigma},\localization{\varsigma})}~\abs{\chi_{\dsys}^{(l^{\prime})}(\localization{\varsigma},\localization{\pi})} \\
	\!\!\!\!\!\!\!\!
	\leq&~ \frac{1}{\dsys^{\abs{\sigma}-\abs{\pi}}}\frac{1}{\denv^{\abs{\sigma} + (k-1)l^{\prime}/2}}~ \constant_{\tilde{\tau}}^{(l,l^{\prime},k)} ~ \delta_{\Gamma_{\sigma} \supseteq\Gamma_{\pi}}
	~\leq~
	 \constant_{\tilde{\tau}}^{(l,l^{\prime},k)} ~\delta_{\Gamma_{\sigma} \supseteq\Gamma_{\pi}}~\left\{\begin{array}{ll}
	\left(\frac{1}{\dsys\denv}\right)^{l/2} & \substack{l>0 \\ l^{\prime} = 0} \\
	\frac{1}{\denv^{k+(k+1)(l-l^{\prime})/2}}\left(\frac{\dsys}{\denv^{k}}\right)^{((l^{\prime}-2)-(l-l^{\prime}))/2} & \substack{l>0 \\ l^{\prime} > 0}
	\end{array} \right. \nonumber\\
	\constant_{\tilde{\tau}}^{(l,l^{\prime},k)} ~=&~ l^{\prime}!~\constant_{\tau}^{(l,k)}~\constant_{\chi}^{(l^{\prime})} = l!^{2(k-1)}~l^{\prime}!~\weingarten[k]{l}~ \catalan[l]{l}^{3k + 4(k-1)}~\catalan[l^{\prime}]{l^{\prime}}^{4} ~,
\end{align}
\begin{align}
	\!\!\!\!\!\!\!\!
	\tilde{\tau}_{\ensemblechaar}^{(l,k)\prime}(\sigma,\pi)
	=&~ \!\!\!\!\!\!\!\!\!\!\sum_{\substack{\varsigma \in \permutations[l]\\\Gamma_{\sigma} = \Gamma_{\varsigma} = \Gamma \supseteq \Gamma^{\prime} = \Gamma_{\pi}}}\!\!\!\!\!\!\!\!\!\! \chi_{\dsys}^{(l)}(\localization{\sigma},\localization{\varsigma})~\tau_{\ensemblechaar}^{(l,k)}(\localization{\varsigma},\localization{\pi})
	~\leq~ \!\!\!\!\!\!\!\!\!\! \max_{\substack{\sigma,\varsigma,\pi \in \permutations[l]\\\Gamma_{\sigma} = \Gamma_{\varsigma} = \Gamma \supseteq \Gamma^{\prime} = \Gamma_{\pi}}}\!\!\!\!\!\!\!\!\!\! !l~ \abs{\chi_{\dsys}^{(l)}(\localization{\sigma},\localization{\varsigma})}~\abs{\tau_{\ensemblechaar}^{(l,k)}(\localization{\varsigma},\localization{\pi})} \\
	\!\!\!\!\!\!\!\!
	\leq&~ \dsys^{\abs{\sigma}-\abs{\pi}}\frac{1}{\denv^{l/2 + (k-1)l^{\prime}/2}}~ \constant_{\tilde{\tau}^{\prime}}^{(l,l^{\prime},k)} ~ \delta_{\Gamma_{\sigma} \supseteq\Gamma_{\pi}}
	~\leq~
	 ~ \constant_{\tilde{\tau}^{\prime}}^{(l,l^{\prime},k)} ~\delta_{\Gamma_{\sigma} \supseteq\Gamma_{\pi}}~\left\{\begin{array}{ll}
	\denv^{l/2-1}\left(\frac{\dsys}{\denv}\right)^{l-1} & \substack{l>0 \\ l^{\prime} = 0} \\
	\frac{1}{\denv^{k+(1-2k)(l-l^{\prime})/2}}\left(\frac{\dsys}{\denv^{k}}\right)^{((l-2)+(l-l^{\prime}))/2} & \substack{l>0 \\ l^{\prime} > 0}
	\end{array} \right. \nonumber\\
	\constant_{\tilde{\tau}^{\prime}}^{(l,l^{\prime},k)} ~=&~ l!~\constant_{\tau}^{(l,k)}~\constant_{\chi}^{(l)} = l!^{2(k-1)}~l!~\weingarten[k]{l}~ \catalan[l]{l}^{3k + 4(k-1)}~\catalan[l^{\prime}]{l}^{4}~,
\end{align}
\noindent where the final bounds follow from considering scaling separately for each case of $l^{\prime}=0,l^{\prime}>0$, and from the fact that $l>0$-locality permutations $\sigma \in \permutations[l]$ have sizes bounded by $l/2 \leq \ceil{l/2} \leq \abs{\sigma} \leq l-1$.
\end{proof}

We now will make use of the derived bounds for the modified $k$-concatenated, $t$-order cHaar moment operator transfer matrix elements to bound its spectra, norm, and trace, and assess its overall spectral properties.

\begin{theorem}[Spectrum of cHaar Moment Operator]\label{theorem:app:spectrum_of_chaar_moment_operator}
	The $t$-th order cHaar moment operator has a single leading eigenvalue $\lambda = 1$, and in the asymptotic limit $\dsys\denv \to \infty$, for fixed $t$, all non-leading eigenvalues are upper bounded by,
	\begin{align}
		0 ~\leq~ \abs{\lambda} ~\leq&~~ \frac{1}{\denv^{l/2}}~\constant_{\lambda}^{(l)} ~<~ 1~,
	\end{align}
	and $\abs{\lambda} < 1$ for sufficiently large $\denv$, given $t$-dependent constants $\constant_{\lambda}^{(l)} = l!^{2}~\weingarten[]{l}~ \catalan[l]{l}^{7}$, and some $2 \leq l \leq t$.
\end{theorem}
\begin{proof}
Here, we use Gershgorin's circle theorem \cite{horn2012matrix}, in the non-orthogonal, block matrix case. Eigenvalues $\lambda$ of the $\Gamma$-support, $l\geq 2$-locality, $!l$-sized diagonal blocks of the cHaar moment operator, satisfy
\begin{align}
	\!\!\!\!
	\abs{\lambda} ~=~ \abs{\lambda^{k}}^{\frac{1}{k}}
	~\leq&~~ \left[!l~\max_{\substack{\sigma,\pi \in \permutations[l]\\\Gamma_{\sigma} = \Gamma_{\pi} = \Gamma}}
	\!\! \abs{\tilde{\tau}_{\ensemblechaar}^{(l,k)}(\sigma,\pi)}\right]^{\frac{1}{k}}
	~\leq~ \left[\frac{1}{\denv^{k}}\left(\frac{\dsys}{\denv^{k}}\right)^{\frac{l-2}{2}}\!\!\constant_{\lambda}^{(l,k)}\right]^{\frac{1}{k}} ~\overset{k \to \infty}{\leq}~\frac{1}{\denv^{l/2}}~\constant_{\lambda}^{(l)}\\
	\constant_{\lambda}^{(l,k)} ~=&~~ l!~ \constant_{\tilde{\tau}}^{(l,l,k)}
	~=~ l!^{2k}~\weingarten[k]{l}~ \catalan[l]{l}^{7k}
	~=~ \constant_{\lambda}^{(l)k} \quad\quad , \quad \quad \constant_{\lambda}^{(l)} ~=~ l!^{2}~\weingarten[]{l}~ \catalan[l]{l}^{7}~, \!\!
\end{align}
given eigenvalues of $k$-concatenations of a matrix are equal to $k$ powers of the eigenvalues, given the bounds on the modified $k$-concatenated transfer matrix elements $\tilde{\tau}_{\ensemblechaar}^{(l,k)}$, and given we can take the $k \to \infty$ limit for fixed $t,\dsys,\denv$.
\end{proof}

\begin{theorem}[Norm and Trace of Moment Operators]\label{theorem:app:norm_and_trace_of_moment_operators}
	The $k$-concatenated Haar, cHaar, and Depolarize moment operators have norms and traces, in the asymptotic limit $\dsys\denv \to \infty$ for fixed $t,k$, upper bounded by,
	\begin{align}
		\norm{\widehat{\Tau}_{\ensemblehaar}^{(t)k}}^{2}
		=&~ \Tr\!\left[\widehat{\Tau}_{\ensemblehaar}^{(t)k}\right]
		= t!
		\quad \quad , \quad \quad
		\norm{\widehat{\Tau}_{\ensembledep}^{(t)k}}^{2}
		= \Tr\!\left[\widehat{\Tau}_{\ensembledep}^{(t)k}\right]
		= 1 \\[6pt]
		\norm{\widehat{\Tau}_{\ensemblechaar}^{(t)k}}^{2}
		\leq&~ 1 ~+~ \frac{1}{\denv^{2}}~\constant_{\norm{\tau}}^{(t,k)}
		\quad \quad , \quad \quad
		\Tr\!\left[\widehat{\Tau}_{\ensemblechaar}^{(t)k}\right]
		\leq 1 ~+~\frac{1}{\denv^{k}}~ \constant_{\textnormal{Tr}[\tau]}^{(t,k)} ~,
	\end{align}
	given $t,k$-dependent constants $\constant_{\norm{\tau}}^{(t,k)} = \frac{t!^{2}-1}{t!^{2}}~\constant_{\lambda}^{(t)2k}$,~ $\constant_{\textnormal{Tr}[\tau]}^{(t,k)} = (t!^{2}-1)~\constant_{\lambda}^{(t)k}$, ~and~ $\constant_{\lambda}^{(t)} = t!^{2}~\weingarten[]{t}~ \catalan[t]{t}^{7}$.
\end{theorem}

\begin{proof}\label{proof:app:spectrum_norm_and_trace_of_chaar_moment_operator}
The exact norm and trace of the $k$-concatenated $t$-th order Haar and Depolarize moment operators are given by
\begin{align}
	\norm{\widehat{\Tau}_{\ensemblehaar}^{(t)k}}^{2}
	=&~ \norm{\widehat{\Tau}_{\ensemblehaar}^{(t)}}^{2}
	= \Tr\!\left[\widehat{\Tau}_{\ensemblehaar}^{(t)}\right]
	= \sum_{\sigma,\pi \in \permutations[t]} W_{\dsys}^{(t)-1}(\sigma,\pi)~\chi_{\dsys}^{(t)}(\pi,\sigma)
	= \abs{\permutations[t]} \\
	\norm{\widehat{\Tau}_{\ensembledep}^{(t)k}}^{2}
	=&~ \norm{\widehat{\Tau}_{\ensembledep}^{(t)}}^{2}
	= \Tr\!\left[\widehat{\Tau}_{\ensembledep}^{(t)}\right]
	= 1~.
\end{align}
The exact norm and trace of $t$-th order cHaar moment operator do not appear to simplify, however can be bounded as,
\begin{align}
\norm{\widehat{\Tau}_{\ensemblechaar}^{(t)k}}^{2}
	=&~ \!\!\!\!\!\!\sum_{\substack{\Gamma^{\prime} \subseteq \Gamma \subseteq [t]\\\abs{\Gamma}=l \geq l^{\prime}=\abs{\Gamma^{\prime}}}} \sum_{\substack{\sigma,\pi \in \permutations[l]\\\Gamma_{\sigma}=\Gamma \supseteq \Gamma^{\prime}=\Gamma_{\pi}}} \!\!\!\!\!\! \tilde{\tau}_{\ensemblechaar}^{(l,k)\prime}(\sigma,\pi)~\tilde{\tau}_{\ensemblechaar}^{(l,k)}(\sigma,\pi) \\
	\leq&~ 1 ~+~ (t!^{2}-1) \max_{\substack{l^{\prime} \leq l \leq t\\l \leq 2}} \frac{1}{\denv^{l + (k-1)l^{\prime}}}~ \constant_{\tilde{\tau}}^{(l,l^{\prime},k)}~\constant_{\tilde{\tau}^{\prime}}^{(l,l^{\prime},k)}
	~\leq~ 1 ~+~ \frac{1}{\denv^{2}}\constant_{\norm{\tau}}^{(t,k)}
	 \\[12pt]
	\Tr\!\left[\widehat{\Tau}_{\ensemblechaar}^{(t)k}\right]
	=&~ \sum_{s \in [t!]}\lambda_{s}^{k}
	~\leq~ 1 ~+~ (t!^{2}-1) \max_{\substack{2 \leq l \leq t}} \frac{1}{\denv^{kl/2}} ~\constant_{\lambda}^{(l)k}
	~\leq~ 1 ~+~ \frac{1}{\denv^{k}}\constant_{\textnormal{Tr}[\tau]}^{(t,k)}
	\\[12pt]
	\!\!\!\!
	\constant_{\norm{\tau}}^{(t,k)} = (t!^{2}-1)&~~\constant_{\tilde{\tau}}^{(t,t,k)2} = \frac{t!^{2}-1}{t!^{2}}~\constant_{\lambda}^{(t)2k}
	\quad , \quad
	\constant_{\textnormal{Tr}[\tau]}^{(t,k)} = (t!^{2}-1)~\constant_{\lambda}^{(t)k}
	\quad , \quad
	\constant_{\lambda}^{(t)} = t!^{2}~\weingarten[]{t}~ \catalan[t]{t}^{7} ~,
\end{align}
given the bounds on the modified $k$-concatenated transfer matrix elements $\tilde{\tau}_{\ensemblechaar}^{(l,k)},\tilde{\tau}_{\ensemblechaar}^{(l,k)\prime}$.
\end{proof}

For sufficiently large $\denv$, all non-leading cHaar moment operator eigenvalues are $\abs{\lambda} < 1$, which agrees with the fact that all channels have $\abs{\lambda} \leq 1$ \cite{chruscinski2022dynamical}. For $l=2$, we have exactly $\lambda = \denv(\dsys^{2}-1)/(\dsys^{2}\denv^{2}-1)$.\\

Next, we can find that all trace-preserving moment operators over ensembles $\ensemblechannel$ can be expressed as the Depolarize moment operator $\widehat{\Tau}_{\ensembledep}^{(t)}$, plus deviations $\widehat{\Delta}_{\ensemblechannel}^{(t)}$, and the cHaar $\ensemblechaar$ ensemble is invariant under the Haar $\ensemblehaar$ ensemble,
	\begin{align}
		\!\!\!\!\!\!
		\widehat{\Tau}_{\ensemblechannel}^{(t)} =&~ \widehat{\Tau}_{\ensembledep}^{(t)} ~+~ \widehat{\Delta}_{\ensemblechannel}^{(t)} \\
		\widehat{\Tau}_{\ensemblechaar}^{(t)} =&~ \idsys^{\otimes 2t} \otimes \bbra{\idsys^{\otimes t}} ~\widehat{\Tau}_{\ensembleunitary(\dsys\denv)}^{(t)} ~\idsys^{\otimes 2t} \otimes \kket{\nu_{\spaceenv}^{\otimes t}} \\[12pt]
		\!\!\!\!\!\!\!\!\!
		\widehat{\Tau}_{\ensembledep}^{(t)}\widehat{\Tau}_{\ensemblechannel}^{(t)} = \widehat{\Tau}_{\ensemblechannel}^{(t)\dagger}\widehat{\Tau}_{\ensembledep}^{(t)} = \widehat{\Tau}_{\ensembledep}^{(t)}
		 ~~, ~~
		\widehat{\Tau}_{\ensemblehaar}^{(t)}\widehat{\Tau}_{\ensemblechaar}^{(t)} \!= \widehat{\Tau}_{\ensemblechaar}^{(t)}&~	\widehat{\Tau}_{\ensemblehaar}^{(t)} = \widehat{\Tau}_{\ensemblechaar}^{(t)}
		 ~~ , ~~
		\widehat{\Tau}_{\ensemblehaar}^{(t)k} = \widehat{\Tau}_{\ensemblehaar}^{(t)}
		~~ , ~~
		\widehat{\Tau}_{\ensemblehaar}^{(t)\dagger} = \widehat{\Tau}_{\ensemblehaar}^{(t)} ~. \!\!\!\!\!
	\end{align}~\\
\noindent From these relationships, and the exact forms of the moment operators obtained above, the following relationships hold:
\begin{corollary}[cHaar Moment Operator Limits]\label{corol:app:chaar_moment_operator_limits}
	The $k$-concatenated $t$-th order cHaar moment operator satisfies,
	\begin{align}
		\lim_{\dsys\denv \to \dsys} \widehat{\Tau}_{\ensemblechaar}^{(t)k} = \widehat{\Tau}_{\ensemblehaar}^{(t)}
		\quad \quad ,&~ \quad \quad
		\lim_{\dsys\denv \to \infty} \widehat{\Tau}_{\ensemblechaar}^{(t)k} = \widehat{\Tau}_{\ensembledep}^{(t)} ~.
	\end{align}
\end{corollary}

\begin{figure}[ht]
	\centering
	\hspace{-0.7cm}
	\includegraphics[width=0.825\columnwidth]{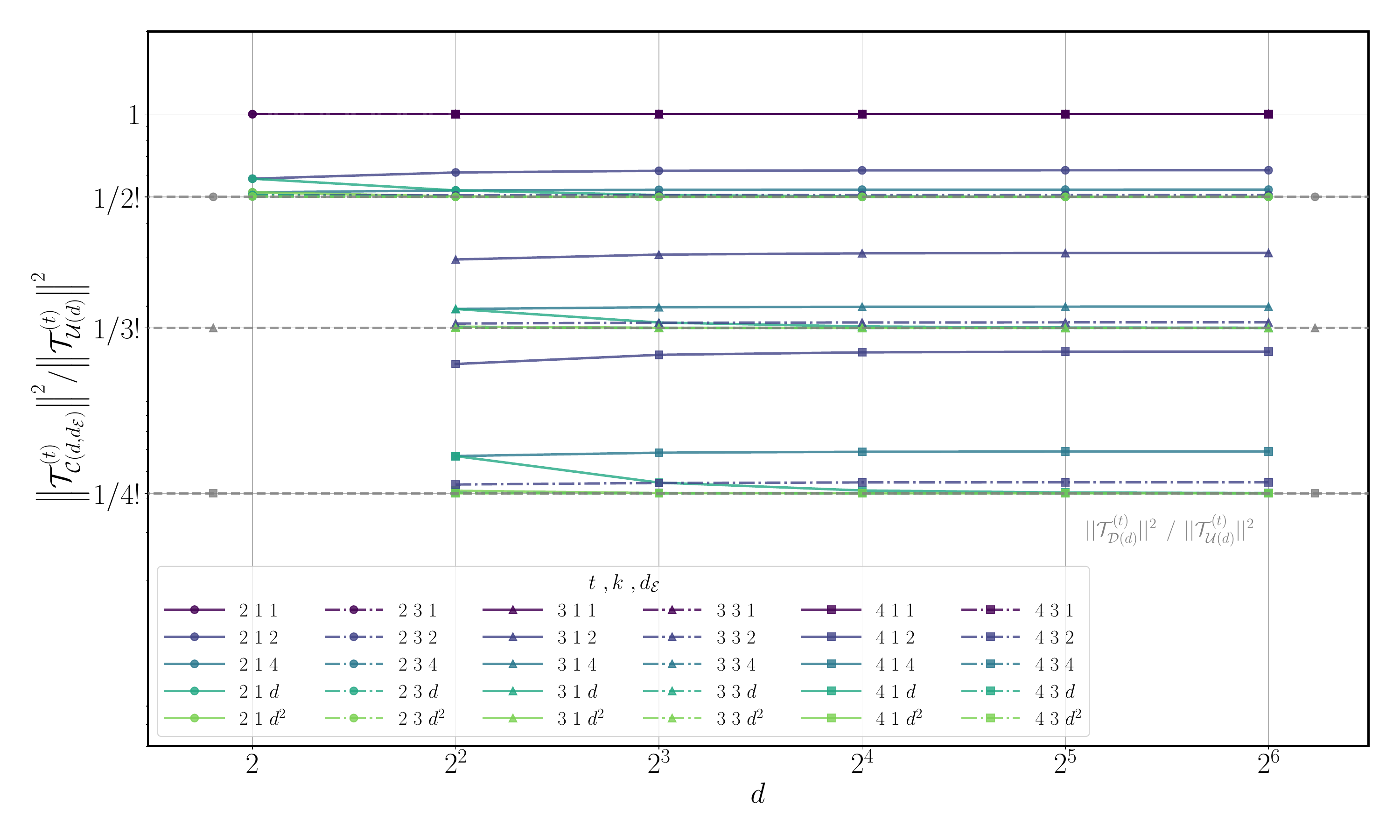}
	\vspace{-0.2cm}
	\caption{\textbf{cHaar moment operator norm scaling.} Norm of the $k$-concatenated $t$-th order cHaar moment operator $\tnorm{\widehat{\Tau}_{\ensemblechaar}^{(t,k)}}^{2}$ as a function of the system dimension $\dsys$, for environment dimensions $\denv$, concatenations $k$, and orders $t$. Norms are computed exactly using the derived closed form expressions for $\widehat{\Tau}_{\ensemblechaar}^{(t,k)}$, and symbolic methods \cite{cardin2026haarpy}. Norms converge to their leading-order, system dimension $\dsys$-independent term, and decay monotonically with environment dimension $\denv$ or with concatenation $k$, for all $t$, from $\tnorm{{\widehat{\Tau}_{\ensemblehaar}^{(t,k)}}}^{2} = t!$ ($1$ on scaled axis) to $\tnorm{\widehat{\Tau}_{\ensembledep}^{(t,k)}}^{2} = 1$ ($1/t!$ gray dashed lines on scaled axis). The depicted scalings confirm the bounded behavior, even for constant environment dimension $\denv < \dsys$, of the cHaar moment operator as it interpolates between unitary and depolarizing behavior. Further, such results suggest that the conjecture that $\tnorm{\widehat{\Tau}_{\ensemblechaar}^{(t,k)}}^{2} \leq \tnorm{\widehat{\Tau}_{\ensemblehaar}^{(t,k)}}^{2}$ holds absolutely.}
	\label{fig:moment_operator_norm}
\end{figure}

\begin{corollary}[cHaar Moment Operator Converges to Depolarizing and Non-Unital Channel]\label{conjecture:app:chaar_moment_operator_converge_to_depolarizing_and_non_unital_channel}
	The $k$-concatenated $t$-th order cHaar moment operator is inherently depolarizing, with next-leading-order concatenation-independent non-unital behavior, and concatenation-dependent unital terms. For all dimensions, given traceless operators $T^{(t,k)}_{\dsys},T^{(t,k)\prime}_{\dsys} \in \mathcal{B}[\spacesys^{\otimes t}]$,
	\begin{align}
		\lim_{\substack{\dsys\denv \to \infty}} \widehat{\Tau}_{\ensemblechaar}^{(t)k} =&~ \frac{1}{\dsys^{t}} \kket{\idsys^{\otimes t}}\bbra{\idsys^{\otimes t}}
		~+~ \frac{1}{\dsys^{t}} \mathcal{O}\!\left(\frac{1}{\dsys\denv}\right) ~ \kket{T^{(t,k)}_{\dsys}}\bbra{\idsys^{\otimes t}}
		~+~ \frac{1}{\dsys^{t}} \mathcal{O}\!\left(\frac{1}{\dsys^{2}\denv^{k}}\right)\kket{T^{(t,k)}_{\dsys}}\bbra{T^{(t,k)\prime}_{\dsys}}~.
	\end{align}
\end{corollary}
\noindent Indeed, in the localized permutation basis, the projector Haar moment operators have an invariant block-diagonal structure, and the non-projector cHaar moment operators have a block-lower-triangular structure, whose elements decay with dimension and concatenations. The average behavior of random quantum channels is thus a useful pedagogical tool, given it interpolates with dimension, between trivial environments and unitarity, and large dimensions and depolarization. In Fig. \ref{fig:moment_operator_norm}, we numerically illustrate this bounded behavior for moment operator norms, for different $t$, $k$, and $\denv$.

The previous results indicate that as we concatenate cHaar random channels $k$ times, the resulting composite transfer matrix elements are suppressed, which intuitively means the moment operator norm should also decrease until it converges to the norm of a single projector with a unit eigenvalue. In the large $\dsys\denv \to \infty$ limit, this intuition is verified, but given that the cHaar moment operator has no closed form, we instead postulate that the following conjecture should hold.

\begin{conjecture}[Hierarchy of Haar, cHaar, Depolarize Ensembles]\label{conjecture:app:hierarchy_of_haar_chaar_depolarize_ensembles}
	Given there exists an exact hierarchy between the $k$-concatenated, $t$-th order Haar, cHaar, and Depolarize moment operators in the asymptotic limit $\dsys\denv \to \infty$,
	\begin{align}
		1 = \norm{\widehat{\Tau}_{\ensembledep}^{(t)}}^{2} \leq \norm{\widehat{\Tau}_{\ensemblechaar}^{(t)k}}^{2} \leq \norm{\widehat{\Tau}_{\ensemblehaar}^{(t)}}^{2} = t!~,
	\end{align}
	it is conjectured that such a hierarchy holds for all concatenations $k$, orders $t$, and dimensions $\dsys,\denv$. Further, it is conjectured that the cHaar norm is monotonically non-increasing with environment dimension and concatenations,
	\begin{align}
		\norm{\widehat{\Tau}_{\ensemblechannel(\dsys,\denv^{\prime})}^{(t)k}}^{2}
		\overset{?}{\geq}
		\norm{\widehat{\Tau}_{\ensemblechaar}^{(t)k}}^{2}
		\quad \quad ~\forall ~ \denv^{\prime} \geq \denv
		\quad \quad ,&~ \quad \quad
		\norm{\widehat{\Tau}_{\ensemblechaar}^{(t)k^{\prime}}}^{2}
		\overset{?}{\leq}
		\norm{\widehat{\Tau}_{\ensemblechaar}^{(t)k}}^{2}
		\quad \quad ~\forall ~ k^{\prime} \geq k
		~.
	\end{align}
\end{conjecture}
Given the relationship between trace-preserving moment operators and the invariant Depolarize moment operator,
\begin{align}
	\norm{\widehat{\Tau}_{\ensemblechannel}^{(t)} - \widehat{\Tau}_{\ensembledep}^{(t)}}^{2} = \norm{\widehat{\Tau}_{\ensemblechannel\vphantom{\ensemblehaar}{}}^{(t)}}^{2} ~-~ \norm{\widehat{\Tau}_{\ensembledep}^{(t)}}^{2} \geq 0
	\quad\quad \to \quad\quad
	\norm{\widehat{\Tau}_{\ensemblechannel\vphantom{\ensemblehaar}{}}^{(t)}}^{2} \geq \norm{\widehat{\Tau}_{\ensembledep}^{(t)}}^{2}~,
\end{align}
and given the partial invariance of the Haar and cHaar ensembles, the conjecture is true if
\begin{align}
	\norm{\widehat{\Tau}_{\ensemblechaar}^{(t)k} - \widehat{\Tau}_{\ensemblehaar}^{(t)}}^{2}
	\geq 0
	\quad \to&~ \quad
	\norm{\widehat{\Tau}_{\ensemblehaar}^{(t)}}^{2} - \norm{\widehat{\Tau}_{\ensemblechaar}^{(t)k}}^{2} \geq 2\left(\Tr\!\left[\widehat{\Tau}_{\ensemblechaar}^{(t)k}\right] - \norm{\widehat{\Tau}_{\ensemblechaar}^{(t)k}}^{2}\right) \overset{?}{\geq} 0 ~.
\end{align}
This conjecture, and this bound on the difference of Haar and cHaar moment operator norms, holds in the asymptotic limit $\dsys\denv \to \infty$, where both norms converge to $1$. However, our analysis is limited in part by the lack of closed forms for the Weingarten functions, and from our approaches of (possibly loosely) bounding quantities in terms of maximum cHaar transfer matrix elements. Different, possibly representation-theoretic based approaches are necessary to prove this non-trivial, but important hierarchy conjecture for general $\dsys,\denv$.

\section{Moment Operators of Noisy Ensembles}\label{app:moment_operators_of_noisy_ensembles}
In these appendices, we calculate moment operators for ensembles of noisy unitary channels, and investigate their scaling with respect to concatenations, dimension, and noise parameters.

\subsection{Composite Noise Channels}\label{subsec:composite_noise_channels}
Here, we seek to calculate moment operators for composite ensembles of channels $\channelnoise \circ U$, where the unitary component is sampled as $U \sim \ensembleunitary$, and where the noise component is a fixed noise channel $\channelnoise$. The $t$-th order composite moment operator, $\widehat{\Tau}_{\ensembleunitary\ensemblenoise}^{(t)} = \channelnoise \circ \widehat{\Tau}_{\ensembleunitary}^{(t)}$, then may be concatenated $k$ times,
\begin{align}
	\widehat{\Tau}_{\ensembleunitary\ensemblenoise}^{(t)k}
	=&~ \prod_{i \in [k]}\widehat{\channelnoise^{\otimes t}}\circ\widehat{\Tau}_{\ensembleunitary}^{(t)}
	~=~ \channeldepolarize[\widehat]{\dsys}^{\otimes t} ~+~ \widehat{\Delta}_{\ensembleunitary\ensemblenoise}^{(t,k)}
	\end{align}
with trace-preserving depolarization $\channeldepolarize[\widehat]{\dsys}^{\otimes t}$ and extraneous $\widehat{\Delta}_{\ensembleunitary\ensemblenoise}^{(t,k)}$ terms of
\begin{equation}
	\channeldepolarize[\widehat]{\dsys}^{\otimes t} = \displaystyle{\frac{1}{\dsys^{t}}\kket{\idsys^{\otimes t}}\bbra{\idsys^{\otimes t}}}
	\quad\textnormal{and}\quad
	\widehat{\Delta}_{\ensembleunitary\ensemblenoise}^{(t,k)} = \displaystyle{\frac{1}{\dsys^{t}} \sum_{\substack{P \in \mathcal{S}_{\ensemblenoise}^{(t)} \backslash \{\idsys^{\otimes t}\},~S \in \mathcal{S}_{\ensembleunitary}^{(t)}}} \tau_{\ensembleunitary\ensemblenoise}^{(t,k)}(P,S)~ \kket{P}\bbra{S}}~.
\end{equation}
The concatenated composite transfer matrices $\tau_{\ensembleunitary\ensemblenoise}^{(t,k)}$ can be expressed in terms of the component transfer matrices $\tau_{\ensemblenoise}^{(t)},\tau_{\ensembleunitary}^{(t)}$, with respective bases $\mathcal{S}_{\ensembleunitary}^{(t)},\mathcal{S}_{\ensemblenoise}^{(t)}$ for each component
\begin{align}
\tau_{\ensembleunitary\ensemblenoise}^{(t)}(P,S) =&~ \hspace{-0.6cm} \sum_{\substack{Q \in \mathcal{S}_{\ensemblenoise}^{(t)},~T \in \mathcal{S}_{\ensembleunitary}^{(t)}}} \hspace{-0.6cm} \tau_{\ensemblenoise}^{(t)}(P,Q)~\chi_{\dsys}^{(t)}(Q,T) ~\tau_{\ensembleunitary}^{(t)}(T,S)
	\quad \to \quad
	\tau_{\ensembleunitary\ensemblenoise}^{(t,k)}(P,S) = \hspace{-0.6cm} \sum_{\substack{Q \in \mathcal{S}_{\ensembleunitary}^{(t)},~ T \in \mathcal{S}_{\ensemblenoise}^{(t)}}} \hspace{-0.6cm} \tau_{\ensembleunitary\ensemblenoise}^{(t)}(P,Q)~\chi_{\dsys}^{(t)}(Q,T)~ \tau_{\ensembleunitary\ensemblenoise}^{(t,k-1)}(T,S)~.\!\!\!
\end{align}
The most appropriate basis for each component, and whether $\mathcal{S}_{\ensembleunitary}^{(t)} \neq \mathcal{S}_{\ensemblenoise}^{(t)}$, greatly affects their interplay, and the ease of analysis of the scaling of moment operator norms $\norm{\widehat{\Delta}_{\ensembleunitary\ensemblenoise}^{(t,k)}} = \norm{\widehat{\Tau}_{\ensembleunitary\ensemblenoise}^{(t)k} - \channeldepolarize[\widehat]{\dsys}^{\otimes t}}$, with variables $t,k,\dsys$.

\subsection{Unital and Non-unital Noise and Haar Random Unitary Channels}\label{subsec:unital_and_non_unital_noise_and_haar_random_unitary_channels}

Here, we consider random unitary channels from any Haar distributed unitary ensemble with group structure $\ensembleunitary$, followed by fixed diagonal unital or non-unital noise $\channelnoise = \channelnoise_{\gamma},\channelnoise_{\gamma\eta}$, and derive the scaling of their $k$-concatenated $t$-th order composite moment operators with $t,k,\dsys,\gamma,\eta$.\\

\noindent For the unitary component, $t$-th order moment operators over $\ensembleunitary$ are, in an orthogonal basis $\mathcal{S}_{\ensembleunitary}^{(t)} = \mathcal{P}_{\dsys}^{t}$,
\begin{align}
		\widehat{\Tau}_{\ensembleunitary}^{(t)}
		=&~ \int_{\ensembleunitary}dU~ U^{\otimes t} \otimes U^{\otimes t ~\! *}
		= \frac{1}{\dsys^{t}}\sum_{\substack{P,S \in \mathcal{P}_{\dsys}^{t}}}~ \tau_{\ensembleunitary}^{(t)}(P,S) ~ \kket{P}\bbra{S}
		\quad : \quad
		\tau_{\ensembleunitary}^{(t)}(P,S) ~\propto~ \delta_{\Gamma_{P}\Gamma_{S}}~.
\end{align}
This expansion in a strictly orthogonal basis can be derived, under the assumption the ensemble $\ensembleunitary$, being a Haar distributed group, has projector moment operators that are support-preserving. The orthogonal basis transfer matrix coefficients $\tau_{\ensembleunitary}^{(t)}(P,S)$ thus also have support-dependent and concatenation-invariant block structure,
\begin{align}
	\sum_{Q \in \mathcal{S}_{\ensembleunitary}^{(t)}} \tau_{\ensembleunitary}^{(t)}(P,Q)~\tau_{\ensembleunitary}^{(t)}(Q,S) ~=~ \tau_{\ensembleunitary}^{(t)}(P,S)  ~\propto~ \delta_{\Gamma_{P}\Gamma_{S}} ~.
\end{align}

\noindent For the noise component, the noise model $\channelnoise = \channelnoise_{\gamma\eta}$ moment operator has the special form of diagonal unital noise and non-unital noise components $\gamma_{\dsys}(P) \geq \gamma~,~ \eta_{\dsys}(P) \leq \eta$ for $P \in \mathcal{S}_{\ensemblenoise_{\gamma\eta}}$, in an orthogonal basis $\mathcal{S}_{\ensemblenoise_{\gamma\eta}} = \mathcal{P}_{\dsys}$,
\begin{align}
	\widehat{\channelnoise}_{\gamma\eta} =&~
	\frac{1}{\dsys}\left[\begin{array}{cccc}
		1 & 0 & \cdots & 0 \\
		\eta_{\dsys}(S) & 1-\gamma_{\dsys}(S) & \cdots & 0 \\
		\vdots & \vdots & \ddots & \vdots \\
		\eta_{\dsys}(P) & 0 & \cdots & 1-\gamma_{\dsys}(P) \\
	\end{array}\right]_{P,S \in \mathcal{P}_{\dsys} \backslash \{I_{\dsys}\}}
	\!\! \to \quad
	\widehat{\channelnoise^{\otimes t}_{\gamma\eta}} = \frac{1}{\dsys^{t}}\sum_{P,S \in \mathcal{S}_{\ensemblenoise_{\gamma\eta}}^{(t)}} \tau_{\ensemblenoise_{\gamma\eta}}^{(t)}(P,S) ~ \kket{P}\bbra{S}~,
\end{align}
and appropriate additional constraints on $\gamma_{\dsys},\eta_{\dsys}$ such that $\channelnoise$ is completely positive \cite{greenbaum2015introduction}.

The $t$-th order noise component transfer matrix $\tau_{\ensemblenoise_{\gamma\eta}}^{(t)}$, for basis operators $P,S \in \mathcal{S}_{\ensemblenoise_{\gamma\eta}}^{(t)} = \mathcal{P}_{\dsys}^{t}$, is
\begin{align}
	\tau_{\ensemblenoise_{\gamma\eta}}^{(t)}(P,S)
	=&~ \prod_{i \in \Gamma_{S}} (1-\gamma_{\dsys}(P_{i})) \prod_{i \in \Gamma_{P} \backslash \Gamma_{S}} \!\!\!\! \eta_{\dsys}(P_{i})~ \delta_{P_{\Gamma_{S}}S_{\Gamma_{S}}}
	~\in~ \mathcal{O}\!\left((1 - \gamma)^{\abs{S}}~\eta^{\abs{P}-\abs{S}}\right)~\delta_{P_{\Gamma_{S}}S_{\Gamma_{S}}} ~.
\end{align}
We restrict to this special form (a diagonal component arising from Pauli-like-noise, and a dense first column arising from a non-unital component) as we seek to understand exact theoretical properties. More general noise models with other non-diagonal elements, even if initially sparse, become dense and intractable to analyze upon concatenations.\\

\noindent Finally, let us assess the leading-order behavior of the composite and concatenated transfer matrices,
\begin{align}
	\widehat{\Tau}_{\ensembleunitary\ensemblenoise_{\gamma\eta}}^{(t)}
	= \widehat{\channelnoise^{\otimes t}_{\gamma\eta}} \circ \widehat{\Tau}_{\ensembleunitary}^{(t)}
	= \frac{1}{\dsys^{t}}\sum_{\substack{P \in \mathcal{S}_{\ensemblenoise_{\gamma\eta}}^{(t)}\\S \in \mathcal{S}_{\ensembleunitary}^{(t)}}} \tau_{\ensembleunitary\ensemblenoise_{\gamma\eta}}^{(t)}(P,S) ~ \kket{P}\bbra{S}
	\quad \to&~ \quad
	\widehat{\Tau}_{\ensembleunitary\ensemblenoise_{\gamma\eta}}^{(t)k}
	= \frac{1}{\dsys^{t}}\sum_{\substack{P \in \mathcal{S}_{\ensemblenoise_{\gamma\eta}}^{(t)}\\S \in \mathcal{S}_{\ensembleunitary}^{(t)}}} \tau_{\ensembleunitary\ensemblenoise_{\gamma\eta}}^{(t,k)}(P,S) ~ \kket{P}\bbra{S}~.
\end{align}
In particular, given the forms of the individual component transfer matrices, we seek to understand the competition that arises between the unitary components preserving locality, and non-unital noise components increasing locality.

\begin{lemma}[Unital and Non-unital Noise and Haar Random Unitary Channel Moment Operators]\label{lemma:app:unital_and_non_unital_noise_and_haar_random_unitary_channel_moment_operators}
	Composite ensembles, consisting of a random unitary component $\ensembleunitary$ from a Haar distributed unitary group, with a basis $S \in \mathcal{S}_{\ensembleunitary}^{(t)}=\mathcal{P}_{\dsys}^{t}$, and a noise component $\ensemblenoise_{\gamma\eta}$ with a basis $P \in \mathcal{S}_{\ensemblenoise_{\gamma\eta}}^{(t)} = \mathcal{P}_{\dsys}^{t}$, and with diagonal unital, and general non-unital noise coefficients, $\gamma_{\dsys}(P) \geq \gamma~,~\eta_{\dsys}(P) \leq \eta$ for $P \in \mathcal{P}_{\dsys}^{t} \backslash \{\idsys^{\otimes t}\}$, have $k$-concatenated $t$-th order composite transfer matrices of,
	\begin{align}
		\tau_{\ensembleunitary\ensemblenoise_{\gamma\eta}}^{(t,k)}(P,S)
		=&~ \mathcal{O}\!\left((1-\gamma)^{k\abs{S}}\eta^{\abs{P}-\abs{S}}\right)~\tau_{\ensembleunitary}^{(t)}(P_{\Gamma_{S}},S)~\delta_{\Gamma_{P} \supseteq \Gamma_{S}} ~,
	\end{align}
	where the leading-order $\mathcal{O}(\gamma,\eta)$ behavior is strictly due to the unspecified noise component in the $\gamma,\eta \to 0$ limit, and uniformly scales the unitary coefficients by their locality, independent of the specific unitary ensemble itself.
\end{lemma}

\begin{proof}\label{proof:app:unital_and_non_unital_noise_and_haar_random_unitary_channel_moment_operators}
	To begin, we note that the unitary component $\ensembleunitary$ is support-preserving, whereas the non-unital noise component $\ensemblenoise_{\gamma\eta}$ is support-non-decreasing, with individual component transfer matrices, for $P,S \in \mathcal{P}_{\dsys}^{t} = \mathcal{S}_{\ensembleunitary}^{(t)} = \mathcal{S}_{\ensemblenoise_{\gamma\eta}}^{(t)}$, of
	\begin{align}
		\!\!\!\! \!\!\!
		\tau_{\ensembleunitary\ensemblenoise_{\gamma\eta}}^{(t)}(P,S)
		=&~ \!\!\!\prod\limits_{i \in \Gamma_{S}}(1-\gamma_{\dsys}(P_{i})) \!\!\!\!\!\prod\limits_{i \in \Gamma_{P} \backslash \Gamma_{S}}\!\!\!\!\eta_{\dsys}(P_{i}) ~\tau_{\ensembleunitary}^{(t)}(P_{\Gamma_{S}},S) ~\delta_{\Gamma_{P} \supseteq \Gamma_{S}}
		\in \mathcal{O}\!\left((1-\gamma)^{\abs{S}}\eta^{\abs{P}-\abs{S}}\right)\tau_{\ensembleunitary}^{(t)}(P_{\Gamma_{S}},S) ~\delta_{\Gamma_{P} \supseteq \Gamma_{S}}~\!. \!\!\!\!\!\!\!
	\end{align}
	From locality preservation of the unitary component, diagonal unital noise scaling always occurs, for all $\Gamma_{P} \supseteq \Gamma_{S}$, whereas non-unital noise scaling only occurs when strictly $\Gamma_{P} \supset \Gamma_{S}$, and concatenated operators are constrained to be support-increasing $\Gamma_{P} \supseteq \cdots \supseteq \Gamma_{T} \supseteq \cdots \supseteq \Gamma_{S}$. For instance, the $k=2$ transfer matrix elements are
	\begin{align}
		\tau_{\ensembleunitary\ensemblenoise_{\gamma\eta}}^{(t,2)}(P,S)
		=&~ \sum_{\substack{T \in \mathcal{P}_{\dsys}^{t} \\ \Gamma_{S} \subseteq \Gamma_{T} \subseteq \Gamma_{P}}} \tau_{\ensembleunitary\ensemblenoise_{\gamma\eta}}^{(t)}(P,T)~\tau_{\ensembleunitary\ensemblenoise_{\gamma\eta}}^{(t)}(T,S)\\
		=&~ \sum_{\substack{T \in \mathcal{P}_{\dsys}^{t}\\ \Gamma_{S} \subseteq \Gamma_{T} \subseteq \Gamma_{P}}} \prod_{i \in \Gamma_{T}}(1-\gamma_{\dsys}(P_{i}))\!\!\!\!\prod_{i \in \Gamma_{P} \backslash \Gamma_{T}}\!\!\!\!\!\eta_{\dsys}(P_{i})\prod_{i \in \Gamma_{S}}(1-\gamma_{\dsys}(T_{i}))\!\!\!\!\prod_{i \in \Gamma_{T} \backslash \Gamma_{S}}\!\!\!\!\!\eta_{\dsys}(T_{i}) ~\tau_{\ensembleunitary}^{(t)}(P_{\Gamma_{T}},T)~\tau_{\ensembleunitary}^{(t)}(T_{\Gamma_{S}},S) \\
		=&~ \mathcal{O}\!\left((1-\gamma)^{2\abs{S}}~\eta^{\abs{P}-\abs{S}}\right)~\left[\sum_{\substack{T \in \mathcal{P}_{\dsys}^{t}\\ \Gamma_{T} = \Gamma_{S} \subseteq \Gamma_{P}}} \tau_{\ensembleunitary}^{(t)}(P_{\Gamma_{S}},T)~\tau_{\ensembleunitary}^{(t)}(T,S) ~+~ \mathcal{O}\!\left((1-\gamma)^{\vphantom{\abs{P}}{}}\right) ~\right] \\
		=&~ \mathcal{O}\!\left((1-\gamma)^{2\abs{S}}~\eta^{\abs{P}-\abs{S}}\right)~\tau_{\ensembleunitary}^{(t)}(P_{\Gamma_{S}},S) ~\delta_{\Gamma_{P}\supseteq \Gamma_{S}} ~,
	\end{align}
	and $k$-concatenated composite transfer matrices are thus by induction,
	\begin{align}
	\tau_{\ensembleunitary\ensemblenoise_{\gamma\eta}}^{(t,k)}(P,S)
		=&~ \mathcal{O}\!\left((1-\gamma)^{k\abs{S}}~\eta^{\abs{P}-\abs{S}}\right)~\tau_{\ensembleunitary}^{(t)}(P_{\Gamma_{S}},S) ~\delta_{\Gamma_{P}\supseteq \Gamma_{S}} ~.
	\end{align}
\end{proof}
\noindent Therefore, the leading-order noise scaling is dictated by the concatenation-invariant unitary components being scaled by strictly unital noise until the last concatenation, at which point non-unital noise increases the support from $\Gamma_{S} \to \Gamma_{P}$. Intermediate non-unital noise at concatenations $0<s<k$ also increases the support $\Gamma_{S} \to \Gamma_{T} \to \Gamma_{P}$, however such intermediate noise contributes next-leading-order unital noise scaling $\mathcal{O}((1-\gamma)^{(k-s)\abs{T}})$. Intermediate non-unital noise, in the case it maps unitary component basis operators to larger-support unitary component basis operators, also contributes dimension-dependent scaling when such concatenations occur between unitary components of differing support.\\
\noindent We now assess the leading-order scaling of moment operator norms for concatenations of composite ensembles.

\begin{theorem}[Unital and Non-unital Noise and Haar Random Unitary Channel Moment Operator Norms]\label{theorem:app:unital_and_non_unital_noise_and_haar_random_unitary_channel_moment_operator_norms}
$k$-concatenations of $t$-th order ensembles of Haar random unitaries, with commutants with locality-$\geq l$ operators, and rank-$r$ moment operators, and diagonal unital and non-unital noise, with noise coefficients $\gamma_{\dsys}(P) \geq \gamma~,~ \eta_{\dsys}(P) \leq \eta $ for $P \in \mathcal{P}_{\dsys} \backslash \{\idsys\}$, have moment operator norms relative to the Depolarize ensemble of,
\begin{align}
	\norm{\widehat{\Tau}_{\ensembleunitary\ensemblenoise_{\gamma}}^{(t)k} - \widehat{\Tau}_{\ensembledep}^{(t)}}^{}_{1}
	= (r-1)~\mathcal{O}\!\left((1-\gamma)^{lk}\right)
	\quad \quad , \quad \quad
	\norm{\widehat{\Tau}_{\ensembleunitary\ensemblenoise_{\gamma\eta}}^{(t)k} - \widehat{\Tau}_{\ensembledep}^{(t)}}^{2}_{2}
	= t~\abs{\mathcal{P}_{\dsys}\backslash \{\idsys\}}~\mathcal{O}\!\left(\eta^{2}\right)
	~.
\end{align}
\end{theorem}
\noindent From the previous theorem it follows that as $k \to \infty$, unital noise decreases moment operator norms for $\gamma > 0$, and approaches a Depolarize $t$-design, whereas non-unital noise increases moment operator norms for $\eta > 0$, and never approaches a Depolarize $t$-design.

\begin{proof}\label{proof:app:unital_and_non_unital_noise_and_haar_random_unitary_channel_moment_operator_norms}
	Moment operator $p$-norms follow directly from their expansion in an orthogonal basis. Composite unital noise and unitary component moment operators may be decomposed into support $\Gamma \subseteq [t]$-dependent unitary component projectors $\widehat{\Tau}_{\ensembleunitary}^{(\Gamma)} = (1/\dsys^{t})\sum^{\substack{P,S \in \mathcal{S}_{\ensembleunitary}^{(t)}}}_{\substack{\Gamma_{P} = \Gamma_{S} = \Gamma}} \tau_{\ensembleunitary}^{(t)}(P,S) ~\kket{P}\bbra{S}$, scaled by unital noise $\gamma_{\dsys} \geq \gamma$, with deviations from the Depolarize term $\channeldepolarize{\dsys}^{\otimes t} = (1/\dsys^{t})\kket{\idsys^{\otimes t}}\bbra{\idsys^{\otimes t}}$, scaled by non-unital noise $\eta_{\dsys} \leq \eta$,
	\begin{align}
		\!\!\!
		\widehat{\Tau}_{\ensembleunitary\ensemblenoise_{\gamma}}^{(t)k}
		~=~ \widehat{\Tau}_{\ensembledep}^{(t)} ~+~ \!\!\!\!\sum_{\substack{\Gamma \subseteq[t],s=\abs{\Gamma}\geq l}} \!\!\! \mathcal{O}\!\left((1-\gamma)^{sk}\right)~\widehat{\Tau}_{\ensembleunitary}^{(\Gamma)}
		\quad\quad ,&~ \quad\quad
		\widehat{\Tau}_{\ensembleunitary\ensemblenoise_{\eta}}^{(t)k}
		~=~ \widehat{\Tau}_{\ensembledep}^{(t)} ~+~ \!\!\!\!\sum_{\substack{\Gamma \subseteq[t],s=\abs{\Gamma}\geq l\\P \in \mathcal{P}_{\dsys},\Gamma_{P}=\Gamma}} \!\!\! \mathcal{O}\!\left(\eta^{s}\right)~\kket{P}\bbra{\idsys^{\otimes t}} ~+~ \cdots \!\!\!\!\!\!\!\!\!\!\!\!\!\!\!\!\!\!
		\\
		\!\!\!\!\!\!\!\!\!\!\!\!\!\!\!\!\!
		\norm{\widehat{\Tau}_{\ensembleunitary\ensemblenoise_{\gamma}}^{(t)k}}^{p}_{p}
		~=~ 1 ~+~ \sum_{l \leq s \leq t}\binom{t}{s}~r_{s}~\mathcal{O}\!\left((1-\gamma)^{psk}\right)
		\quad\quad ,&~ \quad\quad
		\norm{\widehat{\Tau}_{\ensembleunitary\ensemblenoise_{\eta}}^{(t)k}}^{2}_{2}
		~=~ 1 ~+~ \sum_{l \leq s \leq t}~\binom{t}{s}~\abs{\mathcal{P}_{\dsys} \backslash \{\idsys\}}^{s}~\mathcal{O}\!\left(\eta^{2s}\right)~.
	\end{align}
	There are $t\abs{\mathcal{P}_{\dsys}\backslash \{\idsys\}}$ leading-order $1$-locality non-unital terms, and $\tbinom{t}{l}$ leading-order $l$-locality unital terms. For example, Haar ensemble moment operator projectors have rank $r_{l}=!l$, and $r = t!$.
\end{proof}

In fact, we can also obtain bounds on norms of generic composite moment operators,
\begin{align}
	\widehat{\Tau}_{\ensembleunitary\ensemblenoise_{\gamma\eta}}^{(t)} \to&~ \widehat{\Tau}_{\ensembleunitary\ensemblenoise_{\gamma\eta}}^{(t)} = \widehat{\channelnoise}^{(t)}_{\gamma\eta}\widehat{\Tau}_{\ensembleunitary}^{(t)}
	= \left[\begin{array}{cc}
		\widehat{\Tau}_{\ensembledep}^{(t)} & 0 \\
		\widehat{\channelnoise}^{(t)}_{\eta} & \widehat{\channelnoise}^{(t)}_{\gamma\eta}\left(\widehat{\Tau}_{\ensembleunitary}^{(t)}-\widehat{\Tau}_{\ensembledep}^{(t)}\right)
	\end{array}\right]~,
\end{align}
with $\varepsilon$-$t$-design unitary components with respect to projector ensembles $\ensembleunitary$,
\begin{align}
	\widehat{\Tau}_{\ensembleunitary} \to \widehat{\Tau}_{\ensembleunitary}^{(t)} = \widehat{\Tau}_{\ensembleunitary^{\prime}}^{(t)} ~+~ \widehat{\Tau}_{\ensembleunitary}^{(t)} ~-~ \widehat{\Tau}_{\ensembleunitary^{\prime}}^{(t)}
	= \left[\begin{array}{cc}
	\widehat{\Tau}_{\ensembledep}^{(t)} & 0 \\
	0 & \widehat{\Tau}_{\ensembleunitary^{\prime}}^{(t)} + \widehat{\Tau}_{\ensembleunitary}^{(t)} - \widehat{\Tau}_{\ensembleunitary^{\prime}}^{(t)} - \widehat{\Tau}_{\ensembledep}^{(t)}
	\end{array}\right]~,
\end{align}
and with generic noise components $\channelnoise_{\gamma\eta}^{(t)}$,
\begin{align}
	\widehat{\channelnoise^{\otimes t}_{\gamma\eta}} \to&~ \widehat{\channelnoise}^{(t)}_{\gamma\eta}
	= \widehat{\Tau}_{\ensembledep}^{(t)} ~+~ \widehat{\channelnoise}^{(t)}_{\eta} ~+~ \widehat{\channelnoise}^{(t)}_{\gamma\eta}
	= \left[\begin{array}{cc}
	\widehat{\Tau}_{\ensembledep}^{(t)} & 0 \\
	\widehat{\channelnoise}^{(t)}_{\eta} & \widehat{\channelnoise}^{(t)}_{\gamma\eta}
	\end{array}\right]	~,
\end{align}
with depolarizing $\widehat{\Tau}_{\ensembledep}^{(t)}$, non-unital $\widehat{\channelnoise}^{(t)}_{\eta}$, and unital components $\widehat{\channelnoise}^{(t)}_{\gamma\eta}$ across the $t$ copies, where,
\begin{align}
	\widehat{\Tau}_{\ensembledep}^{(t)}\widehat{\Tau}_{\ensembleunitary}^{(t)} = \widehat{\Tau}_{\ensembledep}^{(t)}
	\quad , \quad
	\widehat{\channelnoise}^{(t)}_{\eta}\widehat{\Tau}_{\ensembleunitary}^{(t)} = \widehat{\channelnoise}^{(t)}_{\eta}
	\quad , \quad
	\widehat{\Tau}_{\ensembledep}^{(t)}\widehat{\channelnoise}^{(t)}_{\eta} = \widehat{\Tau}_{\ensembledep}^{(t)}\widehat{\channelnoise}^{(t)}_{\gamma\eta} = \widehat{\channelnoise}^{(t)}_{\eta}\widehat{\channelnoise}^{(t)}_{\gamma\eta} = 0~.
\end{align}
For example, we may consider composite $\dsys \to q^{n}$-dimensional spaces, with local noise,
\begin{align}
	\widehat{\channelnoise}_{\gamma\eta} \to \widehat{\channelnoise}_{\gamma\eta}^{\otimes n}
	= \widehat{\left(\channeldepolarize[]{q} + {\channelnoise}^{(1)}_{\eta} + {\channelnoise}^{(1)}_{\gamma\eta}\right)^{\otimes n}}
	= \left[\begin{array}{cc}
	\widehat{\Tau}_{\ensembledep[q]}^{} & 0 \\
	\widehat{\channelnoise}^{(1)}_{\eta} & \widehat{\channelnoise}^{(1)}_{\gamma}
	\end{array}\right]^{\otimes n}
	= \widehat{\Tau}_{\ensembledep[q]}^{\otimes n} ~+~ \widehat{\channelnoise}^{(n)}_{\eta} ~+~ \widehat{\channelnoise}^{(n)}_{\gamma\eta}~,
\end{align}
with depolarizing $\widehat{\Tau}_{\ensembledep[q]}^{\otimes n}$, non-unital $\widehat{\channelnoise}^{(n)}_{\eta}$ and unital $\widehat{\channelnoise}^{(n)}_{\gamma\eta}$ components. For $q=2$ qubits, local noise has, without loss of generality, non-unital $\widehat{\channelnoise}^{(1)}_{\eta}$ and diagonal unital components $\widehat{\channelnoise}^{(1)}_{\gamma\eta}$, up to unitaries \cite{mele2024noise}.\\

\begin{theorem}[Unital and Non-unital Noise and Random Unitary Channel Moment Operator Norms]\label{theorem:app:unital_and_non_unital_noise_and_random_unitary_channel_moment_operator_norms}
$k$-concatenations of $t$-order moment operators for composite ensembles consisting of, $\varepsilon$-$t$-design unitary ensembles $\ensembleunitary$ with respect to a reference unitary ensemble $\ensembleunitary^{\prime}$, and a generic noise $\widehat{\channelnoise}^{(t)}_{\gamma\eta} = \widehat{\Tau}_{\ensembledep}^{(t)} + \widehat{\channelnoise}^{(t)}_{\eta} + \widehat{\channelnoise}^{(t)}_{\gamma\eta}$ component, have norms of,
\begin{align}
	\!\!
	\norm{\widehat{\Tau}_{\ensembleunitary\ensemblenoise_{\gamma\eta}}^{(t)k} - \widehat{\Tau}_{\ensembledep}^{(t)}}_{p}
	~\leq&~~ \norm{\widehat{\channelnoise}^{(t)}_{\eta}}_{\infty}\frac{\norm{\widehat{\channelnoise}^{(t)}_{\gamma\eta}}_{\infty}^{k}\left(\norm{\widehat{\Tau}_{\ensembleunitary^{\prime}}^{(t)}-\widehat{\Tau}_{\ensembledep}^{(t)}}_{p} + \norm{\widehat{\Tau}_{\ensembleunitary}^{(t)} - \widehat{\Tau}_{\ensembleunitary^{\prime}}^{(t)}}_{p}\right)^{k}-1}{\norm{\widehat{\channelnoise}^{(t)}_{\gamma\eta}}_{\infty}\left(\norm{\widehat{\Tau}_{\ensembleunitary^{\prime}}^{(t)}-\widehat{\Tau}_{\ensembledep}^{(t)}}_{p} + \norm{\widehat{\Tau}_{\ensembleunitary}^{(t)} - \widehat{\Tau}_{\ensembleunitary^{\prime}}^{(t)}}_{p}\right) - 1} \\
	&~~~~+~ \norm{\widehat{\channelnoise}^{(t)}_{\gamma\eta}}_{\infty}^{k}\left(\norm{\widehat{\Tau}_{\ensembleunitary^{\prime}}^{(t)}-\widehat{\Tau}_{\ensembledep}^{(t)}}_{p}+\norm{\widehat{\Tau}_{\ensembleunitary}^{(t)} - \widehat{\Tau}_{\ensembleunitary^{\prime}}^{(t)}}_{p}\right)^{k} ~.
\end{align}
In the case of the reference $\ensembleunitary^{\prime}$ having rank-$r$ projector moment operators $\widehat{\Tau}_{\ensembleunitary^{\prime}}^{(t)}$, with locality-$\geq l$ terms, and strictly diagonal unital and non-unital noise, with bounded $\infty$-norms $\norm{\widehat{\Tau}_{\ensembleunitary}^{(t)} - \widehat{\Tau}_{\ensembleunitary^{\prime}}^{(t)}}_{\infty} \leq \varepsilon~,~\norm{\widehat{\channelnoise}^{(t)}_{\gamma}}_{\infty} \leq 1 - \gamma~,~\norm{\widehat{\channelnoise}^{(t)}_{\eta}}_{\infty} \leq \eta$,
\begin{align}
	\norm{\widehat{\Tau}_{\ensembleunitary\ensemblenoise_{\gamma\eta}}^{(t)k} - \widehat{\Tau}_{\ensembledep}^{(t)}}_{p}
	~\leq&~~ \eta ~\frac{\left((1-\gamma)^{l}(r-1) + (1-\gamma)\varepsilon\right)^{k}-1}{\left((1-\gamma)^{l}(r-1) + (1-\gamma)\varepsilon\right)-1} ~+~ \left((1-\gamma)^{l}(r-1) + (1-\gamma)\varepsilon\right)^{k}
	~.
\end{align}
\end{theorem}

\begin{proof}\label{proof:app:unital_and_non_unital_noise_and_random_unitary_channel_moment_operator_norms}
	Given generic composite unitary and noise components, and various norm inequalities,
	\begin{align}
	\!\!\!\!\!\!\!
	\norm{\widehat{\Tau}_{\ensembleunitary\ensemblenoise_{\gamma\eta}}^{(t)k} - \widehat{\Tau}_{\ensembledep}^{(t)}}_{p}
	=&~ \norm{\left(\left(\widehat{\Tau}_{\ensembledep}^{(t)} + \widehat{\channelnoise}^{(t)}_{\eta} + \widehat{\channelnoise}^{(t)}_{\gamma\eta}\right)\left(\widehat{\Tau}_{\ensembledep}^{(t)} +\widehat{\Tau}_{\ensembleunitary^{\prime}}^{(t)}  - \widehat{\Tau}_{\ensembledep}^{(t)} + \widehat{\Tau}_{\ensembleunitary}^{(t)} - \widehat{\Tau}_{\ensembleunitary^{\prime}}^{(t)}\right)\right)^{k} - \widehat{\Tau}_{\ensembledep}^{(t)}}_{p} \!\!\!\\
	\!\!\!\!\!\!\!
	\leq&~ \norm{\sum_{s \in [k]} \left(\widehat{\channelnoise}^{(t)}_{\gamma\eta}\left(\widehat{\Tau}_{\ensembleunitary^{\prime}}^{(t)} + \widehat{\Tau}_{\ensembleunitary}^{(t)} - \widehat{\Tau}_{\ensembleunitary^{\prime}}^{(t)}\right)\right)^{s}\widehat{\channelnoise}^{(t)}_{\eta}}_{p}\!\!\! ~+~ \norm{\vphantom{\sum_{s \in [k]}}{}\left(\widehat{\channelnoise}^{(t)}_{\gamma\eta}\left(\widehat{\Tau}_{\ensembleunitary^{\prime}}^{(t)} + \widehat{\Tau}_{\ensembleunitary}^{(t)} - \widehat{\Tau}_{\ensembleunitary^{\prime}}^{(t)}\right)\right)^{k}}_{p} \!\!\!\\
	\!\!\!\!\!\!\!
	\leq&~ \norm{\widehat{\channelnoise}^{(t)}_{\eta}}_{\infty}\sum_{s \in [k]} \norm{\widehat{\channelnoise}^{(t)}_{\gamma\eta}\left(\widehat{\Tau}_{\ensembleunitary^{\prime}}^{(t)} + \widehat{\Tau}_{\ensembleunitary}^{(t)} - \widehat{\Tau}_{\ensembleunitary^{\prime}}^{(t)}\right)}_{p}^{s} + \norm{\widehat{\channelnoise}^{(t)}_{\gamma\eta}}_{\infty}^{k}\left(\norm{\widehat{\Tau}_{\ensembleunitary^{\prime}}^{(t)}-\widehat{\Tau}_{\ensembledep}^{(t)}}_{p}+\norm{\widehat{\Tau}_{\ensembleunitary}^{(t)} - \widehat{\Tau}_{\ensembleunitary^{\prime}}^{(t)}}_{p}\right)^{k} \!\!\!\!\!\\
	\!\!\!\!\!\!\!
	\leq&~ \norm{\widehat{\channelnoise}^{(t)}_{\eta}}_{\infty}\frac{\norm{\widehat{\channelnoise}^{(t)}_{\gamma\eta}}_{\infty}^{k}\left(\norm{\widehat{\Tau}_{\ensembleunitary^{\prime}}^{(t)}-\widehat{\Tau}_{\ensembledep}^{(t)}}_{p} + \norm{\widehat{\Tau}_{\ensembleunitary}^{(t)} - \widehat{\Tau}_{\ensembleunitary^{\prime}}^{(t)}}_{p}\right)^{k}-1}{\norm{\widehat{\channelnoise}^{(t)}_{\gamma\eta}}_{\infty}\left(\norm{\widehat{\Tau}_{\ensembleunitary^{\prime}}^{(t)}-\widehat{\Tau}_{\ensembledep}^{(t)}}_{p} + \norm{\widehat{\Tau}_{\ensembleunitary}^{(t)} - \widehat{\Tau}_{\ensembleunitary^{\prime}}^{(t)}}_{p}\right) - 1} \\
	&~~~~+~ \norm{\widehat{\channelnoise}^{(t)}_{\gamma\eta}}_{\infty}^{k}\left(\norm{\widehat{\Tau}_{\ensembleunitary^{\prime}}^{(t)}-\widehat{\Tau}_{\ensembledep}^{(t)}}_{p}+\norm{\widehat{\Tau}_{\ensembleunitary}^{(t)} - \widehat{\Tau}_{\ensembleunitary^{\prime}}^{(t)}}_{p}\right)^{k} ~. \nonumber
	\end{align}
	Let us consider the case of the reference unitary ensemble $\ensembleunitary^{\prime}$ having rank-$r$ projector moment operators $\widehat{\Tau}_{\ensembleunitary^{\prime}}^{(t)} = \sum_{\substack{\Gamma \subseteq [t]}}\widehat{\Tau}_{\ensembleunitary^{\prime}}^{(\Gamma)}$, with locality-$\abs{\Gamma} \geq l$ terms, and there being strictly diagonal unital $\widehat{\channelnoise}^{(t)}_{\gamma\eta} = \widehat{\channelnoise}^{(t)}_{\gamma}$ and non-unital $\widehat{\channelnoise}^{(t)}_{\eta}$ noise, with bounded $\infty$-norms $\norm{\widehat{\Tau}_{\ensembleunitary}^{(t)} - \widehat{\Tau}_{\ensembleunitary^{\prime}}^{(t)}}_{\infty} \leq \varepsilon~,~\norm{\widehat{\channelnoise}^{(t)}_{\gamma}}_{\infty} \leq 1 - \gamma~,~\norm{\widehat{\channelnoise}^{(t)}_{\eta}}_{\infty} \leq \eta$. The diagonal unital noise $\widehat{\channelnoise}^{(t)}_{\gamma} = \widehat{\channelnoise}^{\otimes t}_{\gamma}$ thus scales each projective moment operator term according to its locality-$\abs{\Gamma} \geq l$, and for $s \geq 0$,
	\begin{align}
		\!\!\!
		\norm{\left(\widehat{\channelnoise}^{(t)}_{\gamma}\left(\widehat{\Tau}_{\ensembleunitary^{\prime}}^{(t)} - \widehat{\Tau}_{\ensembledep}^{(t)}\right)\right)^{s}}_{p}
		= \norm{\sum_{\substack{\Gamma \subseteq [t]\\\abs{\Gamma}\geq l}}\widehat{\channelnoise}^{(\Gamma)s}_{\gamma}\widehat{\Tau}_{\ensembleunitary^{\prime}}^{(\Gamma)}}_{p}
		\leq \sum_{\substack{\Gamma \subseteq [t]\\\abs{\Gamma}\geq l}}\norm{\widehat{\channelnoise}^{(\Gamma)s}_{\gamma}}_{\infty}\norm{\widehat{\Tau}_{\ensembleunitary^{\prime}}^{(\Gamma)}}_{p} \leq \norm{\widehat{\channelnoise}_{\gamma}}_{\infty}^{ls}\norm{\widehat{\Tau}_{\ensembleunitary^{\prime}}^{(t)}-\widehat{\Tau}_{\ensembledep}^{(t)}}_{p}~, \\
		\!\!\!
		\norm{\left(\widehat{\channelnoise}^{(t)}_{\gamma}\left(\widehat{\Tau}_{\ensembleunitary^{\prime}}^{(t)} - \widehat{\Tau}_{\ensembledep}^{(t)} ~+~ \widehat{\Tau}_{\ensembleunitary}^{(t)} - \widehat{\Tau}_{\ensembleunitary^{\prime}}^{(t)}\right)\right)^{s}}_{p}
		\leq \left(\norm{\widehat{\channelnoise}_{\gamma}}_{\infty}^{l}\norm{\widehat{\Tau}_{\ensembleunitary^{\prime}}^{(t)}-\widehat{\Tau}_{\ensembledep}^{(t)}}_{p}~+~\norm{\widehat{\channelnoise}_{\gamma}}_{\infty}\norm{\widehat{\Tau}_{\ensembleunitary}^{(t)} - \widehat{\Tau}_{\ensembleunitary^{\prime}}^{(t)}}_{p}\right)^{s} \!\!\!\! ~.
	\end{align}
	\begin{align}
	\!\!\!\!\!\!\!
	\norm{\widehat{\Tau}_{\ensembleunitary\ensemblenoise_{\gamma\eta}}^{(t)k} - \widehat{\Tau}_{\ensembledep}^{(t)}}_{p}
	\leq&~ \norm{\sum_{s \in [k]} \left(\widehat{\channelnoise}^{(t)}_{\gamma}\left(\widehat{\Tau}_{\ensembleunitary^{\prime}}^{(t)} + \widehat{\Tau}_{\ensembleunitary}^{(t)} - \widehat{\Tau}_{\ensembleunitary^{\prime}}^{(t)}\right)\right)^{s}\widehat{\channelnoise}^{(t)}_{\eta}}_{p}\!\!\! ~+~ \norm{\vphantom{\sum_{s \in [k]}}{}\left(\widehat{\channelnoise}^{(t)}_{\gamma}\left(\widehat{\Tau}_{\ensembleunitary^{\prime}}^{(t)} + \widehat{\Tau}_{\ensembleunitary}^{(t)} - \widehat{\Tau}_{\ensembleunitary^{\prime}}^{(t)}\right)\right)^{k}}_{p} \!\!\!\\
	\!\!\!\!\!\!\!
	\leq&~ \norm{\widehat{\channelnoise}^{(t)}_{\eta}}_{\infty}\sum_{s \in [k]} \left(\norm{\widehat{\channelnoise}_{\gamma}}_{\infty}^{l}\norm{\widehat{\Tau}_{\ensembleunitary^{\prime}}^{(t)}-\widehat{\Tau}_{\ensembledep}^{(t)}}_{p}~+~\norm{\widehat{\channelnoise}_{\gamma}}_{\infty}\norm{\widehat{\Tau}_{\ensembleunitary}^{(t)} - \widehat{\Tau}_{\ensembleunitary^{\prime}}^{(t)}}_{p}\right)^{s} \\
	&~~~~+~ \left(\norm{\widehat{\channelnoise}_{\gamma}}_{\infty}^{l}\norm{\widehat{\Tau}_{\ensembleunitary^{\prime}}^{(t)}-\widehat{\Tau}_{\ensembledep}^{(t)}}_{p}~+~\norm{\widehat{\channelnoise}_{\gamma}}_{\infty}\norm{\widehat{\Tau}_{\ensembleunitary}^{(t)} - \widehat{\Tau}_{\ensembleunitary^{\prime}}^{(t)}}_{p}\right)^{k} \!\!\!\!\! \nonumber \!\!\!\!\\
	\leq&~ \eta ~\frac{\left((1-\gamma)^{l}(r-1) + (1-\gamma)\varepsilon\right)^{k}-1}{\left((1-\gamma)^{l}(r-1) + (1-\gamma)\varepsilon\right)-1} ~+~ \left((1-\gamma)^{l}(r-1) + (1-\gamma)\varepsilon\right)^{k} ~.
	\end{align}
	If the unitary ensemble is an exact $t$-design with respect to the reference ensemble, there is the further simplification,
	\begin{align}
	\!\!\!\!\!\!\!
	\norm{\widehat{\Tau}_{\ensembleunitary\ensemblenoise_{\gamma\eta}}^{(t)k} - \widehat{\Tau}_{\ensembledep}^{(t)}}_{p}
	\leq&~ \norm{\widehat{\channelnoise}^{(t)}_{\eta}}_{\infty}\sum_{s \in [k]} \norm{\left(\widehat{\channelnoise}^{(t)}_{\gamma}\left(\widehat{\Tau}_{\ensembleunitary}^{(t)} - \widehat{\Tau}_{\ensembledep}^{(t)}\right)\right)^{s}}_{p} ~+~ \norm{\left(\widehat{\channelnoise}^{(t)}_{\gamma}\left(\widehat{\Tau}_{\ensembleunitary}^{(t)} - \widehat{\Tau}_{\ensembledep}^{(t)}\right)\right)^{k}}_{p} \!\!\!\!\\
	\!\!\!\!\!\!\!
	\leq&~ \norm{\widehat{\channelnoise}^{(t)}_{\eta}}_{\infty}\sum_{s \in [k]} \norm{\widehat{\channelnoise}_{\gamma}}_{\infty}^{ls}\norm{\widehat{\Tau}_{\ensembleunitary}^{(t)}-\widehat{\Tau}_{\ensembledep}^{(t)}}_{p} ~+~ \norm{\widehat{\channelnoise}_{\gamma}}_{\infty}^{lk}\norm{\widehat{\Tau}_{\ensembleunitary}^{(t)}-\widehat{\Tau}_{\ensembledep}^{(t)}}_{p} \!\!\!\!\\
	\leq&~ \eta ~(r-1)~\frac{(1-\gamma)^{lk}-1}{(1-\gamma)^{l}-1} ~+~ (r-1)~(1-\gamma)^{lk} ~.
	\end{align}
\end{proof}

\subsection{Unital Noise and Parametrized Random Unitary Channels}\label{subsec:unital_noise_and_parameterized_random_unitary_channels}
Here, we consider parameterized random unitary channels $\ensembleunitary = \ensemblevariable$, and fixed diagonal unital noise $\channelnoise = \channelnoise_{\gamma}$, and derive the scaling of their $k$-concatenated $t$-th order composite moment operators with $t,k,\dsys,\gamma$.\\

For the unitary component $\ensembleunitary = \ensemblevariable$, we assume parameterized unitary channels
\begin{align}
	\ensemblevariable =&~ \{\unitary_{\theta}^{G} = e^{-i\theta G} ~,~ \theta \sim \Theta\} ~,
\end{align}
generated by single Hermitian, involutory operators $G \in \mathcal{P}_{\dsys}$, with a distribution of parameters $\Theta$, and a basis for the generator's commutant of $\mathcal{S}_{G} \subseteq \mathcal{P}_{\dsys}$. $t$ copies of the parameterized unitaries for involutory generators $G$ have the form,
\begin{align}
	\unitary_{\theta}^{G \otimes t} ~~=&~ \sum_{l \in [t+1]}u_{\theta}^{(l)}~G^{(l)}
	\quad : \quad
	G^{(l)} = \sum_{\substack{\Gamma \subseteq [t]\\\abs{\Gamma} = l}} \otimes_{i \in \Gamma} ~ G_{i}
	\quad , \quad
	u_{\theta}^{(l)} = (-i)^{l}~\cos^{t}{\theta}~\tan^{l}{\theta}~,
\end{align}
and the associated $t$-th order moment operators are functions of trigonometric moments of the parameter distribution,
\begin{align}
	\widehat{\Tau}_{G\ensemblevariable}^{(t)}
	~~=&~~ \expval{\unitary_{\theta}^{G~\!\otimes t} \otimes \unitary_{\theta}^{G~\!\otimes t~\! *}}_{\theta \sim \Theta}
	~~=~ \!\!\!\! \sum_{l,l^{\prime} \in [t+1]}\expval{u_{\theta}^{(l)}u_{\theta}^{(l^{\prime})*}}_{\theta \sim \Theta}~G^{(l)} \otimes G^{(l^{\prime})*}~.
\end{align}
Crucial to this analysis is an understanding of the $t$-th order commutant of parameterized random unitaries $\mathcal{S}_{\ensemblevariable}^{(t)}$, which depends non-trivially on the generator commutant $\mathcal{S}_{G}$, and can be described by the locality $l \in [t+1]$ of its operators,
\begin{align}
	\hspace{-0.4cm}
	\mathcal{S}_{\ensemblevariable}^{(t)} =&~ \hspace{-0.2cm} \bigcup_{l \in [t+1]} \mathcal{S}_{\ensemblevariable}^{(t|l)} ~:~ \mathcal{S}_{\ensemblevariable}^{(t|l)} = \{X \in \mathcal{S}_{\ensemblevariable}^{(t)} ~:~ \abs{X} = l\}~.
\end{align}
A complete understanding of the entire commutant is non-trivial, and is left for future work. However, given the form of the moment operator, and that higher $t$-th order commutants include lower $l\leq t$-locality commutants, therefore the $l=1$-locality local component of commutants is the generator commutant at each of the $t$ copies of the space,
\begin{align}
	\mathcal{S}_{\ensemblevariable}^{(t|1)} =&~ \bigcup_{i \in [t]} \mathcal{S}_{G_{i}}~.
\end{align}
We can also evaluate the trigonometric moments involved in this moment operator exactly for $t=1,2$.
\begin{lemma}[$t=1,2$ Twirls of parameterized Random Unitaries]\label{lemma:app:t=12_twirls_of_parameterized_random_unitaries}
	Twirls over $t=1,2$-order parameterized random unitary ensembles $e^{-i\theta G} \sim \ensemblevariable$, given Hermitian involutory generators $G$, uniform parameter distributions $\Theta$, and $X \in \mathcal{B}[\spacesys]$, are
	\begin{align}
		\Tau_{\ensemblevariable}^{(1)}(X) =&~ \frac{1}{2}\left[X + GXG\right] \\
		\Tau_{\ensemblevariable}^{(2)}(X) =&~ \frac{1}{8}\left[3\left(X + \left(G \otimes G\right) X \left(G \otimes G\right)\right) - \left\{G \otimes G,X\right\} + \left(I \otimes G + G \otimes I\right) X \left(I \otimes G + G \otimes I\right)\right]~.
	\end{align}
\end{lemma}

\noindent Given these insights into the local component of the commutant, the $t$-th order $\ensemblevariable$ ensemble moment operator is
\begin{align}
	\widehat{\Tau}_{\ensemblevariable}^{(t)} =&~ \frac{1}{\dsys^{t}} \sum_{\substack{P \in \mathcal{S}_{\ensemblevariable}^{(t|1)}}} \kket{P}\bbra{P} ~+~ \frac{1}{\dsys^{t}}\sum_{\substack{P,S \in \mathcal{S}_{\ensemblevariable}^{(t)} \backslash \mathcal{S}_{\ensemblevariable}^{(t|1)}}} \tau_{\ensemblevariable}^{(t,k)}(P,S) ~ \kket{P}\bbra{S} ~.
\end{align}
\noindent For the noise component, the  moment operator $\channelnoise = \channelnoise_{\gamma}$ has the specific form of diagonal unital noise components $\gamma_{\dsys}(P) \geq \gamma$ for $P \in \mathcal{S}_{\ensemblenoise_{\gamma}}$, in an orthogonal basis $\mathcal{S}_{\ensemblenoise_{\gamma}} = \mathcal{P}_{\dsys}$,
\begin{align}
	\widehat{\channelnoise}_{\gamma} =&~
	\frac{1}{\dsys}\left[\begin{array}{cccc}
		1 & 0 & \cdots & 0 \\
		0 & 1-\gamma_{\dsys}(S) & \cdots & 0 \\
		\vdots & \vdots & \ddots & \vdots \\
		0 & 0 & \cdots & 1-\gamma_{\dsys}(P) \\
	\end{array}\right]_{P,S \in \mathcal{P}_{\dsys} \backslash \{I_{\dsys}\}}
	\!\! \to \quad
	\widehat{\channelnoise^{\otimes t}_{\gamma}} = \frac{1}{\dsys^{t}}\sum_{P,S \in \mathcal{S}_{\ensemblenoise_{\gamma}}^{(t)}} \tau_{\ensemblenoise_{\gamma}}^{(t)}(P,S) ~ \kket{P}\bbra{S}~,
\end{align}
The $t$-th order noise component transfer matrix $\tau_{\ensemblenoise_{\gamma}}^{(t)}$, for basis operators $P,S \in \mathcal{S}_{\ensemblenoise_{\gamma}}^{(t)} = \mathcal{P}_{\dsys}^{t}$, is
\begin{align}
	\tau_{\ensemblenoise_{\gamma}}^{(t)}(P,S)
	=&~ \prod_{i \in \Gamma_{S}} (1-\gamma_{\dsys}(S_{i}))~ \delta_{PS}
	~\in~ \mathcal{O}\!\left((1 - \gamma)^{\abs{S}}\right)~\delta_{PS} ~.
\end{align}
The $t$-th order noise component basis can also be described by the locality $l \in [t+1]$ of its operators,
\begin{align}
	\hspace{-0.4cm}
	\mathcal{S}_{\ensemblenoise_{\gamma}}^{(t)} =&~ \hspace{-0.2cm} \bigcup_{l \in [t+1]} \mathcal{S}_{\ensemblenoise_{\gamma}}^{(t|l)} ~:~ \mathcal{S}_{\ensemblenoise_{\gamma}}^{(t|l)} = \{X \in \mathcal{S}_{\ensemblenoise_{\gamma}}^{(t)} ~:~ \abs{X} = l\}~.
\end{align}
and the local component is the noise component basis at each of the $t$ copies of the space,
\begin{align}
	\mathcal{S}_{\ensemblenoise_{\gamma}}^{(t|1)} = \cup_{i \in [t]}\mathcal{S}_{\ensemblenoise_{\gamma}i}.
\end{align}

\noindent Given the transfer matrices for the unitary and noise components, and that their respective bases satisfy $\mathcal{S}_{G} \subseteq \mathcal{S}_{\ensemblenoise_{\gamma}} = \mathcal{P}_{\dsys} ~\to~ \mathcal{S}_{\ensemblevariable}^{(t|1)} \subseteq \mathcal{S}_{\ensemblenoise_{\gamma}}^{(t|1)} = \cup_{i \in [t]} \mathcal{P}_{{\dsys}_{i}}$, the $k$-concatenated $t$-th order composite moment operator is
\begin{align}
	\widehat{\Tau}_{\ensemblevariable\channelnoise_{\gamma}}^{(t)k} =&~ \frac{1}{\dsys^{t}} \sum_{\substack{P,S \in \mathcal{S}_{\ensemblevariable}^{(t|1)}}} \tau_{\ensemblevariable\channelnoise_{\gamma}}^{(t,k)}(P,S) ~ \kket{P}\bbra{S} ~+~ \frac{1}{\dsys^{t}}\sum_{\substack{P \in \mathcal{S}_{\ensemblenoise_{\gamma}}^{(t)} \backslash \mathcal{S}_{\ensemblenoise_{\gamma}}^{(t|1)} \\ S \in \mathcal{S}_{\ensemblevariable}^{(t)} \backslash \mathcal{S}_{\ensemblevariable}^{(t|1)}}} \tau_{\ensemblevariable\ensemblenoise_{\gamma}}^{(t,k)}(P,S) ~ \kket{P}\bbra{S} ~,
\end{align}
with diagonal $k$-concatenated $t$-th order composite transfer matrices, for the local operators $P,S \in \mathcal{S}_{\ensemblevariable}^{(t|1)}$, of
\begin{align}
	\tau_{\ensemblevariable\ensemblenoise_{\gamma}}^{(t,k)}(P,S)
	~=&~~ \prod_{i \in \Gamma_{S}} (1-\gamma_{\dsys}(S_{i}))^{k}~ \delta_{PS}
	~\in~ \mathcal{O}\!\left((1 - \gamma)^{k\abs{S}}\right)~\delta_{PS} \quad : \quad P,S \in \mathcal{S}_{\ensemblevariable}^{(t|1)} ~.
\end{align}
	Finally, we can assess the leading-order scaling of moment operator norms for $k$-concatenations of composite ensembles.
\begin{theorem}[Unital Noise and Parametrized Random Unitary Channel Moment Operator Norms]\label{theorem:app:unital_noise_and_parameterized_random_unitary_channel_moment_operator_norms}
	$k$-concatenations of composite ensembles, consisting of parameterized random unitaries, generated by involutory operators $G$ with commutants $\mathcal{S}_{G}$, and diagonal unital noise $\channelnoise_{\gamma}$ with noise coefficients $\gamma_{\dsys}(P) \geq \gamma$ for $P \in \mathcal{S}_{\ensemblenoise_{\gamma}} \backslash \{\idsys\}$, have moment operator norms of,
	\begin{align}
		\norm{\widehat{\Tau}_{\ensemblevariable\ensemblenoise_{\gamma}}^{(t)k} - \widehat{\Tau}_{\ensembledep}^{(t)}}^{2}_{2}
		=&~ t~\abs{\mathcal{S}_{G} \backslash \{\idsys\}}~\mathcal{O}\!\left((1-\gamma)^{2k}\right) ~.
	\end{align}
	Therefore unital noise decreases moment operator norms for $\gamma > 0$, and as $k \to \infty$, approaches a Depolarize $t$-design.
\end{theorem}

\begin{proof}\label{proof:app:unital_noise_and_parameterized_random_unitary_channel_moment_operator_norms}
	Leading-order behavior of the composite moment operators is associated with local operators in the commutant, $\mathcal{S}_{\ensemblevariable}^{(t|1)} = \cup_{i} \mathcal{S}_{G_{i}}$. Under the assumption of the intersection of the unitary and noise component bases of $\mathcal{S}_{\ensemblevariable}^{(t|1)} \subseteq \mathcal{S}_{\ensemblenoise_{\gamma}}^{(t|1)}$, the local operator components of the associated moment operator transfer matrices are invariant under $k$-concatenations, up to scaling by diagonal noise coefficients. The moment operator norm is thus,
	\begin{align}
		\norm{\widehat{\Tau}_{\ensemblevariable\ensemblenoise_{\gamma}}^{(t)k} - \widehat{\Tau}_{\ensembledep}^{(t)}}^{2}_{2}
		=&~ \sum_{P \in \mathcal{S}_{\ensemblevariable}^{(t|1)} \backslash \{\idsys^{\otimes t}\}}\mathcal{O}\!\left((1-\gamma)^{2k}\right) ~+~ \hspace{-0.5cm}
		\sum_{\substack{P,T \in \mathcal{S}_{\ensemblenoise_{\gamma}}^{(t)} \backslash \mathcal{S}_{\ensemblenoise_{\gamma}}^{(t|1)} \\ S,Q \in \mathcal{S}_{\ensemblevariable}^{(t)} \backslash \mathcal{S}_{\ensemblevariable}^{(t|1)}}} \chi_{\dsys}^{(t)}(P,T)\chi_{\dsys}^{(t)}(S,Q)\tau_{\ensemblevariable\ensemblenoise_{\gamma}}^{(t,k)}(P,Q)\tau_{\ensemblevariable\ensemblenoise_{\gamma}}^{(t,k)}(T,S) \\
		=&~ t~\abs{\mathcal{S}_{G} \backslash \{\idsys\}}~\mathcal{O}\!\left((1-\gamma)^{2k}\right) ~+~ \mathcal{O}\!\left((1-\gamma)^{2k+2}\right)~.
	\end{align}
\end{proof}

\newpage

\section{Moments and Expectation Values}\label{app:designs_and_trainability}
In these appendices, we generalize previous results \cite{holmes2021connecting}, and show how to use channel moment operators to bound biases and variances of expectation values of functions over channel ensembles. Such bounds allow us to study concentration properties of expectation values, leading to notions of channel-design-induced concentration phenomena.

\subsection{Statistics of Ensembles}\label{subsec:statistics_of_ensembles}
In what follows, given an operator $X \in \mathcal{B}[\spacesys^{\otimes t}]$, we will denote as $X_{\channel} = \channel^{\otimes t}(X)$ the action of the $t$-th fold tensor product action of a channel $\channel$. Further, for $\rho,O \in \mathcal{B}[\spacesys]$, the following property holds $\Tr\!\left[\rho_{\channel}O\right] = \Tr\!\left[\rho ~ O_{\channel^{\dagger}}\right]$.\\

\noindent Given a channel-dependent function $\mathcal{F}(\channel)$, for $\channel \sim \ensemblechannel$, its $t$-th order moments with respect to an ensemble $\ensemblechannel$ are
\begin{align}
	\expval{\mathcal{F}(\channel)^{t}}_{\ensemblechannel} =&~ \int_{\ensemblechannel} d\channel ~ \mathcal{F}(\channel)^{t}~,
\end{align}
and we denote its mean as $\mu_{\ensemblechannel}^{}= \expval{\mathcal{F}(\channel)}_{\ensemblechannel}$, and its variance as $\sigma_{\ensemblechannel}^{2} = \expval{\mathcal{F}(\channel)^{2}}_{\ensemblechannel} - \expval{\mathcal{F}(\channel)}_{\ensemblechannel}^{2}$~. \\

\noindent Given that $\mathcal{F}(\channel)$ is a random variable over the ensemble $\ensemblechannel$, Chebyshev's inequality allows us to upper bound the probability that $\mathcal{F}(\channel)$ deviates from its mean, in terms of its variance, as
\begin{align}
	p\left(\abs{ \mathcal{F}(\channel) - \mu_{\ensemblechannel}^{} } \geq \epsilon\right)\leq \frac{\sigma_{\ensemblechannel}^{2}}{\epsilon^{2}} ~.
\end{align}
\noindent Here, we study the mean and variance of channel-dependent functions, which generally depend on so-called inherent terms, which are due to taking expectation values over large spaces, and depend on so-called design terms, which are due to design properties of ensembles with respect to reference ensembles. Such expressions are with respect to quantum state inputs $\rho \in \mathcal{B}[\spacesys]$, and Hermitian observables $O \in \mathcal{B}[\spacesys]$ that act non-trivially in a subspace $\spaceobs \subseteq \spacesys$ of dimension $\dobs \leq \dsys$, with complement space $\spaceobsbar \subseteq \spacesys$ of dimension $\dobsbar = \dsys/\dobs$. Such expressions are also with respect to reference ensembles $\ensemblechannel^{\prime}$, such as the cHaar $\ensemblechannel^{\prime}=\ensemblechaar$ or Depolarize $\ensemblechannel^{\prime}=\ensembledep$ ensembles. To bound various quantities, we will use Schatten norms $\norm{A}_{p}^{p} = \Tr\!\left[\abs{A}^{p}\right]~,~ p \in [1,\infty]$ for $A,B \in \mathcal{B}[\spacesys^{\otimes t}]$, which satisfy monotonicity $\norm{A}_{p} \leq \norm{A}_{q} ~,~ q \leq p$, sub-additivity $\norm{A+B}_{p} \leq \norm{A}_{p}+\norm{B}_{p}$, sub-multiplicativity $\norm{AB}_{p} \leq \norm{A}_{p}\norm{B}_{p}$, and Holderms $\norm{AB}_{r} \leq \norm{A}_{p}\norm{B}_{q} ~,~ 1/p + 1/q = 1/r$, and von-Neumann $\abs{\Tr\!\left[AB\right]} \leq \norm{A}_{p}\norm{B}_{q} ~,~ 1/p + 1/q = 1$ inequalities \cite{baumgartner2011inequality}.\\

\noindent Let us also recall the $t=1,2$-order twirls $\expval{\rho_{\channel}^{\otimes t}}_{\ensemblechannel^{\prime}}$ of the specific reference ensembles $\ensemblechannel^{\prime}$ considered, with respect to input states $\rho \in \mathcal{B}[\spacesys]$, given the identity $\idsys$, and swap operators $S_{\dsys}$ acting on $\spacesys$.

\noindent For the cHaar reference ensemble $\ensemblechannel^{\prime}=\ensemblechaar$,
\begin{align}
	\expval{\rho_{\channel}}_{\ensemblechaar}
	=&~ \frac{1}{\dsys}~\Tr\!\left[\rho\right]~ \idsys \\
	\expval{\rho_{\channel}^{\otimes 2}}_{\ensemblechaar}
	=&~ \frac{1}{\dsys^{2}}\frac{1}{1-\frac{1}{\dsys^{2}\denv^{2}}}\left[\left[\Tr^{2}\left[\rho\right] - \frac{1}{\dsys\denv}\Tr\!\left[\rho^{2}\right]\right] \idsys^{\otimes 2} ~+~ \frac{1}{\denv}\left[-\frac{1}{\dsys\denv}\Tr^{2}\left[\rho\right] + \Tr\!\left[\rho^{2}\right]\right]S_{\dsys}\right] ~, \\
\intertext{and for the Depolarize reference ensemble $\ensemblechannel^{\prime}=\ensembledep$,}
	\expval{\rho_{\channel}}_{\ensembledep}
	=&~\frac{1}{\dsys}~\Tr\!\left[\rho\right]~ \idsys \\
	\expval{\rho_{\channel}^{\otimes 2}}_{\ensembledep}
	=&~ \frac{1}{\dsys^{2}} ~\Tr^{2}\left[\rho\right]~\idsys^{\otimes 2}~.
\end{align}
Finally, we recall that $t$-th order twirls of ensembles $\ensemblechannel$, with respect to input states $\rho \in \mathcal{B}[\spacesys]$, can be written in terms of reference ensembles $\ensemblechannel^{\prime}$, as $\expval{\rho_{\channel}^{\otimes t}}_{\ensemblechannel} = \expval{\rho_{\channel}^{\otimes t}}_{\ensemblechannel^{\prime}} + \Delta^{(t)}_{\ensemblechannel\ensemblechannel^{\prime}}(\rho^{\otimes t})$, where the deviation has a $p$-norm of $\Epsilon_{\ensemblechannel\ensemblechannel^{\prime}}^{(t|p)}(\rho^{\otimes t}) = \tnorm{\Delta^{(t)}_{\ensemblechannel\ensemblechannel^{\prime}}(\rho^{\otimes t})}_{p}$.

\subsection{Expectation Value Bias}\label{subsec:expectation_value_bias}
Let us consider the channel-dependent function $\mathcal{F}(\channel) = \mathcal{L}_{\channel}$ of an expectation value, with respect to input states $\rho \in \mathcal{B}[\spacesys]$ and Hermitian observables $O \in \mathcal{B}[\spacesys]$,
\begin{equation}
	\mathcal{L}_{\channel}(\rho,O) = \Tr\!\left[\rho_{\channel}O\right]~.
\end{equation}

First, we can study how much expectation value means $\expval{\mathcal{L}_{\channel}}_\ensemblechannel$ and $\expval{\mathcal{L}_{\channel}}_{\ensemblechannel^{\prime}}$ differ. Concretely we bound the bias $\expval{\abs{\mathcal{L}_{\channel} - \mu_{\ensemblechannel^{\prime}}}}_{\ensemblechannel}$ where $\mu_{\ensemblechannel^{\prime}} = \expval{\mathcal{L}_{\channel}}_{\ensemblechannel^{\prime}}$.

\begin{theorem}[Expectation Value Bias]\label{theorem:app:expectation_value_bias}
Let $\ensemblechannel,\ensemblechannel^{\prime}$ be ensembles of quantum channels, let $\rho$ be a quantum state, and let $O$ be an observable that acts non-trivially on a subspace $\spaceobs \subseteq \spacesys$ of dimension $\dobs$, with complement space $\spaceobsbar$. The following bound on the expectation value $\mathcal{L}_{\channel}$ bias holds
\begin{align}
	\expval{\abs{\mathcal{L}_{\channel}(\rho,O) - \mu_{\ensemblechannel^{\prime}}(\rho,O)}}_{\ensemblechannel}
	~\leq&~~
	\nu_{\ensemblechannel^{\prime}}^{}(\rho,O) ~+~ \alpha_{\ensemblechannel\ensemblechannel^{\prime}}(\rho,O)~,
\end{align}
where, given $S_{\dobs}$ is the swap operator acting on $\spaceobs$, and $\idobsbar$ is the identity operator acting on $\spaceobsbar$, the inherent $\nu_{\ensemblechannel^{\prime}}$ and design $\alpha_{\ensemblechannel\ensemblechannel^{\prime}}$ expectation value bias terms are respectively,
\begin{align}
	\nu_{\ensemblechannel^{\prime}}(\rho,O)
	=&~ \sqrt{
	\Tr\!\left[\left(\expval{\rho_{\channel}^{\otimes 2}}_{\ensemblechannel^{\prime}} - \expval{\rho_{\channel}}_{\ensemblechannel^{\prime}}^{\otimes 2}\right) \left(\vphantom{\left(\left(\expval{\rho_{\channel}^{\otimes 2}}_{\ensemblechannel^{\prime}} - \expval{\rho_{\channel}}_{\ensemblechannel^{\prime}}^{\otimes 2}\right) \otimes \idobs \right)} S_{\dobs} \otimes \idobsbar^{\otimes 2} \right)\right]} ~\sqrt{\dobs}~\norm{O}_{\infty} \\
	\alpha_{\ensemblechannel\ensemblechannel^{\prime}}(\rho,O)
	=&~ \sqrt{\Tr\!\left[\left(\Delta^{(2)}_{\ensemblechannel\ensemblechannel^{\prime}}~(\rho^{\otimes 2})
	-\Delta_{\ensemblechannel\ensemblechannel^{\prime}}(\rho) ~\otimes~ \expval{\rho_{\channel}}_{\ensemblechannel^{\prime}}^{}
	-\expval{\rho_{\channel}}_{\ensemblechannel^{\prime}}^{} \otimes~ \Delta_{\ensemblechannel\ensemblechannel^{\prime}}(\rho)
	\right) \left(\vphantom{\expval{\rho_{\channel|\ensemblechannel^{\prime}}^{\otimes 2}}_{\ensemblechannel\ensemblechannel^{\prime}}^{(2)}} S_{\dobs} \otimes \idobsbar^{\otimes 2} \right)\right]}~\sqrt{\dobs}~\norm{O}_{\infty} ~.
	\end{align}
\end{theorem}

\begin{proof}\label{proof:app:expectation_value_bias}
To simplify our analysis of the bias of expectation values, we will denote the deviation of a transformed input $\rho \to \rho_{\channel} \in \mathcal{B}[\spacesys]$, where $\channel \sim \ensemblechannel$, from its mean with respect to a reference ensemble $\ensemblechannel^{\prime}$, as
\begin{align}
	\tilde{\rho}_{\channel|\ensemblechannel^{\prime}} =&~ \rho_{\channel} - \expval{\rho_{\channel}}_{\ensemblechannel^{\prime}}~,
\end{align}
and the $t=1,2$-order moments of these deviations, with respect to an ensemble $\ensemblechannel$, are
\begin{align}
	\expval{\tilde{\rho}_{\channel|\ensemblechannel^{\prime}}}_{\ensemblechannel}
	=&~ \Delta_{\ensemblechannel\ensemblechannel^{\prime}}(\rho)\\
	\expval{\tilde{\rho}_{\channel|\ensemblechannel^{\prime}}^{\otimes 2}}_{\ensemblechannel}
	=&~ \expval{\rho_{\channel}^{\otimes 2}}_{\ensemblechannel^{\prime}} - \expval{\rho_{\channel}}_{\ensemblechannel^{\prime}}^{\otimes 2} + \Delta^{(2)}_{\ensemblechannel\ensemblechannel^{\prime}}(\rho^{\otimes 2}) - \Delta_{\ensemblechannel\ensemblechannel^{\prime}}(\rho) \otimes \expval{\rho_{\channel}}_{\ensemblechannel^{\prime}} - \expval{\rho_{\channel}}_{\ensemblechannel^{\prime}} \otimes \Delta_{\ensemblechannel\ensemblechannel^{\prime}}(\rho) ~.
\end{align}
The average expectation value bias can then be bounded using the following norm inequalities,
\begin{align}
	\hspace{-8cm}
	\expval{\abs{\mathcal{L}_{\channel}(\rho,O) - \mu_{\ensemblechannel^{\prime}}(\rho,O)}}_{\ensemblechannel}
	= \expval{\abs{\Tr\!\left[\rho_{\channel} O\otimes I\right] - \expval{\Tr\!\left[\rho_{\channel} O\otimes I\right]}_{\ensemblechannel^{\prime}}}}_{\ensemblechannel}
\end{align}
\vspace{-24pt}
\begin{align}
	=&~ \expval{\abs{\Tr\!\left[\tilde{\rho}_{\channel|\ensemblechannel^{\prime}} O\otimes I\right]}}_{\ensemblechannel} \\
	=&~ \expval{\abs{\Tr_{\spaceobs}\left[\Tr_{\spaceobsbar}\left[\tilde{\rho}_{\channel|\ensemblechannel^{\prime}}\right] O\right]}}_{\ensemblechannel} \\
	\leq&~ \expval{\norm{\Tr_{\spaceobsbar}\left[\tilde{\rho}_{\channel|\ensemblechannel^{\prime}}\right]}_{1}~\norm{O}_{\infty}}_{\ensemblechannel} \\
	\leq&~ \expval{\norm{\idobs}_{2}~\norm{\Tr_{\spaceobsbar}\left[\tilde{\rho}_{\channel|\ensemblechannel^{\prime}}\right]}_{2}~\norm{O}_{\infty}}_{\ensemblechannel} \\
	=&~ \expval{\sqrt{\norm{\Tr_{\spaceobsbar}\left[\tilde{\rho}_{\channel|\ensemblechannel^{\prime}}\right]}_{2}^{2}}}_{\ensemblechannel}~\norm{\idobs}_{2}~\norm{O}_{\infty}
	\\
	\leq&~ \sqrt{\expval{\norm{\Tr_{\spaceobsbar}\left[\tilde{\rho}_{\channel|\ensemblechannel^{\prime}}\right]}_{2}^{2}}_{\ensemblechannel}} ~\norm{\idobs}_{2}~\norm{O}_{\infty} \\
	=&~ \sqrt{\expval{\Tr_{\spaceobs}\left[\Tr^{2}_{\spaceobsbar}\left[\tilde{\rho}_{\channel|\ensemblechannel^{\prime}}\right]\right]}_{\ensemblechannel}}~\norm{\idobs}_{2}~\norm{O}_{\infty} \\
	=&~ \sqrt{\expval{\Tr_{\spaceobs^{\otimes 2}}\left[\Tr_{\spaceobsbar}\left[\tilde{\rho}_{\channel|\ensemblechannel^{\prime}}\right]^{\otimes 2} S_{\dobs} \right]}_{\ensemblechannel}}~\norm{\idobs}_{2}~\norm{O}_{\infty} \\
	=&~ \sqrt{\expval{\Tr_{(\spaceobs \otimes \spaceobsbar)^{\otimes 2}}\left[\tilde{\rho}_{\channel|\ensemblechannel^{\prime}}^{\otimes 2} ~S_{\dobs} \otimes \idobsbar^{\otimes 2} \right]}_{\ensemblechannel}}~\norm{\idobs}_{2}~\norm{O}_{\infty} \\
	=&~ \sqrt{\Tr\!\left[\expval{\tilde{\rho}_{\channel|\ensemblechannel^{\prime}}^{\otimes 2}}_{\ensemblechannel} \left(\vphantom{\left[\tilde{\rho}_{\channel|\ensemblechannel^{\prime}}\right]^{\otimes 2}} S_{\dobs} \otimes \idobsbar^{\otimes 2}\right)\right]}~\norm{\idobs}_{2}~\norm{O}_{\infty} \\
	=&~ \sqrt{\Tr\!\left[\left(\expval{\rho_{\channel}^{\otimes 2}}_{\ensemblechannel^{\prime}} - \expval{\rho_{\channel}}_{\ensemblechannel^{\prime}}^{\otimes 2} + \Delta^{(2)}_{\ensemblechannel\ensemblechannel^{\prime}}(\rho^{\otimes 2}) - \Delta_{\ensemblechannel\ensemblechannel^{\prime}}(\rho) \otimes \expval{\rho_{\channel}}_{\ensemblechannel^{\prime}}^{} - \expval{\rho_{\channel}}_{\ensemblechannel^{\prime}}^{} \otimes \Delta_{\ensemblechannel\ensemblechannel^{\prime}}(\rho)\right) \left( S_{\dobs} \otimes \idobsbar^{\otimes 2} \right)\right]}~\norm{\idobs}_{2}~\norm{O}_{\infty}
	\\
	\leq&~ \left(\sqrt{\Tr\!\left[\left(\expval{\rho_{\channel}^{\otimes 2}}_{\ensemblechannel^{\prime}} - \expval{\rho_{\channel}}_{\ensemblechannel^{\prime}}^{\otimes 2}\right) \left(\vphantom{\left[\tilde{\rho}_{\channel|\ensemblechannel^{\prime}}\right]^{\otimes 2}} S_{\dobs} \otimes \idobsbar^{\otimes 2} \right)\right]} \right. \\
	&~~~+ \left.\sqrt{\Tr\!\left[\left(\Delta^{(2)}_{\ensemblechannel\ensemblechannel^{\prime}}(\rho^{\otimes 2}) - \Delta_{\ensemblechannel\ensemblechannel^{\prime}}(\rho) \otimes \expval{\rho_{\channel}}_{\ensemblechannel^{\prime}}^{} - \expval{\rho_{\channel}}_{\ensemblechannel^{\prime}}^{} \otimes \Delta_{\ensemblechannel\ensemblechannel^{\prime}}(\rho)\right) \left(\vphantom{\Delta^{(2)}_{\ensemblechannel\ensemblechannel^{\prime}}(\rho^{\otimes 2})} S_{\dobs} \otimes \idobsbar^{\otimes 2} \right)\right]}~\right)~\norm{\idobs}_{2}~\norm{O}_{\infty} \nonumber \\
	=&~ \nu_{\ensemblechannel^{\prime}}(\rho,O) ~+~ \alpha_{\ensemblechannel\ensemblechannel^{\prime}}(\rho,O)~.
\end{align}
\end{proof}

\begin{corollary}[Expectation Value Bias for Reference Ensembles]\label{corol:app:expectation_value_bias_for_reference_ensembles}
For specific observables $O$ and reference ensembles $\ensemblechannel^{\prime}$, the expectation value mean is
\begin{align}
	\mu_{\ensemblechannel^{\prime}}^{}(\rho,O)
	=&~ \frac{\Tr\!\left[\rho\right]\Tr\!\left[O\right]}{\dsys}
	= \left\{ \begin{array}{ll}
	0 & \Tr[O]=0~,~\ensemblechannel^{\prime}=\ensemblechaar \\
	\frac{1}{\dsys} & \Tr[O]=1~,~\ensemblechannel^{\prime}=\ensemblechaar \\
	0 & \Tr[O]=0~,~\ensemblechannel^{\prime}=\ensembledep \\
	\frac{1}{\dsys} & \Tr[O]=1~,~\ensemblechannel^{\prime}=\ensembledep
	\end{array}\right. ~,
\end{align}
and the inherent $\nu_{\ensemblechannel^{\prime}}$ and design $\alpha_{\ensemblechannel\ensemblechannel^{\prime}}$ expectation value bias terms are respectively,
\begin{align}
	\nu_{\ensemblechannel^{\prime}}(\rho,O)	=&~ \left\{ \begin{array}{ll}
		\mathcal{O}\!\left(\sqrt{\frac{\dobs^{2}}{\dsys\denv}}\right)~\norm{\rho}_{2}\norm{O}_{\infty} & \ensemblechannel^{\prime}=\ensemblechaar\\
		0 & \ensemblechannel^{\prime}=\ensembledep
	\end{array} \right. \\
	\alpha_{\ensemblechannel\ensemblechannel^{\prime}}(\rho,O)
	\leq&~ \sqrt{\Epsilon^{(2|\diamond)}_{\ensemblechannel\ensemblechannel^{\prime}}\norm{\rho}_{1}^{2} + 2~\Epsilon^{(1|\diamond)}_{\ensemblechannel\ensemblechannel^{\prime}}\norm{\rho}_{1}}~\norm{O}_{\infty}~,\!\!\!\!
\end{align}
under the assumption that the $t=1$-order twirl for a reference ensemble $\ensemblechannel^{\prime}$ is proportional to the identity, $\expval{\rho_{\channel}}_{\ensemblechannel^{\prime}} = \frac{\Tr\!\left[\rho\right]}{\dsys}I$.
\end{corollary}

\begin{proof}\label{proof:app:expectation_value_bias_for_reference_ensembles}
The expectation value mean follows directly from its definition $\mu_{\ensemblechannel^{\prime}}^{}(\rho,O) = \expval{\mathcal{L}_{\channel}(\rho,O)}_{\ensemblechannel^{\prime}} = \Tr\!\left[\expval{\rho_{\channel}}_{\ensemblechannel^{\prime}}O\right] $, and substituting in the assumption that the $t=1$-order twirl for the reference ensemble $\ensemblechannel^{\prime}$ is $\expval{\rho_{\channel}}_{\ensemblechannel^{\prime}} =~ \frac{\Tr\!\left[\rho\right]}{\dsys}I$. \\

\noindent The expectation value bias inherent term follows from the deviation $\expval{\rho_{\channel}^{\otimes 2}}_{\ensemblechannel^{\prime}} - \expval{\rho_{\channel}}_{\ensemblechannel^{\prime}}^{\otimes 2}$ of the $t=2$ moment for specific reference ensembles $\ensemblechannel^{\prime}$, given the forms of their respective $t=1,2$-order twirls $\expval{\rho_{\channel}^{\otimes 2}}_{\ensemblechannel^{\prime}},\expval{\rho_{\channel}}_{\ensemblechannel^{\prime}}^{\otimes 2}$.\\

\noindent For the cHaar reference ensemble $\ensemblechannel^{\prime}=\ensemblechaar$,
\begin{align}
	\expval{\rho_{\channel}^{\otimes 2}}_{\ensemblechaar} - \expval{\rho_{\channel}}_{\ensemblechaar}^{\otimes 2}
	=&~ \frac{1}{\dsys^{2}}\frac{1}{1-\frac{1}{\dsys^{2}\denv^{2}}}\left[\left[\frac{1}{\dsys^{2}\denv^{2}}\Tr^{2}\left[\rho\right] - \frac{1}{\dsys\denv}\Tr\!\left[\rho^{2}\right]\right] \idsys^{\otimes 2} ~+~ \frac{1}{\denv}\left[-\frac{1}{\dsys\denv}\Tr^{2}\left[\rho\right] + \Tr\!\left[\rho^{2}\right]\right]S_{\dsys}\right] \\
	=&~ \frac{1}{\dsys^{2}}~\norm{\rho}_{2}^{2}~\left[\mathcal{O}\!\left(\frac{1}{\dsys^{}\denv^{}}\right)~ \idsys^{\otimes 2} ~+~ \mathcal{O}\!\left(\frac{1}{\denv^{}}\right)~ S_{\dsys}\right] ~,\\
\intertext{and for the Depolarize reference ensemble $\ensemblechannel^{\prime}=\ensembledep$,}
	\expval{\rho_{\channel}^{\otimes 2}}_{\ensembledep} - \expval{\rho_{\channel}}_{\ensembledep}^{\otimes 2}
	=&~ \frac{1}{\dsys^{2}} ~\Tr^{2}\left[\rho\right]~\idsys^{\otimes 2} ~-~ \left(\frac{1}{\dsys} ~\Tr\!\left[\rho\right]~\idsys^{}\right)^{\otimes 2}
	~=~ 0 ~.
\end{align}
It follows that the expectation value bias inherent term is,
\begin{align}
	\nu_{\ensemblechannel^{\prime}}(\rho,O)
	=&~ \sqrt{
	\Tr\!\left[\left(\expval{\rho_{\channel}^{\otimes 2}}_{\ensemblechannel^{\prime}} - \expval{\rho_{\channel}}_{\ensemblechannel^{\prime}}^{\otimes 2}\right) \left(\vphantom{\left(\left(\expval{\rho_{\channel}^{\otimes 2}}_{\ensemblechannel^{\prime}} - \expval{\rho_{\channel}}_{\ensemblechannel^{\prime}}^{\otimes 2}\right) \otimes \idobs \right)} S_{\dobs} \otimes \idobsbar^{\otimes 2} \right)\right]} ~\norm{\idobs}_{2}~\norm{O}_{\infty} \\
	=&~ \left\{ \begin{array}{ll}
	\sqrt{\frac{1}{\dsys^{2}}~\left[\mathcal{O}\!\left(\frac{1}{\dsys^{}\denv^{}}\right)~ \Tr\!\left[S_{\dobs} \otimes \idobsbar^{\otimes 2}\right] ~+~ \mathcal{O}\!\left(\frac{1}{\denv^{}}\right)~ \Tr\!\left[\idobs^{\otimes 2} \otimes S_{\dobsbar}\right]\right]} ~\sqrt{\dobs}~\norm{\rho}_{2}~ \norm{O}_{\infty} & \ensemblechannel^{\prime}=\ensemblechaar\\
		0 & \ensemblechannel^{\prime}=\ensembledep
	\end{array} \right. \\
	=&~ \left\{ \begin{array}{ll}
		\mathcal{O}\!\left(\sqrt{\frac{\dobs^{2}}{\dsys\denv}}\right)~\norm{\rho}_{2}\norm{O}_{\infty} & \ensemblechannel^{\prime}=\ensemblechaar\\
		0 & \ensemblechannel^{\prime}=\ensembledep
	\end{array} \right.~.
\end{align}
\noindent The expectation value bias design term follows from the following norm inequalities,
\begin{align}
	\!\!\!
	\alpha_{\ensemblechannel\ensemblechannel^{\prime}}(\rho,O)
	=&~ \! \sqrt{\Tr\!\left[\left(\Delta^{(2)}_{\ensemblechannel\ensemblechannel^{\prime}}(\rho^{\otimes 2}) - \Delta_{\ensemblechannel\ensemblechannel^{\prime}}(\rho) \otimes \expval{\rho_{\channel}}_{\ensemblechannel^{\prime}}^{} - \expval{\rho_{\channel}}_{\ensemblechannel^{\prime}}^{} \otimes \Delta_{\ensemblechannel\ensemblechannel^{\prime}}(\rho)\right) \left(\vphantom{\expval{\rho_{\channel|\ensemblechannel^{\prime}}^{\otimes 2}}_{\ensemblechannel\ensemblechannel^{\prime}}^{(2)}} S_{\dobs} \otimes \idobsbar^{\otimes 2} \right)\right]}~\norm{\idobs}_{2}~\norm{O}_{\infty} \\
	\leq&~ \min_{\frac{1}{p} + \frac{1}{q} = 1}\sqrt{\Epsilon^{(2|p)}_{\ensemblechannel\ensemblechannel^{\prime}}~\tnorm{\rho}_{p}^{2} ~+~ 2~{\Epsilon^{(1|p)}_{\ensemblechannel\ensemblechannel^{\prime}}}~\tnorm{\expval{\rho_{\channel}}_{\ensemblechannel^{\prime}}^{}}_{p}~\tnorm{\rho}_{p}}~\norm{\idobs}_{q}~\tnorm{\idsys}_{q}~\norm{O}_{\infty}\\
	\leq&~ \sqrt{\Epsilon^{(2|\diamond)}_{\ensemblechannel\ensemblechannel^{\prime}}\norm{\rho}_{1}^{2} ~+~ 2~\Epsilon^{(1|\diamond)}_{\ensemblechannel\ensemblechannel^{\prime}}\norm{\rho}_{1}}~\norm{O}_{\infty}~\!.\!\!
\end{align}
\end{proof}

\subsection{Expectation Value Variance}\label{subsec:expectation_value_variance}

Here we study the variance of expectation values,
\begin{align}
	\sigma_{\ensemblechannel}^{2}(\rho,O)
	~=&~~ \expval{\mathcal{L}_{\channel}(\rho,O)^{2}}_{\ensemblechannel} - \expval{\mathcal{L}_{\channel}(\rho,O)}_{\ensemblechannel}^{2}~.
\end{align}

\begin{theorem}[Expectation Value Variance]\label{theorem:app:expectation_value_variance}
Let $\ensemblechannel,\ensemblechannel^{\prime}$ be ensembles of quantum channels, let $\rho$ be a quantum state, and let $O$ be an observable. The expectation value $\mathcal{L}_{\channel}$ variance is,
\begin{align}
	\sigma_{\ensemblechannel}^{2}(\rho,O)
	~=&~~
	\sigma_{\ensemblechannel^{\prime}}^{2}(\rho,O) ~+~ \beta_{\ensemblechannel\ensemblechannel^{\prime}}(\rho,O)~,
\end{align}
where the inherent $\sigma_{\ensemblechannel^{\prime}}^{2}$ and design $\beta_{\ensemblechannel\ensemblechannel^{\prime}}$ expectation value variance terms are respectively,
\begin{align}
	\sigma_{\ensemblechannel^{\prime}}^{2}(\rho,O)
	=&~ \Tr\!\left[\expval{\rho_{\channel}^{\otimes 2}}_{\ensemblechannel^{\prime}} O^{\otimes 2}\right] - \Tr\!\left[\expval{\rho_{\channel}}^{\otimes 2}_{\ensemblechannel^{\prime}} O^{\otimes 2}\right] \\
	\beta_{\ensemblechannel\ensemblechannel^{\prime}}(\rho,O)
	=&~ \Tr\!\left[\Delta_{\ensemblechannel\ensemblechannel^{\prime}}^{(2)}(\rho^{\otimes 2}) O^{\otimes 2}\right] ~-~ \Tr^{2}\!\left[\Delta_{\ensemblechannel\ensemblechannel^{\prime}}(\rho) O\right] ~-~ 2~\!\Tr\!\left[\expval{\rho_{\channel}}_{\ensemblechannel^{\prime}}O\right]~\!\Tr\!\left[\Delta_{\ensemblechannel\ensemblechannel^{\prime}}(\rho) O\right] ~.
	\end{align}
\end{theorem}

\begin{proof}\label{proof:app:expectation_value_variance}
The expectation value variance can be derived from its definition,
\begin{align}
	\sigma_{\ensemblechannel}^{2}(\rho,O)
	=&~ \expval{\mathcal{L}_{\channel}(\rho,O)^{2}}_{\ensemblechannel} - \expval{\mathcal{L}_{\channel}(\rho,O)}_{\ensemblechannel}^{2} \\
	=&~ \Tr\!\left[\expval{\rho_{\channel}^{\otimes 2}}_{\ensemblechannel} O^{\otimes 2}\right] - \Tr\!\left[\expval{\rho_{\channel}}^{\otimes 2}_{\ensemblechannel} O^{\otimes 2}\right] \\
	=&~ \left(\Tr\!\left[\expval{\rho_{\channel}^{\otimes 2}}_{\ensemblechannel^{\prime}} O^{\otimes 2}\right] + \Tr\!\left[\Delta_{\ensemblechannel\ensemblechannel^{\prime}}^{(2)}(\rho^{\otimes 2}) O^{\otimes 2}\right]\right) - \Tr\!\left[\left(\expval{\rho_{\channel}}_{\ensemblechannel^{\prime}} + \Delta_{\ensemblechannel\ensemblechannel^{\prime}}(\rho)\right)^{\otimes 2} O^{\otimes 2}\right] \\
	=&~ \left(\Tr\!\left[\expval{\rho_{\channel}^{\otimes 2}}_{\ensemblechannel^{\prime}} O^{\otimes 2}\right] -  \Tr\!\left[\expval{\rho_{\channel}}_{\ensemblechannel^{\prime}}^{\otimes 2} O^{\otimes 2}\right]\right) ~+~ \left(\Tr\!\left[\Delta_{\ensemblechannel\ensemblechannel^{\prime}}^{(2)}(\rho^{\otimes 2}) O^{\otimes 2}\right] - \Tr\!\left[\Delta_{\ensemblechannel\ensemblechannel^{\prime}}(\rho)^{\otimes 2} O^{\otimes 2}\right]\right)  \\
	&~~-~ \left(\Tr\!\left[\left(\expval{\rho_{\channel}}_{\ensemblechannel^{\prime}} \otimes \Delta_{\ensemblechannel\ensemblechannel^{\prime}}(\rho)\right) O^{\otimes 2}\right] + \Tr\!\left[\left(\Delta_{\ensemblechannel\ensemblechannel^{\prime}}(\rho) \otimes \expval{\rho_{\channel}}_{\ensemblechannel^{\prime}} \right) O^{\otimes 2}\right] \right) \nonumber \\
	=&~ \sigma_{\ensemblechannel^{\prime}}^{2}(\rho,O) ~+~ \Tr\!\left[\Delta_{\ensemblechannel\ensemblechannel^{\prime}}^{(2)}(\rho^{\otimes 2}) O^{\otimes 2}\right] ~-~ \Tr^{2}\!\left[\Delta_{\ensemblechannel\ensemblechannel^{\prime}}(\rho) O\right] ~-~ 2~\!\Tr\!\left[\Delta_{\ensemblechannel\ensemblechannel^{\prime}}(\rho) O\right]~\!\mu_{\ensemblechannel^{\prime}}^{}(\rho,O)~,
\end{align}
given twirls over ensembles $\ensemblechannel$ can be written in terms of reference ensembles $\ensemblechannel^{\prime}$ as, $\expval{\rho_{\channel}^{\otimes t}}_{\ensemblechannel} = \expval{\rho_{\channel}^{\otimes t}}_{\ensemblechannel^{\prime}} + \Delta_{\ensemblechannel\ensemblechannel^{\prime}}^{(t)}(\rho^{\otimes t})$.
\end{proof}

\begin{corollary}[Expectation Value Variance for Reference Ensembles]\label{corol:app:expectation_value_variance_for_reference_ensembles}
For specific observables $O$ with $\abs{\Tr[O]} \leq 1$, and reference ensembles $\ensemblechannel^{\prime}$, the inherent expectation value variance is
\begin{align}
	\!\!\!\!\!
	\sigma_{\ensemblechannel^{\prime}}^{2}(\rho,O)
	=&~ \left\{\begin{array}{ll}
	\frac{1}{\dsys^{2}}\frac{1}{1-\frac{1}{\dsys^{2}\denv^{2}}}\left[\left[\frac{1}{\dsys^{2}\denv^{2}}\Tr^{2}\left[\rho\right] - \frac{1}{\dsys\denv}\Tr\!\left[\rho^{2}\right]\right] \Tr^{2}\left[O\right] + \frac{1}{\denv}\left[-\frac{1}{\dsys\denv}\Tr^{2}\left[\rho\right] + \Tr\!\left[\rho^{2}\right]\right]\Tr\!\left[O^{2}\right] \right] & \ensemblechannel^{\prime} = \ensemblechaar \\
	0 & \ensemblechannel^{\prime} = \ensembledep
	\end{array}\right. \!\!\!\!\!,\!\!\!\! \\
	=&~ \left\{ \begin{array}{ll}
		\mathcal{O}\!\left(\frac{1}{\dsys^{2}\denv}\right)~\norm{\rho}_{2}^{2}~\norm{O}_{2}^{2} & \ensemblechannel^{\prime}=\ensemblechaar \\
		0 & \ensemblechannel^{\prime}=\ensembledep
		\end{array} \right. ~.
\end{align}
and the design $\beta_{\ensemblechannel\ensemblechannel^{\prime}}$ expectation value variance term is bounded by,
\begin{align}
	\!\!\!\!\!
	\beta_{\ensemblechannel\ensemblechannel^{\prime}}(\rho,O)
	\leq&~ \abs{\Tr\!\left[\Delta^{(2)}_{\ensemblechannel\ensemblechannel^{\prime}}(\rho^{\otimes 2}) O^{\otimes 2}\right]} ~+~ 2~\!\abs{\Tr\!\left[\vphantom{\Delta^{(2)}_{\ensemblechannel}}{}\expval{\rho_{\channel}}_{\ensemblechannel^{\prime}}O\right]\Tr\!\left[\vphantom{\Delta^{(2)}_{\ensemblechannel}}{}\Delta_{\ensemblechannel\ensemblechannel^{\prime}}(\rho) O\right]} \\
	\leq&~ \min \left\{\norm{O}_{\infty}\left(\norm{\rho}_{1}^{2}\norm{O}_{\infty}\Epsilon_{\ensemblechannel\ensemblechannel^{\prime}}^{(2|\diamond)} + 2\frac{1}{\dsys}\Epsilon_{\ensemblechannel\ensemblechannel^{\prime}}^{(1|\diamond)}\right) ~,~ \norm{O}_{2}\left(\norm{\rho}_{2}^{2}\norm{O}_{2}\Epsilon_{\ensemblechannel\ensemblechannel^{\prime}}^{(2|2)} + 2\frac{1}{\dsys}\Epsilon_{\ensemblechannel\ensemblechannel^{\prime}}^{(1|2)}\right) \right\} ~.
\end{align}
\end{corollary}

\begin{proof}\label{proof:app:expectation_value_variance_for_reference_ensembles}
The expectation value variance inherent term follows, for specific reference ensembles $\ensemblechannel^{\prime}$, given the forms of the $t=1,2$-order twirls $\expval{\rho_{\channel}^{\otimes t}}_{\ensemblechannel^{\prime}}$, and given $\abs{\Tr[O]} \leq 1$,
\begin{align}
	\!\!\!\!\!
	\sigma_{\ensemblechannel^{\prime}}^{2}(\rho,O)
	=&~ \left\{\begin{array}{ll}
	\frac{1}{\dsys^{2}}\frac{1}{1-\frac{1}{\dsys^{2}\denv^{2}}}\left[\left[\frac{1}{\dsys^{2}\denv^{2}}\Tr^{2}\left[\rho\right] - \frac{1}{\dsys\denv}\Tr\!\left[\rho^{2}\right]\right] \Tr^{2}\left[O\right] + \frac{1}{\denv}\left[-\frac{1}{\dsys\denv}\Tr^{2}\left[\rho\right] + \Tr\!\left[\rho^{2}\right]\right]\Tr\!\left[O^{2}\right] \right] & \ensemblechannel^{\prime} = \ensemblechaar \\
	0 & \ensemblechannel^{\prime} = \ensembledep
	\end{array}\right. \!\!\!\!\!,\!\!\!\! \\
	=&~ \left\{ \begin{array}{ll}
		\mathcal{O}\!\left(\frac{1}{\dsys^{2}\denv}\right)~\norm{\rho}_{2}^{2}~\norm{O}_{2}^{2} & \ensemblechannel^{\prime}=\ensemblechaar \\
		0 & \ensemblechannel^{\prime}=\ensembledep
		\end{array} \right. ~.
\end{align}

\noindent The expectation value variance design term follows from the following norm inequalities, given the forms of the $t=1,2$-order twirls $\expval{\rho_{\channel}^{\otimes t}}_{\ensemblechannel^{\prime}}$, and given $\abs{\Tr[O]} \leq 1$, and $\ensemblechannel^{\prime} \in \{\ensemblechaar,\ensembledep\}$,
\begin{align}
	\beta_{\ensemblechannel\ensemblechannel^{\prime}}(\rho,O)
	=&~ \Tr\!\left[\Delta_{\ensemblechannel\ensemblechannel^{\prime}}^{(2)}(\rho^{\otimes 2}) O^{\otimes 2}\right] ~-~ \Tr^{2}\!\left[\Delta_{\ensemblechannel\ensemblechannel^{\prime}}(\rho) O\right] ~-~ 2~\!\Tr\!\left[\expval{\rho_{\channel}}_{\ensemblechannel^{\prime}}O\right]~\!\Tr\!\left[\Delta_{\ensemblechannel\ensemblechannel^{\prime}}(\rho) O\right] \\
	\leq &~ \abs{\Tr\!\left[\Delta^{(2)}_{\ensemblechannel\ensemblechannel^{\prime}}(\rho^{\otimes 2}) O^{\otimes 2}\right]} ~+~ 2~\!\abs{\Tr\!\left[\vphantom{\Delta^{(2)}_{\ensemblechannel}}{}\expval{\rho_{\channel}}_{\ensemblechannel^{\prime}}O\right]\Tr\!\left[\vphantom{\Delta^{(2)}_{\ensemblechannel}}{}\Delta_{\ensemblechannel\ensemblechannel^{\prime}}(\rho) O\right]} \\
	\leq&~ \min_{\frac{1}{p} + \frac{1}{q} = 1}~\Epsilon_{\ensemblechannel\ensemblechannel^{\prime}}^{(2|p)}(\rho^{\otimes 2})~\norm{O}_{q}^{2} ~+~ \min_{\frac{1}{p} + \frac{1}{q} = 1}~2~\!\abs{\Tr\!\left[\vphantom{\Delta^{(2)}_{\ensemblechannel}}{}\expval{\rho_{\channel}}_{\ensemblechannel^{\prime}}O\right]}~\Epsilon_{\ensemblechannel\ensemblechannel^{\prime}}^{(1|p)}(\rho)~\norm{O}_{q} \\
	\leq&~ \min_{\frac{1}{p} + \frac{1}{q} = 1}~\norm{\rho}_{p}^{2}~\norm{O}_{q}^{2}~\Epsilon_{\ensemblechannel\ensemblechannel^{\prime}}^{(2|p)} ~+~ \min_{\frac{1}{p} + \frac{1}{q} = 1}~2~\!\abs{\Tr\!\left[\vphantom{\Delta^{(2)}_{\ensemblechannel}}{}\expval{\rho_{\channel}}_{\ensemblechannel^{\prime}}O\right]}~\norm{\rho}_{p}~\norm{O}_{q}~\Epsilon_{\ensemblechannel\ensemblechannel^{\prime}}^{(1|p)} \\
	\leq&~ \min_{\substack{\frac{1}{p_{1}} + \frac{1}{q_{1}} = 1\\\frac{1}{p_{2}} + \frac{1}{q_{2}} = 1}}~\norm{\rho}_{p_{2}}^{2}~\norm{O}_{q_{2}}^{2}~\Epsilon_{\ensemblechannel\ensemblechannel^{\prime}}^{(2|p_{2})} ~+~ 2~\frac{1}{\dsys}~\abs{\Tr\!\left[\rho\right]}~\abs{\Tr\!\left[O\right]}~\norm{\rho}_{p_{1}}~\norm{O}_{q_{1}}~\Epsilon_{\ensemblechannel\ensemblechannel^{\prime}}^{(1|p_{1})} \\
	\leq&~ \min \left\{\vphantom{\abs{\Tr\!\left[\Delta^{(2)}_{\ensemblechannel\ensemblechannel^{\prime}}(\rho^{\otimes 2}) O^{\otimes 2}\right]}}{} \right.\\
	&\quad\quad\norm{\rho}_{1}^{2}~\norm{O}_{\infty}^{2}~\Epsilon_{\ensemblechannel\ensemblechannel^{\prime}}^{(2|\diamond)} ~+~ 2~\frac{1}{\dsys}~\abs{\Tr\!\left[\rho\right]}~\abs{\Tr\!\left[O\right]}~\norm{O}_{\infty}~\Epsilon_{\ensemblechannel\ensemblechannel^{\prime}}^{(1|\diamond)} ~, \nonumber \\
	&\quad\quad\norm{\rho}_{1}^{2}~\norm{O}_{\infty}^{2}~\Epsilon_{\ensemblechannel\ensemblechannel^{\prime}}^{(2|\diamond)} ~+~ 2~\frac{1}{\dsys}~\abs{\Tr\!\left[\rho\right]}~\abs{\Tr\!\left[O\right]}~\norm{O}_{2}~\Epsilon_{\ensemblechannel\ensemblechannel^{\prime}}^{(1|2)} ~, \nonumber \\
	&\quad\quad\norm{\rho}_{2}^{2}~\norm{O}_{2}^{2}~\Epsilon_{\ensemblechannel\ensemblechannel^{\prime}}^{(2|2)} ~~+~ 2~\frac{1}{\dsys}~\abs{\Tr\!\left[\rho\right]}~\abs{\Tr\!\left[O\right]}~\norm{O}_{\infty}~\Epsilon_{\ensemblechannel\ensemblechannel^{\prime}}^{(1|\diamond)} ~, \nonumber \\
	&\quad\quad\norm{\rho}_{2}^{2}~\norm{O}_{2}^{2}~\Epsilon_{\ensemblechannel\ensemblechannel^{\prime}}^{(2|2)} ~~+~ 2~\frac{1}{\dsys}~\abs{\Tr\!\left[\rho\right]}~\abs{\Tr\!\left[O\right]}~\norm{O}_{2}~\Epsilon_{\ensemblechannel\ensemblechannel^{\prime}}^{(1|2)} , \nonumber \\
	&~~~~~~\left. \vphantom{\abs{\Tr\!\left[\Delta^{(2)}_{\ensemblechannel\ensemblechannel^{\prime}}(\rho^{\otimes 2}) O^{\otimes 2}\right]}}{} \right\}  \nonumber \\
	\leq&~ \min \left\{\vphantom{\abs{\Tr\!\left[\Delta^{(2)}_{\ensemblechannel\ensemblechannel^{\prime}}(\rho^{\otimes 2}) O^{\otimes 2}\right]}}{} \right.\\
	&\quad\quad\norm{O}_{\infty}\left(\norm{\rho}_{1}^{2}~\norm{O}_{\infty}\Epsilon_{\ensemblechannel\ensemblechannel^{\prime}}^{(2|\diamond)} ~+~ 2~\frac{1}{\dsys}~\Epsilon_{\ensemblechannel\ensemblechannel^{\prime}}^{(1|\diamond)}\right) ~, \nonumber \\
	&\quad\quad\norm{O}_{\infty}^{2}~\Epsilon_{\ensemblechannel\ensemblechannel^{\prime}}^{(2|\diamond)} ~+~ 2~\frac{1}{\dsys}~\norm{O}_{2}~\Epsilon_{\ensemblechannel\ensemblechannel^{\prime}}^{(1|2)} ~, \nonumber \\
	&\quad\quad\norm{\rho}_{2}^{2}~\norm{O}_{2}^{2}~\Epsilon_{\ensemblechannel\ensemblechannel^{\prime}}^{(2|2)} ~~+~ 2~\frac{1}{\dsys}~\norm{O}_{\infty}~\Epsilon_{\ensemblechannel\ensemblechannel^{\prime}}^{(1|\diamond)} ~, \nonumber \\
	&\quad\quad\norm{O}_{2}\left(\norm{\rho}_{2}^{2}~\norm{O}_{2}~\Epsilon_{\ensemblechannel\ensemblechannel^{\prime}}^{(2|2)} ~~+~ 2~\frac{1}{\dsys}~\Epsilon_{\ensemblechannel\ensemblechannel^{\prime}}^{(1|2)}\right) , \nonumber \\
	&~~~~~~\left. \vphantom{\abs{\Tr\!\left[\Delta^{(2)}_{\ensemblechannel\ensemblechannel^{\prime}}(\rho^{\otimes 2}) O^{\otimes 2}\right]}}{} \right\} ~. \nonumber
\end{align}
\end{proof}

\newpage

\section{Relationships between Additive and Relative Errors}\label{app:proof-lemma-errors}

In these appendices, we derive relationships between additive and relative errors for quantum channel ensemble twirls.\\

Given an ensemble of quantum channels $\ensemblechannel$, let an additive $\varepsilon_{\textnormal{add}}$-approximate Depolarize $t$-design in diamond norm be
\begin{equation}
	\norm{{\Tau}_{\ensemblechannel}^{(t)} - {\Tau}_{\ensembledep}^{(t)}}_{\diamond}\leq \varepsilon_{\textnormal{add}}~,
\end{equation}
and a relative $\varepsilon_{\textnormal{add}}$-approximate Depolarize $t$-design be
\begin{equation}\label{eq:-relative-op}
	(1-\varepsilon_{\textnormal{rel}}){\Tau}_{\ensembledep}^{(t)}
	~\preceq~ {\Tau}_{\ensemblechannel}^{(t)}
	~\preceq~
	(1+\varepsilon_{\textnormal{rel}}){\Tau}_{\ensembledep}^{(t)}
	\quad\Longleftrightarrow \quad
	-\varepsilon_{\textnormal{rel}}{\Tau}_{\ensembledep}^{(t)}
	~\preceq~
	{\Tau}_{\ensemblechannel}^{(t)}-{\Tau}_{\ensembledep}^{(t)}
	~\preceq~
	\varepsilon_{\textnormal{rel}}{\Tau}_{\ensembledep}^{(t)}~,
\end{equation}
or alternatively, using the Choi-Jamiolkowski isomorphism, given the $d$-dimensional normalized maximally entangled state $\ket{\Phi} = \frac{1}{\sqrt{\dsys}}\sum_{\alpha \in [\dsys]}\ket{\alpha\alpha} \in \spacesys \otimes \spacesys \to \Phi = \ket{\Phi}\bra{\Phi} \in \mathcal{B}[\spacesys \otimes \spacesys]$,
\begin{equation}
	-\varepsilon_{\textnormal{rel}}({\Tau}_{\ensembledep}^{(t)}\otimes \idsys^{\otimes t})(\Phi^{\otimes t})
	~\preceq~
	({\Tau}_{\ensemblechannel}^{(t)} \otimes \idsys^{\otimes t})(\Phi^{\otimes t})-({\Tau}_{\ensembledep}^{(t)}\otimes \idsys^{\otimes t})(\Phi^{\otimes t})
	~\preceq~
	\varepsilon_{\textnormal{rel}}({\Tau}_{\ensembledep}^{(t)}\otimes \idsys^{\otimes t})(\Phi^{\otimes t})~. \label{eq:-relative-Choi}
\end{equation}

Using these definitions, we will prove that the following Lemma holds.

\begin{lemma}[Additive versus relative error]
Any approximate Depolarize $t$-design with relative error $\varepsilon_{\textnormal{rel}}$ is a $\varepsilon_{\textnormal{add}}$-approximate Depolarize $t$-design with an additive error $\varepsilon_{\textnormal{add}}=2\varepsilon_{\textnormal{rel}}$. Conversely, any approximate Depolarize $t$-design with additive error $\varepsilon_{\textnormal{add}}$ is a $\varepsilon_{\textnormal{rel}}$-approximate Depolarize $t$-design with a relative error $\varepsilon_{\textnormal{rel}}=\dsys^{2t} \varepsilon_{\textnormal{add}}$.
\end{lemma}

\begin{proof}\label{app:proof:proof-lemma-errors}
To simplify our derivation, let us define the difference of moment operators as,
	\begin{equation}
		{\Delta}_{\ensemblechannel\ensembledep}^{(t)} = {\Tau}_{\ensemblechannel}^{(t)} - {\Tau}_{\ensembledep}^{(t)} ~.
	\end{equation}
Then, by definition of the diamond norm as a maximization over pure states $\ket{\Psi} \in \mathcal{H}^{\otimes t} \otimes \mathcal{H}^{\otimes t}$ and the one norm in terms of projectors $\Pi_{\pm}$ onto the positive and negative eigenspaces of its arguments, we have
\begin{align}
	\norm{{\Delta}_{\ensemblechannel\ensembledep}^{(t)}}_{\diamond}&=\max_{\Psi}\norm{({\Delta}_{\ensemblechannel\ensembledep}^{(t)}\otimes \idsys^{\otimes t})(\Psi)}_1\\
	&=\max_{0\leq \Pi_{\pm} \leq \idsys^{\otimes 2t}}\max_{\Psi}\Big[\Tr[\Pi_{+}({\Delta}_{\ensemblechannel\ensembledep}^{(t)}\otimes \idsys^{\otimes t}(\Psi)]
	- \Tr[\Pi_{-}({\Delta}_{\ensemblechannel\ensembledep}^{(t)}\otimes \idsys^{\otimes t}(\Psi)]\Big]~.
\end{align}
If we assume relative error $\varepsilon_{\textnormal{rel}}$, then using Eq.~\eqref{eq:-relative-op} along with the fact that $\Pi_{\pm} ~\preceq~ \idsys^{\otimes 2t}$ we obtain
\begin{equation}
	\Tr[\Pi_{\pm}({\Delta}_{\ensemblechannel\ensembledep}^{(t)}\otimes \idsys^{\otimes t}(\Psi)]
	~\leq~
	\varepsilon_{\textnormal{rel}} \Tr[\Pi_{\pm}({\Tau}_{\ensembledep}^{(t)}\otimes \idsys^{\otimes t}(\Psi)]
	~\leq~
	\varepsilon_{\textnormal{rel}} \Tr[({\Tau}_{\ensembledep}^{(t)}\otimes \idsys^{\otimes t}(\Psi)]=\varepsilon_{\textnormal{rel}}~.
\end{equation}
From this bound, we see that,
\begin{equation}
	\norm{{\Delta}_{\ensemblechannel\ensembledep}^{(t)}}_{\diamond}\leq 2\varepsilon_{\textnormal{rel}}~.
\end{equation}
and hence relative error $\varepsilon_{\textnormal{rel}}$ implies additive error
\begin{equation}
	\varepsilon_{\textnormal{add}}=2\varepsilon_{\textnormal{rel}}~.
\end{equation}

Next, let us define the states as those output by twirls over the quantum channel ensembles acting on $\Phi$,
\begin{equation}
	\Phi_{\ensemblechannel}=({\Tau}_{\ensemblechannel}^{(t)} \otimes \idsys^{\otimes t})(\Phi^{\otimes t})~,\quad\quad \Phi_{\ensembledep}=({\Tau}_{\ensembledep}^{(t)}\otimes \idsys^{\otimes t})(\Phi^{\otimes t})~.
\end{equation}
One can readily verify that the depolarized state is
\begin{equation}\label{eq:rho-depol-proof}
	\Phi_{\ensembledep}=\frac{1}{\dsys^{2t}}\idsys^{\otimes 2t}~,
\end{equation}
with a smallest eigenvalue of
\begin{equation}\label{eq:-smallest-eig}
	\lambda_{\textnormal{min}}(\Phi_{\ensembledep})=\frac{1}{\dsys^{2t}}~.
\end{equation}~\\~\\

Hence, if we have an approximate Depolarize $t$-design with additive error
$\varepsilon_{\textnormal{add}}$
\begin{equation}
	\norm{{\Delta}_{\ensemblechannel\ensembledep}^{(t)}}_{\diamond}\leq \varepsilon_{\textnormal{add}}~,
\end{equation}
this implies
\begin{equation}
	\varepsilon_{\textnormal{add}} \geq \norm{{\Delta}_{\ensemblechannel\ensembledep}^{(t)}}_{\diamond}=\max_{\Psi}\norm{({\Delta}_{\ensemblechannel\ensembledep}^{(t)}\otimes \idsys^{\otimes t})(\Psi)}_1\geq \norm{\Phi_{\ensemblechannel}-\Phi_{\ensembledep}}_1\geq \norm{\Phi_{\ensemblechannel}-\Phi_{\ensembledep}}_\infty~.
\end{equation}
Then, using the minimum eigenvalue of the depolarized state in Eq.~\eqref{eq:-smallest-eig}, we find
\begin{equation}
	-\varepsilon_{\textnormal{add}}\idsys^{\otimes 2t}
	~\preceq~
	\Phi_{\ensemblechannel}-\Phi_{\ensembledep}
	~\preceq~
	\varepsilon_{\textnormal{add}}\idsys^{\otimes 2t}~,
\end{equation}
which can be combined with the equality $\idsys^{\otimes 2t}=\dsys^{2t}\Phi_{\ensembledep}$ (from Eq.~\eqref{eq:rho-depol-proof}), to obtain
\begin{equation}
	-\varepsilon_{\textnormal{add}}\dsys^{2t}\Phi_{\ensembledep}
	~\preceq~
	\Phi_{\ensemblechannel}-\Phi_{\ensembledep}
	~\preceq~
	\varepsilon_{\textnormal{add}}\dsys^{2t}\Phi_{\ensembledep}~.
\end{equation}
As such, additive error $\varepsilon_{\textnormal{add}}$ implies relative  error
\begin{equation}
	\varepsilon_{\textnormal{rel}} = \dsys^{2t}\varepsilon_{\textnormal{add}}~.
\end{equation}

\end{proof}

\end{document}